\pdfoutput=1
\documentclass[cernpreprint, atlasdraft=true, texlive=2016, USenglish, texmf]{atlasdoc}
 
\usepackage{atlaspackage}
\usepackage{adjustbox}
\usepackage{atlascover}
 
\usepackage{atlasbiblatex}
 
\usepackage[jetetmiss=true, hepparticle=true]{atlasphysics}

\graphicspath{{logos/}{figures/}}
 
\usepackage{ANA-EXOT-2019-07-PAPER-defs}

\AtlasTitle{
Search for single production of a vector-like $T$~quark decaying into
a Higgs boson and top quark with fully~hadronic final states using the ATLAS detector
}
\AtlasJournal{PRD}
\PreprintIdNumber{CERN-EP-2021-249}
\AtlasAbstract{
A search is made for a vector-like $T$~quark decaying into a Higgs boson and a top quark
in 13~\TeV\ proton--proton collisions using the ATLAS detector at the Large Hadron Collider
with a data sample corresponding to an integrated luminosity of 139~fb$^{-1}$.
The Higgs-boson and top-quark candidates are identified in the all-hadronic decay mode, where
$H\to b\bar{b}$\ and $t\to b W \to b q \bar{q}^\prime$ are reconstructed as large-radius jets.
The candidate Higgs boson, top quark, and associated B-hadrons are identified
using tagging algorithms.
No significant excess is observed above the background, so
limits are set on the production cross-section of a singlet $T$ quark
at 95\% confidence level, depending on the mass, $\mTVLQ$, and
coupling, $\kappaT$, of the vector-like $T$~quark to Standard Model particles.
In the considered mass range between 1.0 and 2.3~\TeV, the upper limit on the allowed coupling
values increases with \mTVLQ\ from a minimum value of 0.35 for $1.07 < \mTVLQ < 1.4$~\TeV\ to
1.6 for $\mTVLQ = 2.3$~\TeV.
}
 
\author{The ATLAS Collaboration}
 
\AtlasRefCode{EXOT-2019-07}

\AtlasRefCode{EXOT-2019-07}
 
\AtlasNote{EXOT-2019-07}

\AtlasJournalRef{Phys. Rev. D 105 (2022) 092012}
 
\AtlasDOI{10.1103/PhysRevD.105.092012}
\hypersetup{pdftitle={ATLAS document},pdfauthor={The ATLAS Collaboration}}
 
\begin{document}
 
\maketitle
 
\tableofcontents

\section{Introduction}
\label{sec:intro}
 
The discovery of the Higgs boson~\cite{HIGG-2012-27,CMS-HIG-12-028}
and measurements of the Higgs-boson
couplings~\cite{
HIGG-2018-57,
CMS-HIG-17-031,
CMS-HIG-19-001,
CMS-HIG-19-002}\
by the ATLAS and CMS collaborations
confirm that the Standard Model of particle physics (SM) is an accurate
description of nature at currently accessible energy scales.
However,
the SM still leaves many questions unanswered and is therefore not a complete
theory.
For example, radiative corrections to the Higgs-boson
propagator from top-quark loops lead to a quadratic divergence in the
mass of the Higgs boson~\cite{Susskind:1978ms}.
The mechanism to cancel out the contribution from the top quark requires an
unreasonable
degree of fine-tuning to produce the observed 125~\GeV\ Higgs boson.
This so-called hierarchy problem is often considered to indicate
that new physics naturally
cancels out the divergent contributions to the Higgs-boson mass.
 
Vector-like quarks are hypothetical spin-1/2 particles that arise in
various models that address problems in the SM such as the hierarchy
problem.
Vector-like quarks are color-triplets whose left- and right-handed chiralities
transform in the same way under
weak-isospin~\cite{delAguila:1982fs,AguilarSaavedra:2009es}.
In Little Higgs~\cite{ArkaniHamed:2002qy,Schmaltz:2005ky} and
Composite Higgs~\cite{Kaplan:1983sm,Agashe:2004rs} models, the Higgs
boson is naturally light because it is a pseudo Nambu--Goldstone boson
arising from a spontaneously broken global
symmetry~\cite{Hill:2002ap}.
Vector-like quarks arise naturally in such models.
Unlike the chiral current of SM quarks, vector-like quarks
have a pure vector current in the
Lagrangian.
In addition, vector-like quarks do not acquire mass by interacting with
the Higgs field, so they are not excluded by measurements of Higgs-boson properties.
 
In these models, vector-like quarks are expected to couple preferentially to
third-generation quarks~\cite{delAguila:1982fs,Aguilar-Saavedra:2013wba} and can have
both neutral-current and charged-current decays.
An up-type vector-like $T$~quark with charge $+2/3$ can decay into $Wb$,
$Zt$, or \Ht, while a down-type $B$~quark with charge $-1/3$ can
decay into $Wt$, $Zb$, or $Hb$\ (and the charge conjugate states).
To be consistent with
results from precision electroweak measurements, the mass-splitting
between vector-like quarks belonging to the same SU(2) multiplet should be small~\cite{Aguilar-Saavedra:2013qpa}, preventing cascade decays such
as $T\to WB$.
Couplings between the vector-like quarks and the first- and
second-generation quarks are not excluded~\cite{Atre:2008iu,Atre:2011ae}, but
they are expected to be small.
 
Vector-like quarks can be produced singly or in pairs in proton--proton ($pp$) collisions.
There have been numerous searches for the pair production of vector-like quarks~\cite{
CMS-EXO-11-005,
CMS-EXO-11-099,
EXOT-2011-22,
CMS-B2G-13-005,
CMS-B2G-13-006,
CMS-B2G-12-015,
CMS-B2G-14-002,
EXOT-2014-10,
EXOT-2013-18,
EXOT-2016-15,
CMS-B2G-16-024,
CMS-B2G-17-003,
CMS-B2G-17-011,
CMS-B2G-17-012,
EXOT-2016-14,
EXOT-2016-13,
EXOT-2017-34,
EXOT-2017-17,
CMS-B2G-18-005}\
that have excluded $T$-quark
masses below 1.37~\tev\ at 95\%\ confidence level (CL) for a variety of decay modes.
For $T$-quark masses above  $\sim$1~\tev, vector-like quarks would mainly be produced singly
if the couplings to SM particles were sufficiently large.
Searches for single production of $T$~quarks have placed limits on $T$-quark
production cross-sections for $T$-quark masses between 1 and 2 \TeV\
at 95\%\ CL for various SM couplings~\cite{
EXOT-2014-12,
EXOT-2014-13,
CMS-B2G-15-008,
CMS-B2G-16-001,
CMS-B2G-16-005,
CMS-B2G-16-013,
CMS-B2G-17-007,
CMS-B2G-17-018,
CMS-B2G-18-003}.
For these higher masses, where single vector-like quark production is expected to
dominate~\cite{Aguilar-Saavedra:2013qpa},
the cross-section depends on the vector-like quark mass scale as
well as the couplings to SM particles.
 
\begin{figure}[htbp]
\centering
\includegraphics[width=0.4\linewidth]{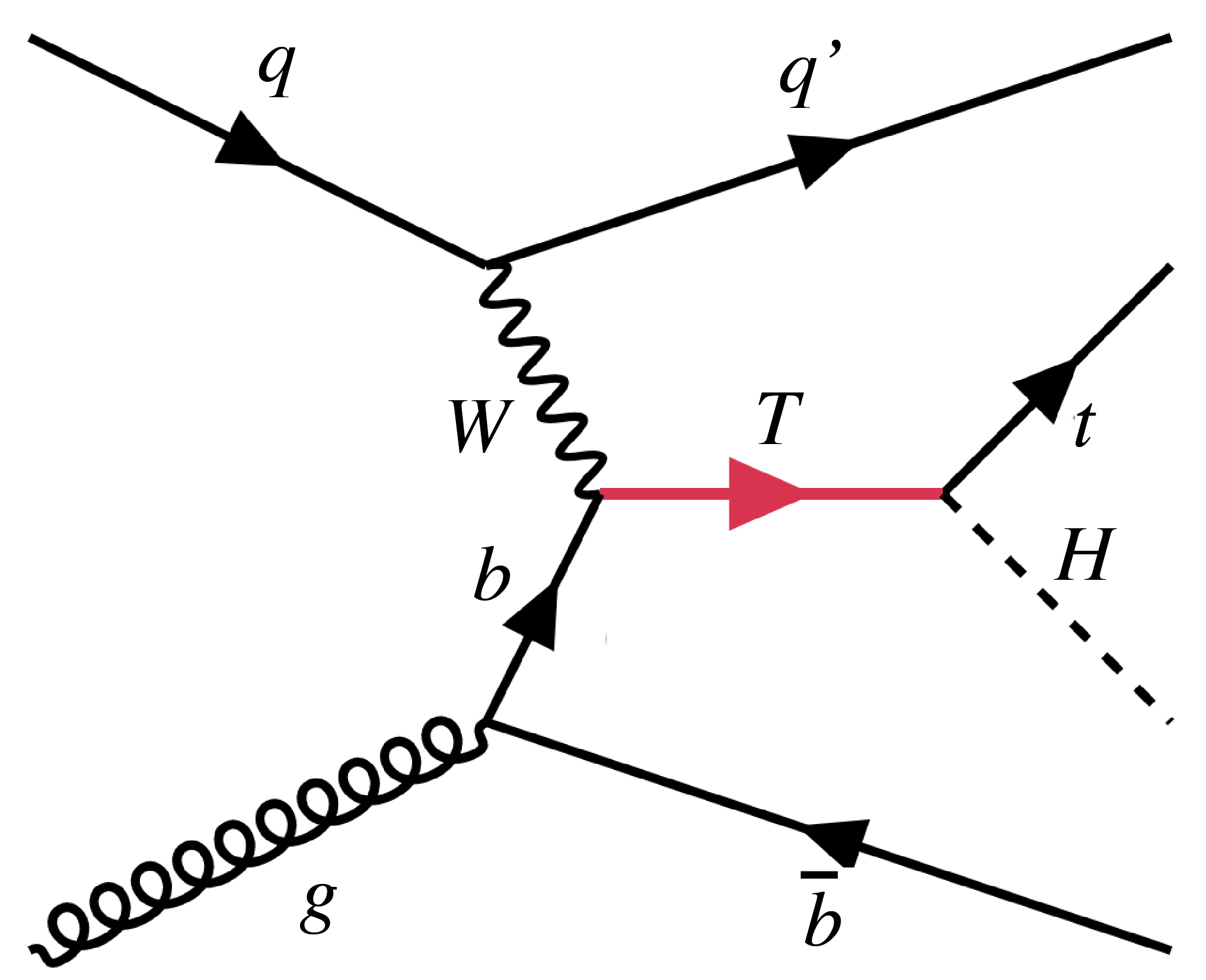}
\caption{{A Feynman diagram illustrating the $W$-mediated production of a single $T$~quark,
shown in red,
decaying into a top quark and Higgs boson.}}
\label{fig:Feynman}
\end{figure}

This paper reports a search for single production of a singlet vector-like $T$~quark
in 13~\TeV\ $pp$ collisions produced at the Large Hadron Collider (LHC) and recorded
by the ATLAS detector in a 139~fb$^{-1}$\ data sample.
The search targets $T$~quarks decaying into a SM Higgs boson and a top quark,
$T\to \Ht$, where
both the Higgs boson and top quark decay hadronically and are reconstructed as
jets of particles.
A Feynman diagram for this process is shown in Figure~\ref{fig:Feynman}.
The mass, \mTVLQ, of the $T$~quark
and the overall coupling factor, \kappaT, to the SM
$W$~boson, $Z$~ boson, and Higgs~boson~\cite{Buchkremer:2013bha}\ are unknown
parameters.
There are also three additional parameters, $\xi_W$, $\xi_Z$, and $\xi_H$,
that determine the $T$-quark branching ratios.
In this analysis, the asymptotic limit of these parameters -- as \mTVLQ\ goes to infinity --
is assumed,
leading to branching ratios of $1/2$, $1/4$, and $1/4$ for
$T\to \Wboson b$,
$T\to \Ht$, and
$T\to \Zboson t$, respectively.
In this model,
the unknown parameters \mTVLQ\ and \kappaT\ define the expected $T$-quark cross-section and
resonance lineshape.
The search assumes this signal model in the interpretation of the data.
 
The $T$~quark is assumed to be a weak singlet state in this analysis;
if additional multiplets of vector-like quarks are assumed, the possible final states
and branching ratios require an approach involving simultaneous
consideration of several final states~\cite{Aguilar-Saavedra:2013qpa}, which is beyond the scope of this paper.
 
The results reported here significantly extend the sensitivity to events in which a singly produced $T$ quark
decays to \Ht\ followed by the hadronic decays $H\rightarrow b\bar{b}$\ and $t\rightarrow bW$.
The use of fully hadronic decays allows the
direct reconstruction of the $T$-quark final state, increasing the expected
signal-to-background ratio
in the signal region defined for the search.
Compared with the most sensitive prior search~\cite{CMS-B2G-18-003},
this search uses ${\sim}4$\ times more integrated luminosity,
its sensitivity is improved by
using tagging techniques resulting in a signal-to-background improvement of ${\sim}3$,
and it uses a data-driven multijet background estimate that reduces the uncertainty in
the background estimate by an order of magnitude.
 
This fully hadronic final state is of particular interest for vector-like quark masses above 1~\tev.
The resulting high-\pT jets from the top-quark and Higgs-boson are ``boosted'',
so that the decay products of the top quark and Higgs boson are collimated and
captured in two large-radius (\largeR) jets.
This final state has the largest branching fraction of all the potential \Ht\ decay modes
and the \largeR jets can be identified as either Higgs-boson or
top-quark candidates through tagging algorithms that use
the substructure within the jet~\cite{JETM-2018-03,FTAG-2018-01}.
In addition, bottom-quark jet identification ($b$-tagging) provides high background
rejection with high efficiency given
the three bottom-quark jets coming from the $H\rightarrow b\bar{b}$\ and $t\rightarrow W b$ decays.
Assuming the existence of single $T$-quark production,
the signal would appear as an excess of events with \Ht\ invariant masses around the $T$-quark
mass for values of $\kappaT \lesssim 0.5$.
Above this $\kappaT$, the $\Ht$\ invariant mass distribution broadens to masses below the $T$-quark mass as \kappaT\
increases due to the convolution of increasing width and partonic densities.
 
The largest backgrounds come from boosted
top-quark pair production and multijet events arising from the
production of lighter high-\pt{} quarks ($u$, $d$, $s$, $c$, and $b$) and/or
gluons.
The
ATLAS~\cite{TOPQ-2011-07,
TOPQ-2012-08,
TOPQ-2013-07,
TOPQ-2014-15,
TOPQ-2015-06,
TOPQ-2015-07,
TOPQ-2016-04,
TOPQ-2016-01,
TOPQ-2016-09}
and CMS~\cite{CMS-TOP-11-013,CMS-TOP-14-008,CMS-TOP-12-028,CMS-TOP-14-012,CMS-TOP-16-008,CMS-TOP-14-013,CMS-TOP-16-007,CMS-TOP-16-008,CMS-TOP-16-014,CMS-TOP-17-002,CMS-TOP-18-004,CMS-TOP-18-013}\
collaborations have published measurements of the $\ttbar{}$
differential cross-sections at center-of-mass energies of
$\sqrt{s}=7$~\TeV{}, 8~\TeV{}, and 13~\TeV{} in
$pp$ collisions.
The measured cross-section for the production of
top quarks with $\pt > 300$~\GeV\ is ${\sim}20\%$\ lower than predicted by
perturbative quantum chromodynamics (QCD) calculations performed at
next-to-leading-order (NLO) in the strong coupling constant \alphas.
A control sample of fully reconstructed high-$\pt$\ top-quark pairs is used with
Monte Carlo (MC) models to normalize the expected background from top-quark pairs in the
\Ht\ candidate sample.
The multijet background is estimated using data-driven techniques
developed for
studies of events containing boosted top quarks~\cite{TOPQ-2016-01}.
 
This paper is organized as follows.
Section~\ref{sec:detector}\ describes the ATLAS detector and
Section~\ref{sec:datamcsamples}\  describes the datasets and
MC samples that are used in this analysis.
Section~\ref{sec:objectselection} describes the object definition
and event selection, while Section~\ref{sec:backgrounds}\ summarizes
the estimation of SM backgrounds to the $T$-quark signal.
The systematic uncertainties are presented in Section~\ref{sec:systematics} and Section~\ref{sec:result}\ presents the results of the search.
Conclusions are drawn in Section~\ref{sec:conclusion}.
 

\section{ATLAS detector}
\label{sec:detector}
The ATLAS detector~\cite{PERF-2007-01} at the
LHC is centered on the $pp$\ collision point and covers nearly the whole $4\pi$\
solid angle.\footnote{ATLAS uses a right-handed coordinate system with
its origin at the nominal interaction point (IP) in the center of
the detector and the $z$-axis along the beam pipe.  The $x$-axis
points from the IP to the center of the LHC ring, and the $y$-axis
points upwards.  Cylindrical coordinates $(r,\phi)$ are used in the
transverse plane, $\phi$ being the azimuthal angle around the
$z$-axis.  The pseudorapidity is defined in terms of the polar angle
$\theta$ as $\eta = -\ln \tan(\theta/2)$.  Angular distance is
measured in units of $\Delta R \equiv \sqrt{(\Delta\eta)^{2} +
(\Delta\phi)^{2}}$.}  It consists of an inner tracking detector
surrounded by a \SI{2}{\tesla} superconducting solenoid,
electromagnetic and hadronic calorimeters, and a muon spectrometer
incorporating three large superconducting toroid magnets.
 
The inner detector, including the insertable B-layer added as a new innermost layer in
2014~\cite{ATLAS-TDR-19,ATLAS_IBL}, provides charged-particle
tracking information from a pixel detector and silicon microstrip detector in
the pseudorapidity range $|\eta| < 2.5$ and a transition radiation
tracker covering $|\eta| < 2.0$.
 
The calorimeter system covers the pseudorapidity range $|\eta| < 4.9$
and measures the positions and energies of electrons, photons, and
charged and neutral hadrons.  Within the region $|\eta|< 3.2$,
electromagnetic calorimetry is provided by barrel and endcap
high-granularity lead and liquid-argon sampling calorimeters.  The
hadronic sampling calorimeter uses either scintillator tiles or liquid
argon as active material and steel, copper or tungsten as absorber.
 
The muon spectrometer comprises separate trigger and high-precision
tracking chambers measuring the tracks of muons in a magnetic field
generated by superconducting air-core toroid magnets.  The precision
chamber system covers the region $|\eta| < 2.7$ , while the muon
trigger system covers the range $|\eta| < 2.4$.
 
A two-level trigger system is used to select which events to save for
offline analysis~\cite{TRIG-2016-01}.  The first level is
implemented in hardware/firmware and uses a subset of the detector
information to reduce the event rate from the \SI{40}{\MHz} proton bunch crossings to less
than \SI{100}{\kHz}.  This is followed by a software-based high-level
trigger that reduces the event rate to
approximately \SI{1}{\kHz}.
An extensive software suite~\cite{ATL-SOFT-PUB-2021-001} is used in the
reconstruction and analysis of real and simulated data, in detector operations,
and in the trigger and data acquisition systems of the experiment.
 

\section{Data and simulated samples}
\label{sec:datamcsamples}

This analysis studies $pp$ collisions with a center-of-mass energy of
$\sqrt{s}=13$\,\TeV{} recorded by the ATLAS detector between 2015 and
2018.  Only data-taking periods in which all the subdetectors were
operational are considered.  The dataset corresponds to an integrated
luminosity of \lumitot{}~\cite{ATLAS-CONF-2019-021}, measured using the LUCID-2
detector~\cite{LUCID2}.
The events used in this analysis were
collected by a set of triggers requiring at least one
anti-$k_{t}$ jet~\cite{akt1,Cacciari:2011ma}\ with a jet radius
parameter of $R=1.0$~\cite{TRIG-2016-01}.
The maximum \pT{} threshold value of these triggers was 480~\GeV{},
which was found to be fully
efficient when requiring the offline reconstruction of at least
one \largeR{} jet with $\pT > 500$~\GeV{} and $|\eta|<2.0$, as
described in Section~\ref{sec:objectselection}.

The main backgrounds for this search are from \ttbar{} and multijet
events.
There are also small contributions from single-top-quark
and \ttbarX{} ($X= W,Z,H$) events.  The multijet background is
estimated using a data-driven method described in
Section~\ref{sec:backgrounds}, while the other backgrounds, as well as
the $T$-quark signal events, are estimated with MC simulations
as described below.
The multijet background estimate also includes backgrounds arising from
electroweak and QCD processes such as $W/Z+\text{jets}$\ production.

The $T$-quark signal samples were produced at leading order,
using the \MGNLO[]\ MC generator~\cite{Alwall:2011uj} to generate the hard
interaction and the \PYTHIA[8]\ generator for parton showering and hadronization.
The parton distribution function (PDF) set used is \NNPDF[3.0nlo]~\cite{Ball:2014uwa}.
Both $W$-mediated and $Z$-mediated production contribute to single
$T$-quark production and were included in the
MC event generation, with the $Z$-mediated process having a cross-section
five times smaller than the $W$-mediated process and comprising
less than 1\%\ of the total yield after the event selection described in
Section~\ref{sec:objectselection}.
The matrix elements were calculated using the phenomenological model given in
Ref.~\cite{Buchkremer:2013bha}.
These include all tree-level processes,
ensuring the inclusion of both resonant and nonresonant single $T$-quark production modes.
The decay channel considered is $T \to Ht$, with \mTVLQ\ and \kappaT\ as unknown parameters.
The three additional parameters, $\xi_W$, $\xi_Z$, and $\xi_H$,
that determine the $T$-quark branching ratios are set to the asymptotic limit in \mTVLQ,
leading to branching ratios of $1/2$, $1/4$, and $1/4$ for
$T\to \Wboson b$,
$T\to \Ht$, and
$T\to \Zboson t$, respectively.
In order to accurately model the change in cross-section and
lineshape as \mTVLQ\ and \kappaT\ are varied,
MC samples were created for a variety of
mass and coupling values, with \mTVLQ\ ranging from 1.0 to 2.3~\TeV\ in steps of 0.1~\TeV\
and $\kappaT$\ ranging from 0.1 to 1.6 in steps of 0.05 for $\kappaT < 0.5$\ and 0.1 for
larger \kappaT.
All signal samples are normalized to cross-sections that have been calculated at NLO
in QCD~\cite{VLQatNLO}.
These cross-sections are computed in a $T$-quark narrow-width approximation and a correction factor is applied~\cite{Roy_2020}\ to account for finite-width effects.

For all MC samples, the masses of the top quark (\mt) and Higgs boson were
set to  172.5\,\GeV{} and 125.0\,\GeV{}, respectively.
The production of \ttbar{} events was modeled using the
\POWHEGBOX[v2]~\cite{Frixione:2007nw,Nason:2004rx,Frixione:2007vw,Alioli:2010xd}\
generator.
This provides matrix elements at NLO with the \NNPDF[3.0nlo]~PDF.
In addition, the \hdamp parameter, which controls the matching of the
matrix element to the parton shower in \POWHEG{} and
effectively regulates the high-\pt\ radiation against which the
\ttbar\ system recoils, was set to $1.5\times \mt$~\cite{ATL-PHYS-PUB-2016-020}.
The functional form of the renormalization and factorization scales was
set to the default scale $\sqrt{m_{\textrm{top}}^2 + p_{T,{\textrm{top}}}^{2}}$.
The \PYTHIA[8.230]~\cite{Sjostrand:2014zea} parton-shower and
hadronization models were employed, using a set of tuned parameter values
called the A14 tune~\cite{ATL-PHYS-PUB-2014-021}, and the \NNPDF[2.3lo]\ set of
PDFs~\cite{Ball:2012cx}.  The decays of bottom and charm hadrons were
simulated using the \EVTGEN[1.6.0] program~\cite{Lange:2001uf}.
 
The \ttbar{} sample is normalized to the cross-section prediction at
next-to-next-to-leading order (NNLO) in QCD including the resummation
of next-to-next-to-leading logarithmic (NNLL) soft-gluon terms
calculated using
\TOPpp[2.0]~\cite{Beneke:2011mq,Cacciari:2011hy,Baernreuther:2012ws,Czakon:2012zr,Czakon:2012pz,Czakon:2013goa,Czakon:2011xx}.
For $pp$\ collisions at a center-of-mass energy
of $\rts = \SI{13}{\TeV}$, this cross-section is
$\sigma(\ttbar)_{\textrm{NNLO+NNLL}} = \mathrm{832\pm51~fb}$.
The cross-section uncertainties due to the PDF and \alphas are calculated using the PDF4LHC
prescription~\cite{Butterworth:2015oua} with the \MSTW[nnlo] 68\% CL~\cite{Martin:2009iq,Martin:2009bu},
\CT[10nnlo]~\cite{Lai:2010vv,Gao:2013xoa} and \NNPDF[2.3lo] 5f FFN~\cite{Ball:2012cx}
PDF sets, and are added in quadrature to the effect of the scale uncertainty.

The uncertainty due to initial-state radiation (ISR) was
estimated by varying the Var3c A14 tune, renormalization scale \muR, factorization scale \muF, and the \hdamp parameter independently.
The Var3c A14 tune variation corresponds to the variation of
\alphas for ISR in the A14 tune.
The renormalization scale and factorization scales were varied by factors of 0.5 and 2.0
corresponding to an increase and decrease in ISR, respectively.
The \hdamp uncertainty is measured by comparing the nominal \ttbar{} sample with a sample using $h_{damp} = 3m_{top}$.
The impact of final-state radiation (FSR) uncertainties was evaluated by
increasing and decreasing the renormalization scale for emissions from the
parton shower by a factor of two.
 
The impact of using a different parton-shower and hadronization model was evaluated
by comparing the nominal \ttbar{} sample with a sample that was also generated by
\POWHEGBOX[v2] but used \HERWIG[7.1.3]~\cite{Bahr:2008pv,Bellm:2015jjp}\ instead of \PYTHIA[8.230]
for parton showering and hadronization. The \HERWIG[7.1] default set
of tuned parameters~\cite{Bellm:2015jjp,Bellm:2017jjp} and the \MMHT[lo] PDF set
\cite{Harland-Lang:2014zoa}\ were employed.
 
To assess the uncertainty in the matching of NLO matrix elements to the
parton shower, the \POWHEG{} \ttbar{} sample was compared with a sample of events
generated with \MGNLO[2.6.0] but retaining the
\PYTHIA[8.230] parton-shower and hadronization models.
The \MGNLO[]{} calculation used the
\NNPDF[3.0nlo]\ set of PDFs, as in the \POWHEG{} sample, and \PYTHIA[8] again used
the A14 tune and
the \NNPDF[2.3lo]\ set of PDFs.

The production of a single top quark in association with a $W$ boson ($tW$) was
modeled using
the \POWHEGBOX[v2] generator~\cite{Re:2010bp,Nason:2004rx,Frixione:2007vw,Alioli:2010xd}
at NLO in QCD with the five-flavor scheme and the
\NNPDF[3.0nlo]\ set of PDFs.
The diagram removal scheme~\cite{Frixione:2008yi} was used to remove interference and
overlap with \ttbar\ production.

The \PYTHIA[8.230] parton-shower and hadronization models
were employed,  using the A14
tune and the \NNPDF[2.3lo] set of
PDFs.
The inclusive cross-section for $tW$
production was corrected to the theory prediction calculated at NLO in
QCD with NNLL soft-gluon
corrections~\cite{Kidonakis:2010ux,Kidonakis:2013zqa}.
 
Single-top-quark $t$-channel production was modeled using the
\POWHEGBOX[v2]~\cite{Frederix:2012dh,Nason:2004rx,Frixione:2007vw,Alioli:2010xd}
generator at NLO in QCD using the four-flavor scheme and the
corresponding \NNPDF[3.0nlo]\ set of PDFs.  Parton showering and
hadronization were performed with \PYTHIA[8.230], using the
A14 tune and the \NNPDF[2.3lo]\ set of
PDFs.  The inclusive cross-section was corrected to
the theory prediction calculated at NLO in QCD
with the \HATHOR[2.1] generator~\cite{Aliev:2010zk,Kant:2014oha}.
Single-top-quark $s$-channel MC events were not generated because the
cross-section for this process is much smaller than that for $tW$ production and the
$t$-channel processes.  However, the $s$-channel process makes a
small contribution to the data-driven multijet background estimate and is therefore
partially accounted for.
The production of SM \tH\ is treated in a similar manner,
as the background yield is negligible due to a combination of small cross-section and
low yield in the high-\pT\ region.
 
The production of \ttbar\ in association with a Higgs boson ($\ttbar+H$) was modeled by the
\POWHEGBOX[v2]~\cite{Frixione:2007nw,Nason:2004rx,Frixione:2007vw,Alioli:2010xd,Hartanto:2015uka}
generator at NLO.
The production of \ttbar\ in association with a \Wboson or \Zzero\ boson was modeled
using the \MGNLO[2.3.3] generator at NLO.
Parton showering and hadronization for these processes was performed by \PYTHIA[8.210]\ and
the decays of bottom and charm hadrons were simulated using \EVTGEN[1.2.0].
The cross-sections for the $\ttbar+W/Z/H$ processes were calculated
using \MGNLO[] at NLO QCD and NLO EW accuracies using
Ref.~\cite{deFlorian:2016spz}.  The $\ttbar+Z$\ cross-section was
corrected to take into account
contributions from off-shell \Zzero\ bosons with masses down to 5~\GeV.
The predicted values of the
cross-sections at 13~\TeV\ are $0.88^{+0.09}_{-0.11}~\text{pb}$,
$0.60^{+0.08}_{-0.07}~\text{pb}$, $0.51^{+0.04}_{-0.05}~\text{pb}$ for
$\ttbar+Z$, $\ttbar+W$, and $\ttbar+H$, respectively, where the uncertainties
reflect QCD scale variations.
 
The effect of multiple interactions in the same and neighboring bunch
crossings (\pileup) was modeled by overlaying the simulated hard-scattering event
with inelastic $pp$\ events generated with
the \PYTHIA[8.186] MC generator~\cite{Sjostrand:2007gs}
using the \NNPDF[2.3lo] set of PDFs
and the A3 set of tuned parameters~\cite{ATL-PHYS-PUB-2016-017}.
 
The detector response was simulated using the \GEANT\
framework~\cite{GEANT4,SOFT-2010-01}, and the data and MC events are
reconstructed with the same software algorithms.


\section{Object reconstruction and event selection}
\label{sec:objectselection}
\label{sec:objects}
\subsection{Object definition}
\label{sec:object definition}
This analysis makes use of jets, electrons, muons, and
event-based quantities formed from their combinations.
 
Electron candidates are identified from high-quality inner-detector tracks matched to
calorimeter energy deposits consistent with an electromagnetic shower~\cite{EGAM-2018-01}.
The calorimeter deposits must form a cluster with $E_{\mathrm
T} > 25$~GeV, $|\eta| < 2.47$, and
be outside the transition region $1.37 \leq |\eta| \leq 1.52$\
between the barrel and endcap calorimeters.
A  likelihood-based requirement is used to suppress misidentified jets,
and calorimeter- and track-based isolation requirements are
imposed using the gradient working point~\cite{EGAM-2018-01}, which provides uniform
rejection in $\eta$\ and improved rejection as \pT\ increases.
 
Muon candidates are reconstructed using high-quality inner-detector tracks
combined with tracks reconstructed in the muon spectrometer~\cite{MUON-2018-03}.
Only muon candidates with $\pt > 25$~\GeV\ and $|\eta| < 2.5$\ are considered.
Calorimeter- and track-based isolation criteria similar to those used for electrons are
used~\cite{PERF-2015-10}.
To reduce the impact of nonprompt leptons, muons within
$\Delta R = 0.4$\ of a jet are removed.
 
The \antikt\ algorithm implemented in the FastJet package~\cite{akt1,Cacciari:2011ma}\ is used to define
three types of jets for this analysis: (1) \VRTrack\ jets with a variable-radius parameter with values
between $R = 0.02$\ and $R = 0.4$~\cite{VRjets}, (2) \smallR\ jets with $R=0.4$, and (3) \largeR\ jets with $R = 1.0$.
These are reconstructed independently of each other.
The \VRTrack\ jets make use of tracking
information from the inner detector, the \largeR\ jets use information from
topological clusters~\cite{JETM-2018-02}\ in the calorimeter, and the \smallR\ jets use both tracking information
and topological clusters~\cite{PERF-2015-09}.
 
Only jets that have $\pt > 25$~\GeV\ and  $|\eta| <$ 2.5 are
considered. To reduce pileup effects, the jet-vertex tagger (JVT)
algorithm~\cite{PERF-2014-03} is used to reject \smallR~jets that do not
originate from the primary interaction vertex.
The primary vertex is selected as the one with the largest \(\Sigma \pT^2\),
where the sum is over all tracks with transverse momentum \(\pT > \SI{0.5}{\GeV}\) that are associated with the vertex.
This JVT algorithm is applied only to \smallR~jets with $\pt<60$~\GeV\ and $|\eta|<$ 2.4.
 
The topological clusters used as input to the \smallR\ and \largeR\ jet reconstruction are calibrated using the local
calibration method~\cite{JETM-2018-05}.
The jet energy scale is energy- and $\eta$-dependent with calibration factors derived from
simulation and in situ
measurements~\cite{JETM-2018-02,JETM-2018-06,ATLAS-CONF-2020-022}.
The \largeR\ jet candidates are required to have $|\eta| < 2.0$\
and $\pt >350$~\GeV.
The $\eta$\ requirement is imposed to optimize the $T$-quark signal-to-background ratio and to
select jets in a kinematic regime where the object tagging is efficient and well-understood.
The $\pt$\ requirement ensures that the \largeR\ jets are sufficiently collimated to contain most of the
decay products of the top quark or Higgs boson.
A trimming algorithm~\cite{PERF-2012-02}\ with parameters $R_\text{sub} = 0.2$ and $f_\text{cut} = 0.05$\ is
applied to suppress gluon radiation and further mitigate pileup effects.
The \smallR\ jets are used to validate the modeling of \largeR\ jets\ arising from the \ttbar\ and
multijet backgrounds and are not used directly in the event selection.
Only \smallR\ jets with $\pt > 25$~\GeV\ and $|\eta| <$ 2.5 are considered,
so as to match the \VRTrack\ jet candidates.
 
\subsection{Higgs boson, top quark, and $b$-jet tagging}
This analysis searches for Higgs bosons, top quarks, and $b$-hadron jets ($b$-jets) to identify
$T$-quark candidates that undergo a $T\rightarrow Ht$ decay,
followed by $H\rightarrow \bbbar{}$, $t\rightarrow Wb$,
and $W\to q\bar{q}^\prime$\ decays.
Distinct tagging algorithms are employed to identify these three objects.
 
Higgs-boson candidates are identified by requiring the \largeR\ jet mass~\cite{JETM-2018-05}\ to be between 100 and 140~\GeV,
along with an upper bound on the
jet-substructure variable \tautwoone{}~\cite{thaler_2011,thaler_2012},
which is a relative measure of whether the jet has a two-body or one-body structure.
The upper bound on \tautwoone{} is chosen as a function of the jet \pt{} in order to achieve
a tagging efficiency of 70\% for Higgs bosons, independent of their \pt.
The tagger provides a rejection factor between five and ten for
light-quark and gluon jets.
 
The top-quark-tagging algorithm uses a deep-neural-network (DNN) scheme~\cite{JETM-2018-03}.
It makes use of jet-substructure variables to discriminate between top-quark jets and jets
arising from \Wboson, \Zzero, Higgs bosons,
gluons, and lighter quarks.
An 80\% efficiency working point is used, which is defined for all top-quark jets whose
decay products are clustered together into the \largeR\ jet.
In addition, only jets with a reconstructed mass between 140 and 225 \GeV\ are considered.
The orthogonal mass window requirements for tagging Higgs bosons and top quarks ensure that a jet can only be identified
as either a top-quark or Higgs-boson candidate.
 
The $b$-tagging algorithm used is known as DL1, a DNN-based tagging scheme that uses the secondary vertex information
and the impact parameters of the charged tracks in a \VRTrack\ jet~\cite{FTAG-2018-01}.
The working point chosen for this algorithm results in 70\%\ tagging efficiency for $b$-jets in
\ttbar\ events, with a rejection of ${\sim}10$\ and ${\sim}400$\ for charm and light quarks,
respectively.
This algorithm is applied to all \VRTrack\  jets that have been geometrically matched to the
\largeR\ jets by requiring that the jet axes have an $\eta\!-\!\phi$\ distance
$\Delta R < 1.0$.
 
\subsection{Event preselection}
\label{sec:event_selection}
 
A preselection is performed to obtain a sample of candidate signal and background events.
Each event is required to have a primary vertex with five or more associated tracks
with $\pT > 0.5$~\GeV~\cite{ATL-PHYS-PUB-2015-026}.
 
To identify the fully hadronic \Ht\ decay topology, events must have at least two \largeR\ jets
with $\pt > 350$~\GeV\ and $|\eta| < 2.0$.
The highest-\pT\ jet is required to have $\pt > 500$~\GeV\ to ensure that the inclusive jet
trigger used to record the events has 100\%\ efficiency.
The two highest-\pT\ \largeR\ jets are referred to as the leading and second-leading jets. All other \largeR\ jets
are ignored.
The \largeR\ jets must have a mass between 100 and 225~GeV.
 
To remove candidates where a \ttbar\ event has resulted in a lepton+jet or dilepton final state,
events are rejected if they have an identified electron or muon candidate, as described in
Section~\ref{sec:object definition}.
 
This preselection defines the data sample used in
the $T$-quark search, which comprises about 4~million events.
 
\subsection{Event classification by tagging states}
\label{sec:event_categorization}
The leading and second-leading \largeR\ jet candidates are examined to determine if either jet satisfies the
Higgs-boson-tagging or
top-quark-tagging criteria.
In addition, each \VRTrack\ jet contained within a \largeR\ jet is examined to determine if it is $b$-tagged.
In what follows, a $b$-tagged \VRTrack\ jet associated with a \largeR\ jet is referred to as a $b$-tag.
 
\begin{figure*}[t]
\centering
\includegraphics[width=0.90\textwidth]{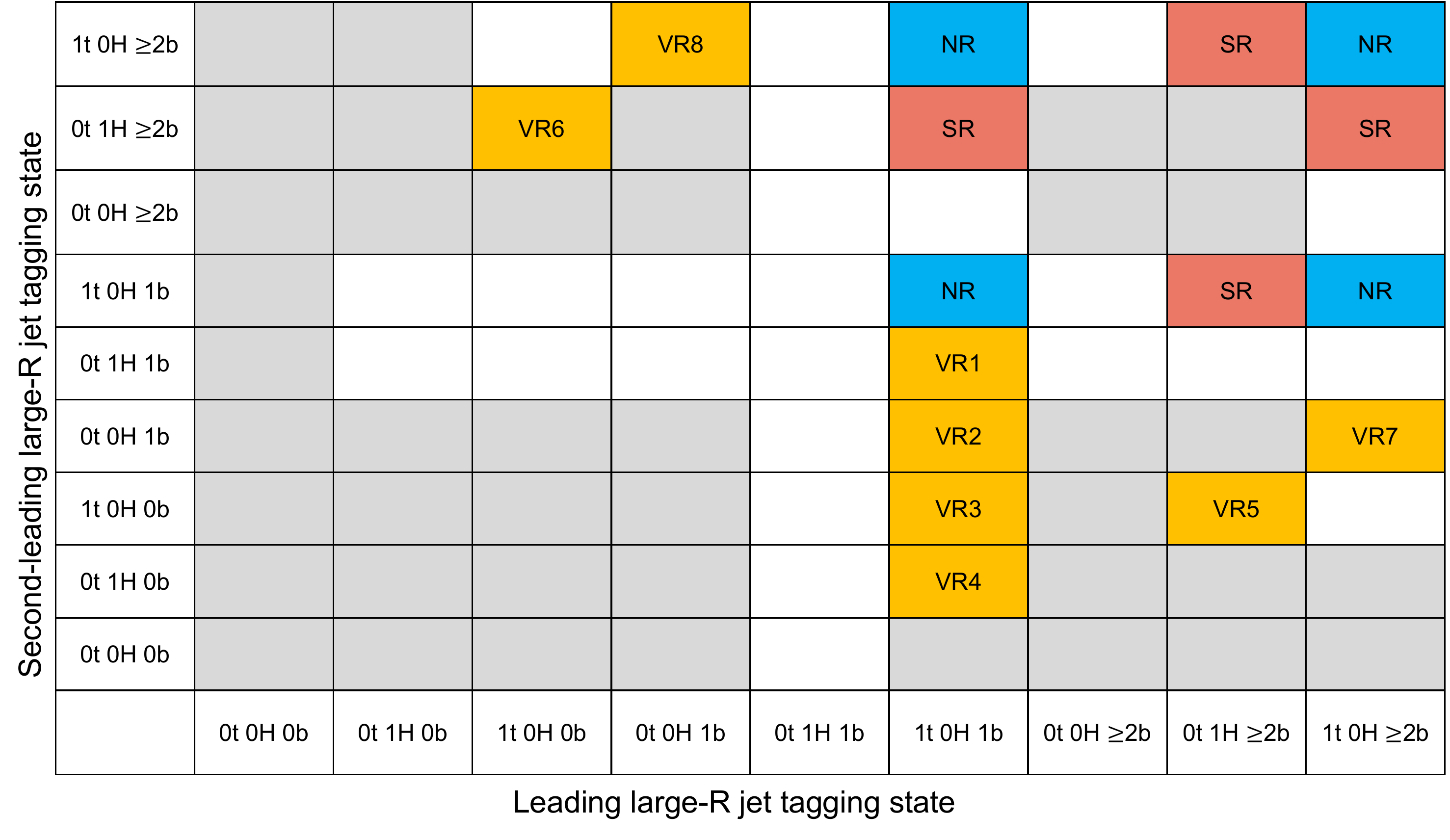}
\caption{The $9\times 9$\ matrix that represents the 81 exclusive event-tagging states
defined by the 9 possible tagging states of each \largeR\ jet.
Each event can be in only one of the 81 event-tagging states.
The red event-tagging states comprise the events in the signal region (\SR),
the blue event-tagging states comprise the \ttbar\ normalization region (\NR), and
the yellow event-tagging states are validation regions labeled $\VR{1}$\ through $\VR{8}$.
The gray event-tagging states are regions used to estimate the multijet background, as described in Section~\ref{subsec: ABCD}.
}
\label{fig: 9by9 matrix}
\end{figure*}

With these tagging definitions, the events are classified according to
the tagging states of each \largeR\ jet:  the jet could be neither Higgs-boson-tagged nor top-quark-tagged,
be Higgs-boson-tagged, or be top-quark-tagged.
The jet also could have no $b$-tags, 1~$b$-tag, or ${\ge}2$~$b$-tags.
Altogether, a \largeR\ jet could be in one of nine different tagging states, so a
$9\times9$\
matrix is defined as shown in Figure~\ref{fig: 9by9 matrix}\ to categorize  all possible
tagging states of the two jets in an event.
 
Three sets of regions are defined: a signal region \SR, a \ttbar\ normalization region \NR,
and eight validation regions, as illustrated in Figure~\ref{fig: 9by9 matrix}.
The signal region consists of those events where one jet
is Higgs-boson-tagged with ${\ge}2$~$b$-tags
and the other jet
is top-quark-tagged with ${\ge}1$~$b$-tag, and comprises four event-tagging states as illustrated
in Figure~\ref{fig: 9by9 matrix}.
 
The \ttbar normalization region is designed to contain the highest-purity sample of
\ttbar\ pair events.
The four event-tagging states that define this region are those with both \largeR\ jets
being top-quark-tagged and each having at least 1~$b$-tag.
Top-quark-tagged jets with ${\ge}2$~$b$-tags are included, as these typically result from
mistagging a charm-quark jet arising from a $W\to c\bar{s}$\ decay.
This region is used to study top-quark-tagging performance
and to validate the top-quark acceptance and background estimates.
 
The validation regions are used to validate the background estimation techniques used in the \SR\
and \NR.
The regions with a leading \largeR\ jet top-quark-tagged with 1~$b$-tag and the
second-leading \largeR\ jet not being top-quark-tagged with 1~$b$-tag (\VR{1}\ and \VR{2}) or
the second-leading \largeR\ jet being either Higgs-boson-tagged or top-quark-tagged
with no $b$-jets (\VR{3}\ and \VR{4}) validate the multijet and
non-all-hadronic \ttbar\ background estimates.
The validation regions defined by the event-tagging states with a
Higgs-boson-tagged \largeR\ jet with ${\ge}2$~$b$-tags
and with the other jet top-quark-tagged with no associated $b$-tags (\VR{5}\ and \VR{6}) are
expected to be dominated by mistagged events
and are used to cross-check the mistagging estimates for the Higgs-boson, top-quark, and
$b$-jet tagging schemes.
The validation regions defined by one jet that is neither Higgs-boson-tagged nor top-quark-tagged
with 1~$b$-tag and the other jet being top-quark-tagged with ${\ge}2$~$b$-tags
(\VR{7}\ and \VR{8}) are considered to validate the
background modeling involving ${\ge}3$~$b$-tags.
 
 
\section{Background estimation and validation}
\label{sec:backgrounds}
 
 
The two largest contributions to the \SR\ and \NR\ are
multijet events and \ttbar\ production, with smaller contributions arising from events with
only one hadronically decaying top quark or from \ttbar\ production in association
with a \Wboson, \Zzero, or Higgs boson.
 
The multijet background in all the regions is estimated using a data-driven technique
employing sidebands and control regions dominated by multijet events and originally developed to study boosted \ttbar\ production \cite{TOPQ-2012-15,TOPQ-2014-15,TOPQ-2016-09}.
This background is found to be the largest source of candidate events in the signal region
and is determined iteratively, as described in
Section~\ref{subsec: ABCD}.
 
The second-largest background in the \SR\ consists of events in which a pair of boosted top quarks
decay hadronically to produce two top-quark jets.
This background is estimated using MC calculations normalized by the event yield
in the \NR\ after subtracting other backgrounds.
As this subtraction requires an estimate of the multijet background, the
background-subtracted event yield is determined iteratively, as described in
Section~\ref{subsec: TTbarDiffXs}.
 
The third-largest contribution in the \SR\
is from the non-all-hadronic \ttbar\ process where one top quark decayed semileptonically
and the other hadronically.
In this case, the final-state leptons are not reconstructed or are misidentified and not rejected
by the electron and muon veto.
The rate of this process, estimated using MC samples, is normalized using the
observed \ttbar\ event yield in the \NR.
 
Other contributions from SM processes with at least one top-quark jet
are estimated using MC samples as described in Section~\ref{subsec: MiscBkg}.
 
\subsection{The data-driven multijet background estimate}
\label{subsec: ABCD}
 
The multijet background is estimated by a data-driven method using events
from specifically chosen event-tagging states to estimate the multijet background event yields in the signal, normalization,
and validation regions.
The event-tagging states used are dominated by
multijet backgrounds and have small contributions from events with one or more top quarks.
These event-tagging states also have potential contributions from $T$-quark production of
less than 1\%\ for all choices of $T$-quark masses and couplings considered in this search.
Contributions from $W/Z$+jets are negligible due to a combination of
a relatively low cross-section for high-\pT\ hadronically-decaying bosons~\cite{STDM-2018-34}\
and the tagging requirements.
For a given event-tagging state, the number of events from all MC backgrounds is subtracted
from the observed number of events with that event-tagging state.
This provides an estimate of the number of multijet events in each event-tagging state.
As noted above, the \ttbar\ background subtracted from each region is normalized to the
event yield in the \NR\ that depends on the multijet estimate in that region.
Hence, the multijet background and \ttbar\ yield are calculated iteratively.
This procedure is similar to the algorithm used in Ref.~\cite{TOPQ-2016-09}.
 
For example, consider the \SR\ event-tagging state defined by requiring that the
leading jet is top-quark-tagged with 1~$b$-tag and the second-leading jet
is Higgs-boson-tagged with ${\ge}2$~$b$-tags.
The method uses the numbers of multijet events $N_{\mathrm{A}}$, $N_{\mathrm{B}}$, $N_{\mathrm{C}}$, and $N_{\mathrm{D}}$,
after MC background subtraction, in four regions A, B, C, and D.
In region A the leading jet is neither top-quark-tagged nor Higgs-boson-tagged with no $b$-tags and the
second-leading jet is neither top-quark-tagged nor Higgs-boson-tagged with no $b$-tags.
In region B the leading jet is neither top-quark-tagged nor Higgs-boson-tagged with no $b$-tags and the
second-leading jet is Higgs-boson-tagged with ${\ge}2$~$b$-tags.
In region C the leading jet is top-quark-tagged with 1~$b$-tag and the
second-leading jet is neither top-quark-tagged nor Higgs-boson-tagged with no $b$-tags.
In region D, which is one of the \SR\ event-tagging states, the leading jet is
top-quark-tagged with 1~$b$-tag and the
second-leading jet is Higgs-boson-tagged with ${\ge}2$~$b$-tags.
If the tagging efficiencies of the two \largeR\ jets are uncorrelated,
then the ratio of the numbers of multijet events in two distinct event-tagging states that differ
only by the tagging state for one of the \largeR\ jets will be independent of the tagging state of the other \largeR\ jet.
In this example, the ratio of $N_{\mathrm{D}}$ to $N_{\mathrm{C}}$
is equal to the ratio of $N_{\mathrm{B}}$ to $N_{\mathrm{A}}$ since the ratios only differ by the tagging state of the leading large-R jet.
Hence, the number of multijet events in region D is
\begin{eqnarray}\nonumber
N_{\mathrm{D}} = {\frac{N_{\mathrm{B}}\times N_{\mathrm{C}}}{N_{\mathrm{A}}}}.
\end{eqnarray}
 
A corresponding method is performed for each of the event-tagging states of the
\SR\ and \NR, and for all the
validation regions using different event-tagging states to define regions A, B, and C.
Since the \ttbar\ background subtraction in regions A, B, and C is normalized to the
\ttbar\ event yield in the \NR, which requires an estimate of the multijet background,
the calculation of the multijet background and the \ttbar\ event yield in each region
is iterated as described in
Section~\ref{subsec: TTbarDiffXs}.
 
The assumption of uncorrelated jet-tagging states is only approximately true.
The level of correlation is determined by examining ratios of the numbers
of events with specific event-tagging states that do not overlap with the \SR, \NR, or
validation regions, shown as the gray event-tagging states in Figure~\ref{fig: 9by9 matrix}.
The observed corrections between the jet-tagging states defined by the
top-quark, Higgs-boson, and $b$-tagging
criteria are applied to the multijet background estimates for each of the event-tagging states that
define the \SR, \NR, and the eight validation regions,
with the total corrections varying from 1.01 to 1.10 with uncertainties ranging from 0.03 to 0.06.
In the calculation of the multijet background for the event-tagging state belonging to the \SR\
illustrated above, there are four corrections applied as a product.
The multijet estimates are calculated independently for each of the four event-tagging states
that make up the \SR\ and \NR, after which they are summed.
This provides a fully data-driven multijet background estimate in each region.
 
For example, to calculate the correlation between the mistagging probabilities when
the leading jet is  top-quark-tagged and the second-leading jet is Higgs-boson-tagged,
the event yields in four regions, E, F, G, and H are considered.
In region E the leading jet is neither top-quark-tagged nor Higgs-boson tagged with no $b$-tags and
the second-leading jet is neither top-quark-tagged nor Higgs-boson-tagged with no $b$-tags.
In region F the leading jet is top-quark-tagged with no $b$-tags and the second-leading jet is neither
top-quark-tagged nor  Higgs-boson-tagged with no $b$-tags.
In region G the leading jet is neither top-quark-tagged nor Higgs-boson-tagged with no $b$-tags and
the second-leading jet is Higgs-boson-tagged with no $b$-tags.
In region H the leading jet is top-quark-tagged with no $b$-tags and
the second-leading jet is Higgs-boson-tagged with no $b$-tags.
The ratio between the number of events in regions E and F is related to the ratio
of events in regions G and H by
\begin{eqnarray}\nonumber{\frac{N_{\mathrm{E}}}{N_{\mathrm{F}}} = K\frac{N_{\mathrm{G}}}{N_{\mathrm{H}}}},
\end{eqnarray}
where $K$ is the measure of the correlation in mistagging probabilities
between the leading jet being top-quark-tagged and the
second-leading jet being Higgs-boson-tagged,
with  both \largeR\ jets having no associated $b$-tags.
The value of $K$\ in this example is $0.976\pm0.004$, where the uncertainty is statistical only,
and is applied as a correction to the multijet background estimate.
 
Each correlation is measured in an analogous way using the numbers of events in pairs of
event-tagging states.
The pairs chosen have MC background contributions less than 8\% of the observed
event yield, thus reducing the
systematic uncertainties arising from the subtraction of the MC background contributions.
The multijet background estimate taking into account the tagging correlations is calculated
bin-by-bin for each distribution
so that the shape of the multijet background distribution is measured as well as the total background event yield.
 
Since the multijet background depends on the \ttbar\ background subtraction, the
two are determined iteratively as described in the next section.
 
\subsection{Top-quark pair yields and multijet backgrounds in the \NR\ and \SR}
\label{subsec: TTbarDiffXs}
 
Previous measurements of the \ttbar differential cross-sections for
highly boosted top quarks~\cite{TOPQ-2016-09}\
show that the observed cross-section is lower than MC predictions by ${\sim}20$\%.
To avoid the uncertainty this would create in the \ttbar\ background contribution and the
multijet estimates in each region, the ratio of the observed rate to the predicted
rate of events in the \NR, \alphanorm, is used to normalize the predicted \ttbar\ background
contributions in the \SR, validation regions, and the event-tagging-states used for the
multijet estimate.
 
The value of $\alphanorm$ is determined after the initial multijet estimate that uses the
nominal \ttbar\ prediction by requiring the
predicted event yield in the \NR\ to match the observed yield.
However, the multijet estimate itself is a function of $\alphanorm$, as the estimation technique
described in the previous section
requires the subtraction of the \ttbar\ background contribution in the
multijet-dominated event-tagging states during its calculation.
Thus, both the multijet estimate and $\alphanorm$ are calculated iteratively using
\begin{equation}\nonumber
\alphanorm_{n+1} = \frac{{{N_{\text{Data}}} - {N_{\text{Multijet}}(\alphanorm_{n})} - {N_{\text{top-related}}}}}{{{N_{\ttbar\text{MC}}}}},
\end{equation}
where
$\alphanorm_n$\ is the value of $\alphanorm$\ resulting from the $n^\text{th}$\ iteration,
$N_{\text{Data}}$\ is the observed event yield in the \NR,
$N_{\text{Multijet}}(\alphanorm_n)$\ is the data-driven multijet background event yield from the
$n^\text{th}$\ iteration in the \NR,
$N_{\text{top-related}}$\ is the sum of the backgrounds from single-top-quark,
$\ttbar + \Wboson$, $\ttbar + \Zzero$, and $\ttbar + H$ production
that are estimated by MC calculations in the \NR, and
$N_{\ttbar\text{MC}}$\ is the sum of the \ttbar\ events with
all-hadronic and non-all-hadronic decays in the \NR.
 
In each iteration of the multijet estimate, $N_{\ttbar\text{MC}}$\ is scaled
by $\alphanorm_{n+1}$ before subtraction.
This calculation converges to subpercent level in four iterations to a value of
$\alphanorm = 0.82 \pm 0.01$, where only
statistical uncertainties are considered.
This is consistent with cross-section measurements of boosted \ttbar\
production~\cite{TOPQ-2016-01}.
The \ttbar\ contribution predicted by the MC calculations in the \SR\ is scaled by \alphanorm.
 
The resulting \ttbar\ yield estimates are $8587 \pm 1369$\ events and $174 \pm 35$\
events in the \NR\ and \SR, respectively, where the uncertainties include the systematic
uncertainties described in Section~\ref{sec:systematics}.
This estimate of the \ttbar\ yield in the \NR\ is used only for the iterative multijet
background estimate.
 
The multijet yields in the \NR\ and \SR\ after this iterative calculation
are  estimated to be $1452 \pm 57$\ and
$316 \pm 9$\ events, respectively.
The uncertainties in the multijet estimates, including the uncertainties in the tagging correlations,
consist of the statistical uncertainties in the event-tagging states used for the calculation
and the systematic uncertainties arising from the MC background subtraction,
as described in Section~\ref{sec:systematics}.

\subsection{Other top-quark backgrounds}
\label{subsec: MiscBkg}
 
Single-top-quark production in the $Wt$ final state and the $t$-channel represent a small contribution to the total background prediction,
which is estimated using the \POWPY{8} MC calculation described
in Section~\ref{sec:datamcsamples}.
The $s$-channel single-top-quark process is not included as an explicit contribution because
of its small cross-section and because a part of it is already taken into account  in the
data-driven multijet estimate.
The uncertainty in the single-top-quark background is increased by 50\%\ to account for
the uncertainty in this contribution.
 
The estimated single-top-quark yields in the \ttbar\ \NR\ and \SR\ are $93\pm52$\
and $8\pm 6$\ events, respectively.
 
The backgrounds from production of a top-quark pair in association with a \Wboson, \Zzero,
or Higgs boson
are also estimated using the MC samples described in Section~\ref{sec:datamcsamples}.
The estimated yields in the \NR\ and \SR\ are $115\pm25$\ events and $9\pm 2$\ events,
respectively.
 
\subsection{Validation of background calculations}
\label{sec: yields}
 
\begin{figure}[t]
\centering
\subfloat[ ]{
\includegraphics[width=0.47\linewidth]{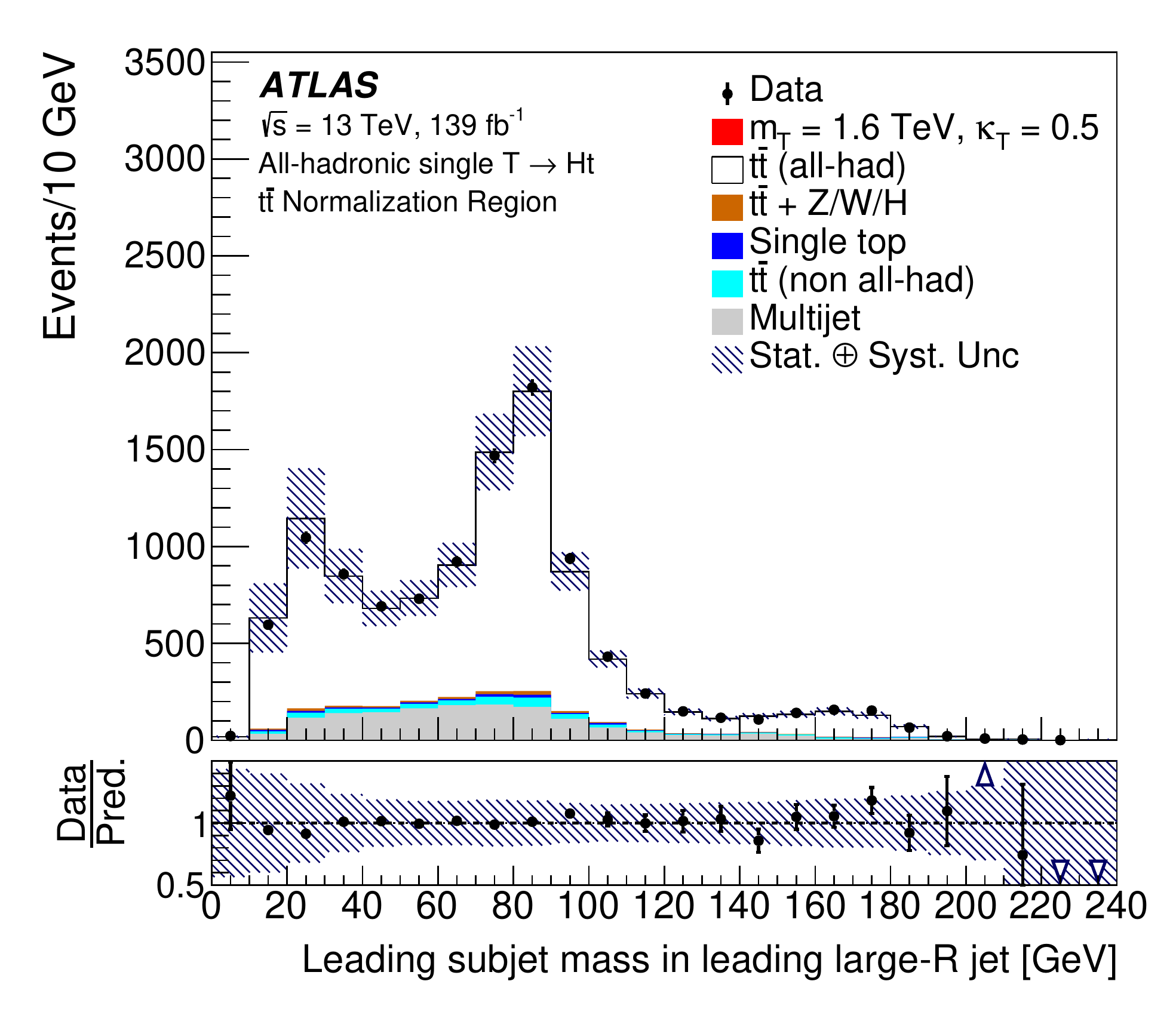}
\label{fig: ttNR subjet mass}
}\quad
\subfloat[ ]{
\includegraphics[width=0.47\linewidth]{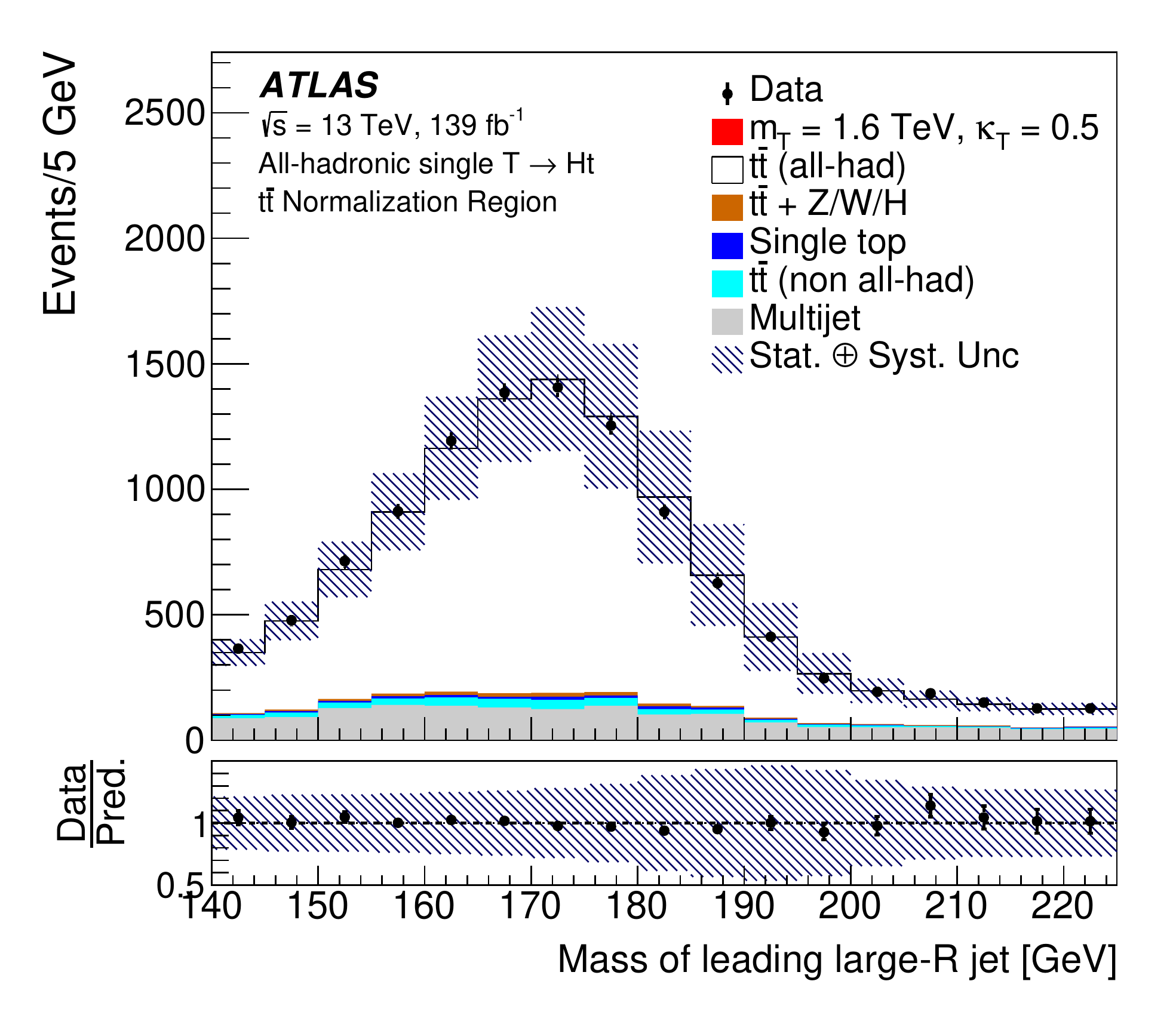}
\label{fig: ttNR lj1 mass}
}
 
\caption{Invariant mass distributions for
(a) the leading \smallR\ jet associated with the
leading \largeR\ jet in the \ttbar\ normalization region where
both the leading and second-leading \largeR\ jets are top-quark-tagged with at least 1~$b$-tag, and
(b) the leading \largeR jet, in the same region.
The predicted distributions include the estimated backgrounds and a hypothetical $T$-quark signal
with $\mTVLQ = 1.6$~\TeV\ and $\kappaT = 0.5$.
The blue hashed lines correspond to the sum in quadrature of the statistical and systematic uncertainties of the prediction in a given bin.
The lower panels show the ratio of the data to the prediction, along with the
uncertainty in the ratio.
A ratio outside the bounds of the axis is represented by a blue arrow.
The last bin includes the event overflows.
Contributions to the predicted yield are stacked in the same order as they appear in the legend.
}
\label{fig: ttNR masses}
\end{figure}

Kinematic variables with the ability to distinguish between \ttbar\ and multijet
contributions in the \NR\ and the validation regions are examined to further validate the
background modeling.
The potential contribution of $T$-quark production to these regions is ${<}1${\%}.
The distributions of the mass of the leading \smallR\ jet associated with the leading \largeR\ jet
events in the \NR\ where both \largeR\ jets are top-quark-tagged and
have ${\ge}1$~$b$-tags\ are shown in Figure~\ref{fig: ttNR subjet mass}.
A \Wboson-mass peak is observed, which arises when the \Wboson-boson decay products are
collimated into a \smallR\ jet, along with a low-mass peak arising from light quarks and bottom quarks.
A shoulder is seen around the top-quark mass, which arises from a small number of
highly boosted top-quark jets where all the decay products of the top quark
are clustered into the \smallR\ jet.
The observed distribution is well-modeled with a large \ttbar\ contribution and a
smaller multijet distribution.
The invariant mass distribution of the leading \largeR\ jet in the same sample, shown in
Figure~\ref{fig: ttNR lj1 mass}, confirms the interpretation that this region is dominated by
\ttbar\ production.
 
\begin{figure}[t]
\centering
\subfloat[ ]{
\includegraphics[width=0.47\linewidth]{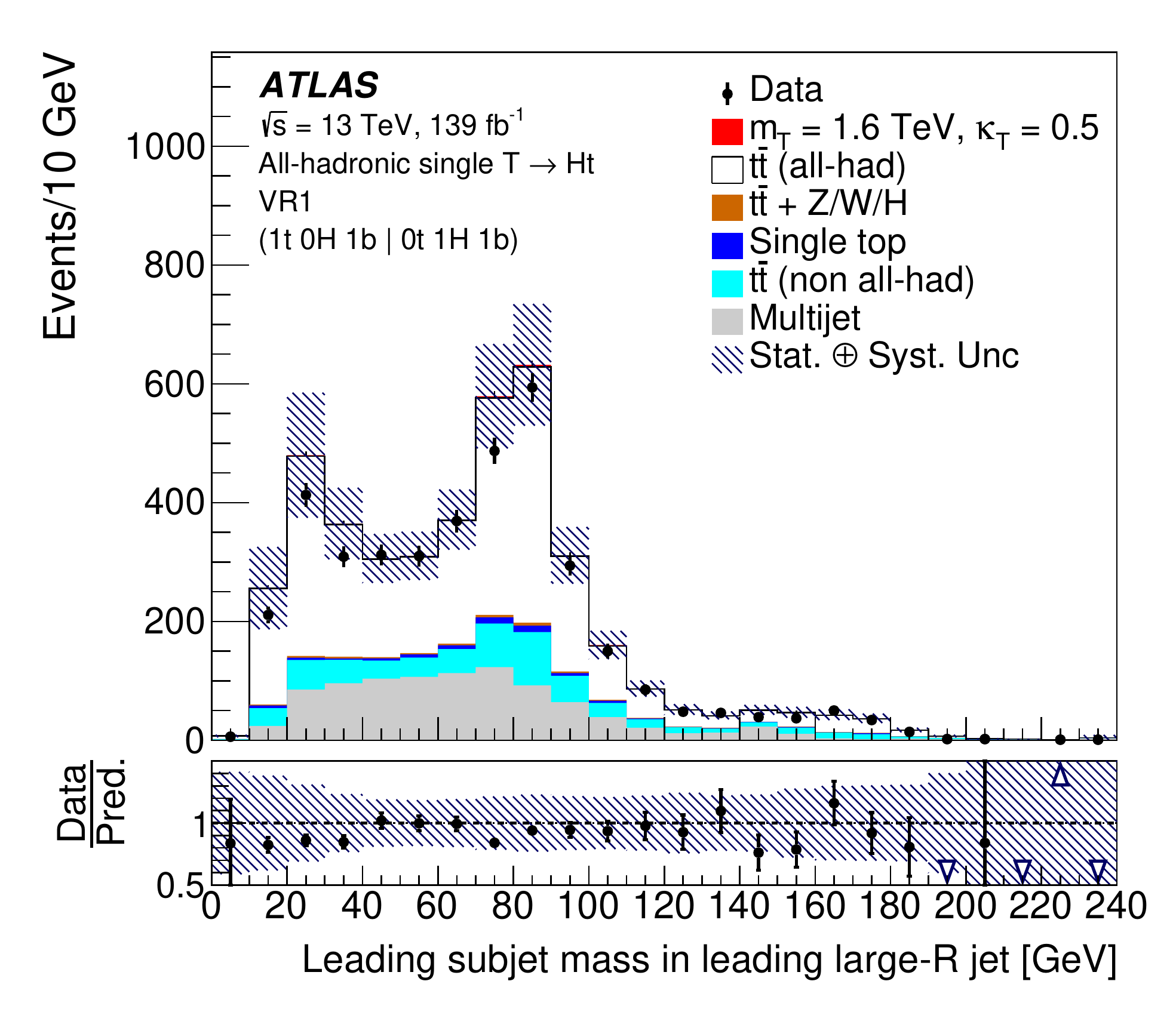}
\label{fig: 65 subjet mass}
}\quad
\subfloat[ ]{
\includegraphics[width=0.47\linewidth]{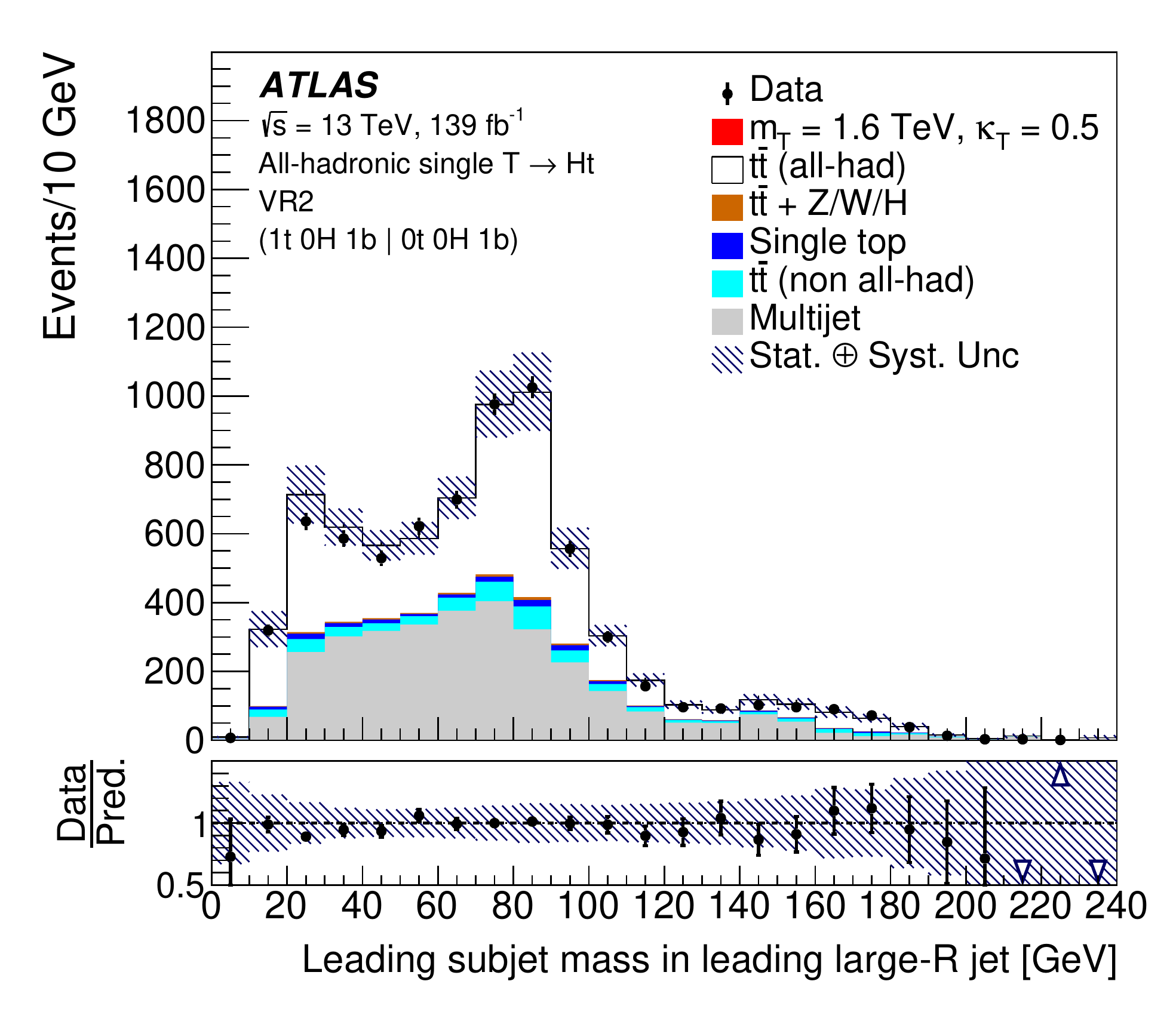}
\label{fig: 64 subjet mass}
}
 
\subfloat[ ]{
\includegraphics[width=0.47\linewidth]{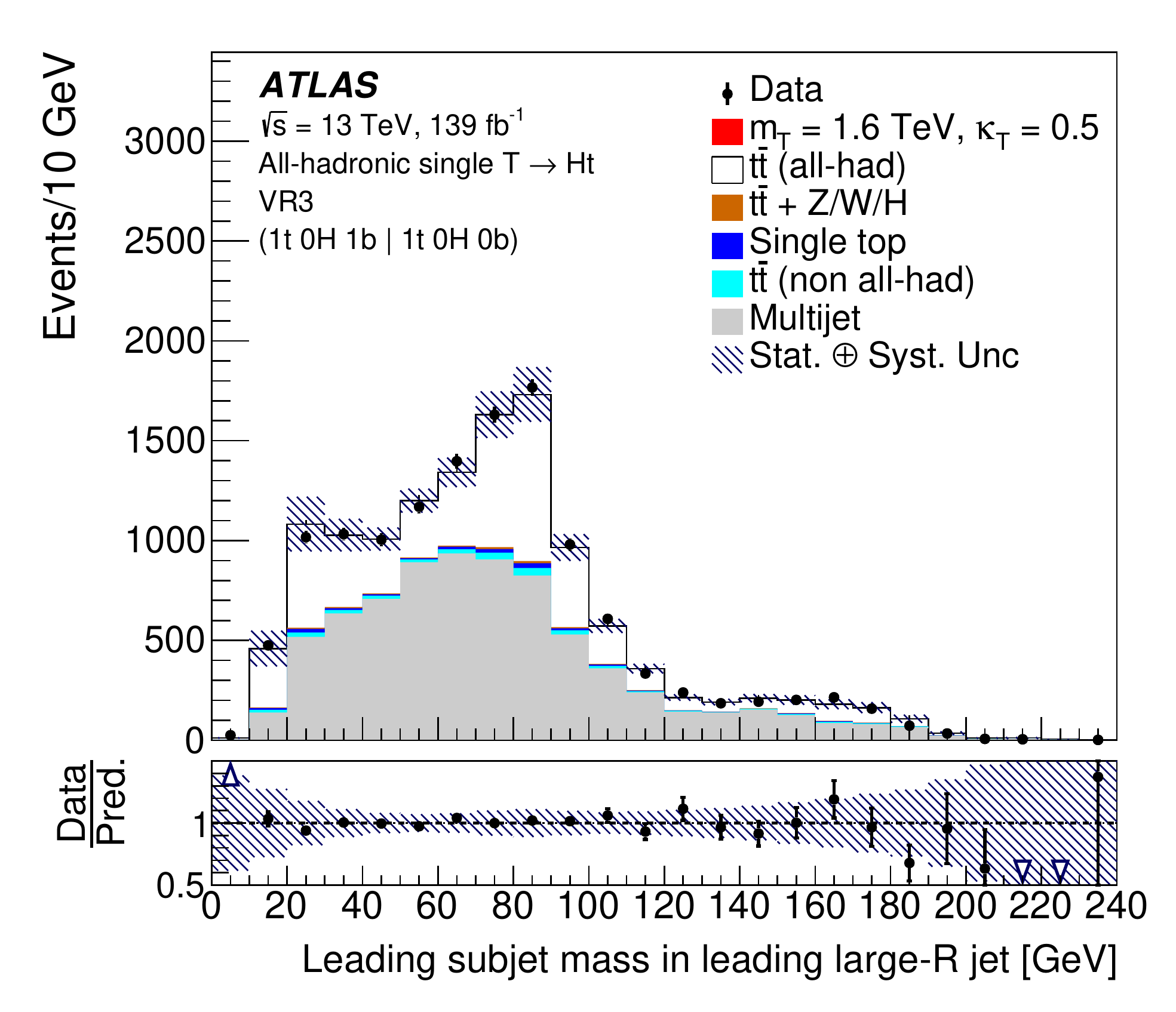}
\label{fig: 63 subjet mass}
}\quad
\subfloat[ ]{
\includegraphics[width=0.47\linewidth]{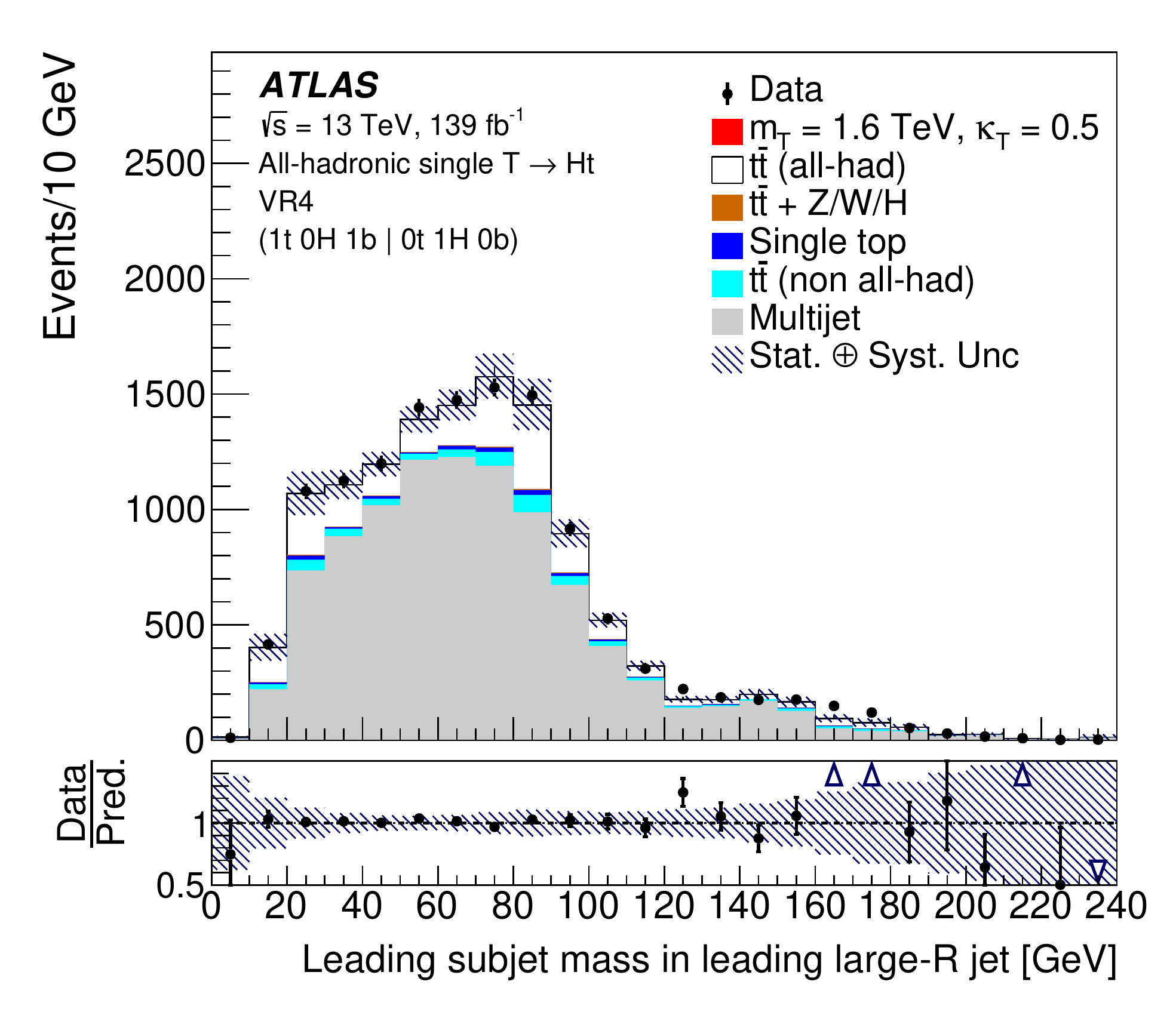}
\label{fig: 62 subjet mass}
}
 
\caption{Invariant mass distributions for the leading \smallR\ jet associated with the
leading \largeR\ jet for
(a) \VR{1}\ defined by requiring the
leading \largeR\ jet be top-quark-tagged with 1~$b$-tag and
the second-leading jet is Higgs-boson-tagged with 1~$b$-tag,
(b) \VR{2}\ defined by requiring the leading \largeR\ jet be top-quark-tagged with
1~$b$-tag and the second-leading jet is neither Higgs-boson-tagged nor top-quark-tagged with 1~$b$-tag,
(c) \VR{3}\ defined by requiring the leading \largeR\ jet be top-quark-tagged with
1~$b$-tag and the second-leading jet is top-quark-tagged with no $b$-tag, and
(d) \VR{4}\ defined by requiring the leading \largeR\ jet be top-quark-tagged with
1~$b$-tag and the second-leading jet is Higgs-boson-tagged with no $b$-tag.
The predicted distribution includes the estimated backgrounds and a hypothetical $T$-quark signal
with $\mTVLQ = 1.6$~\TeV\ and $\kappaT = 0.5$.
The blue hashed lines correspond to the sum in quadrature of the statistical and systematic uncertainties of the prediction in a given bin.
The lower panels show the ratio of the data to the prediction, along with the
uncertainty in the ratio.
A ratio outside the bounds of the axis is represented by a blue arrow.
The last bin includes the event overflows.
Contributions to the predicted yield are stacked in the same order as they appear in the legend.
}
\label{fig: subjet masses VR1 to VR4}
\end{figure}

The distributions of the jet mass for the leading \smallR\ jet associated with the
leading \largeR\ jet are shown in Figure~\ref{fig: subjet masses VR1 to VR4}\
for validation regions \VR{1}\ through \VR{4}.
The relative sizes of the \ttbar\ and multijet contributions vary between these validation
regions, further testing the robustness of their modeling and normalization.
There is agreement between the observed and predicted distributions
in both normalization and shape, except for a small excess in the prediction of
events for \VR{1}.
This is further discussed in Section~\ref{sec:modeling uncertainties}.
 
\begin{figure}[t]
\centering
\subfloat[ ]{
\includegraphics[width=0.47\linewidth]{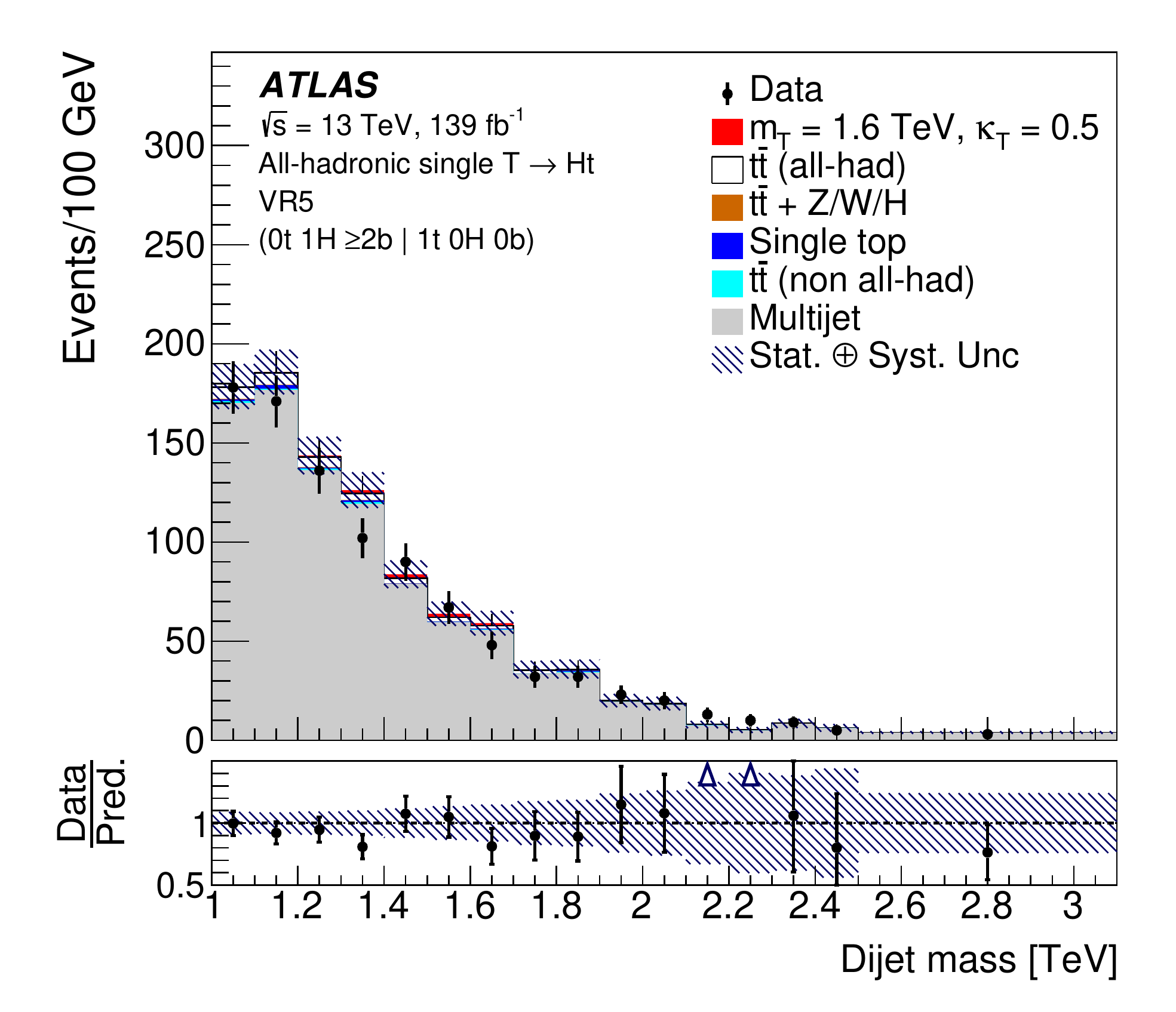}
\label{fig: 83 VLQ_m}
}\quad
\subfloat[ ]{
\includegraphics[width=0.47\linewidth]{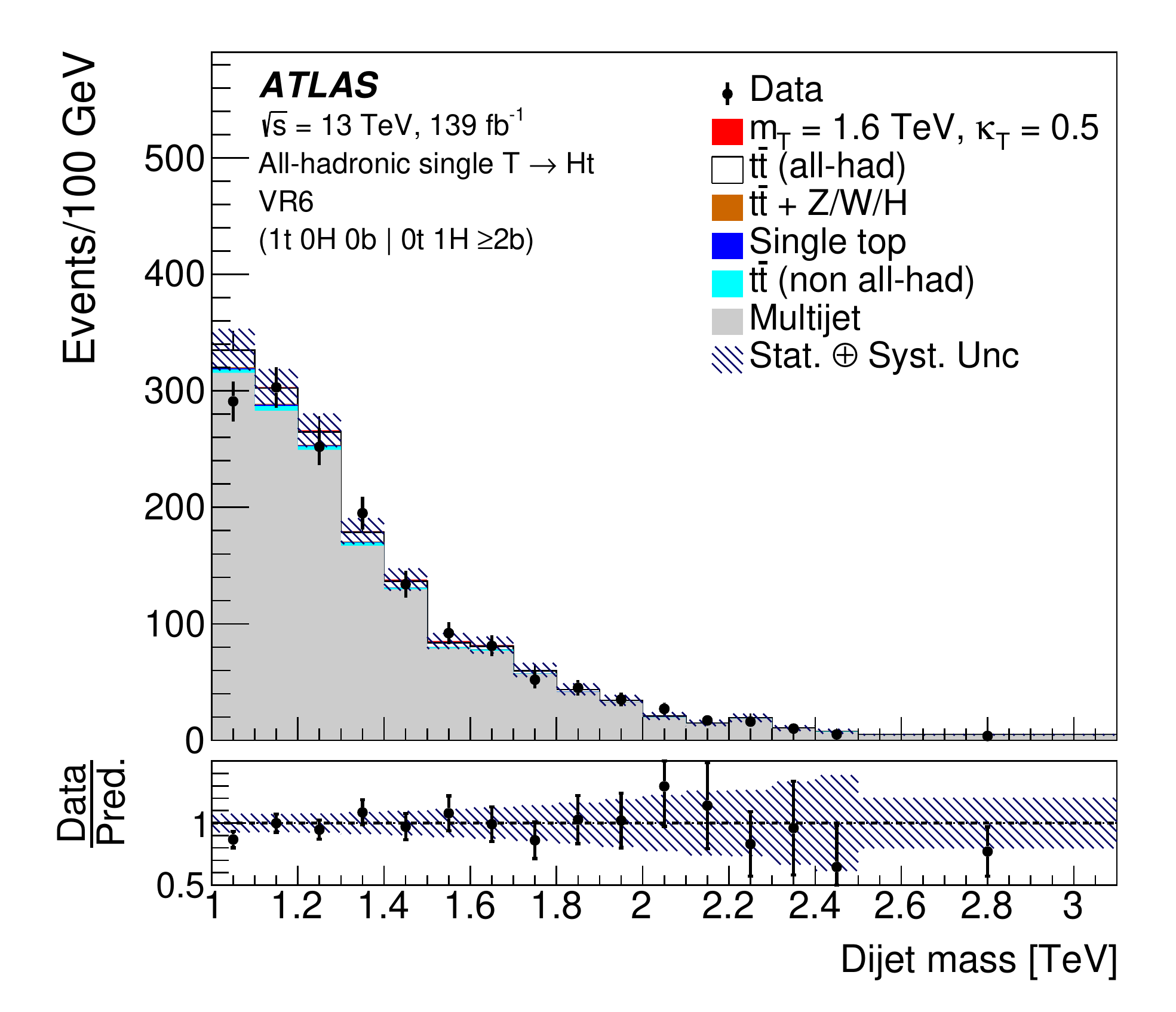}
\label{fig: 38 VLQ_m}
}\quad
\subfloat[ ]{
\includegraphics[width=0.47\linewidth]{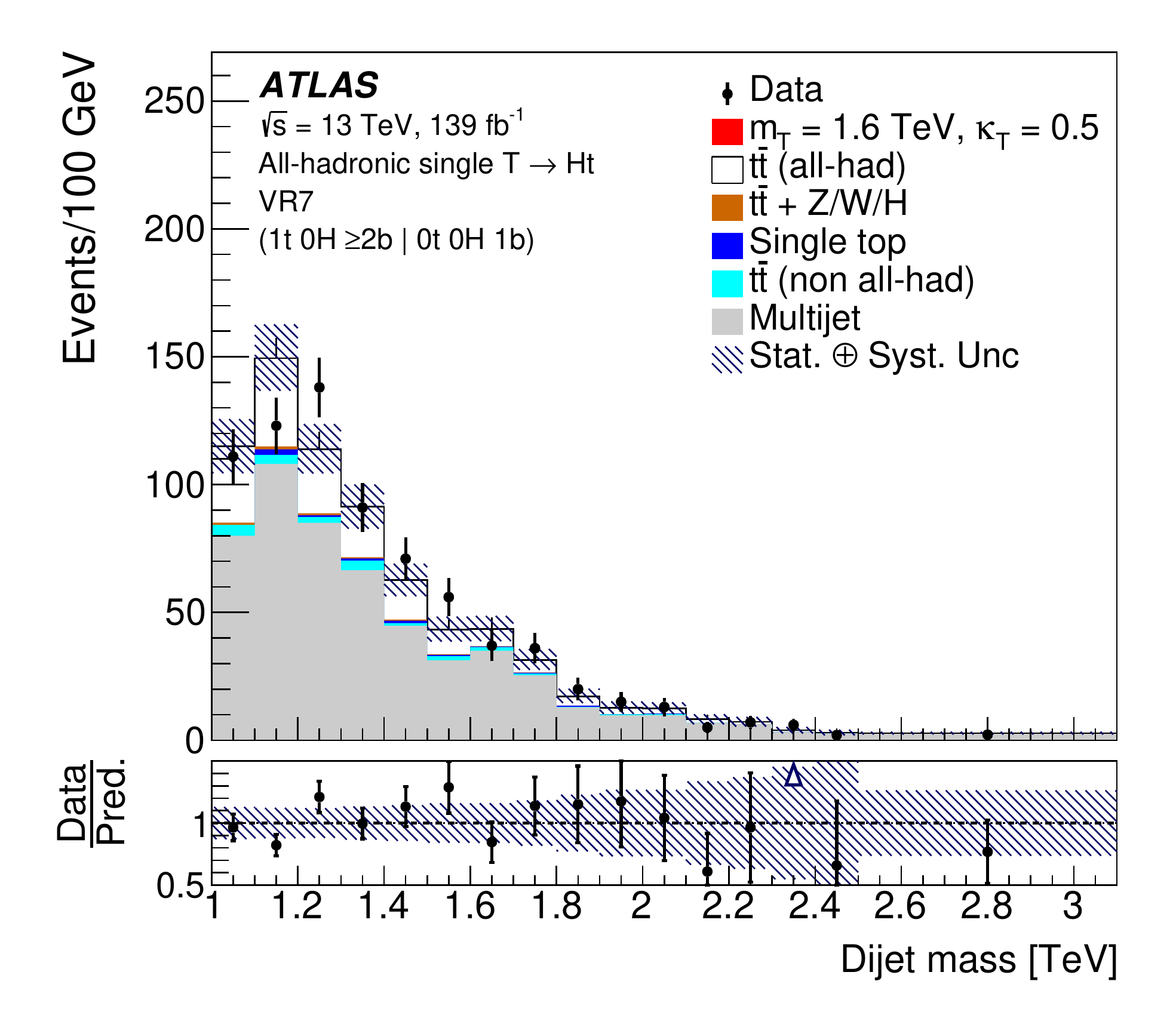}
\label{fig: 94 VLQ_m}
}\quad
\subfloat[ ]{
\includegraphics[width=0.47\linewidth]{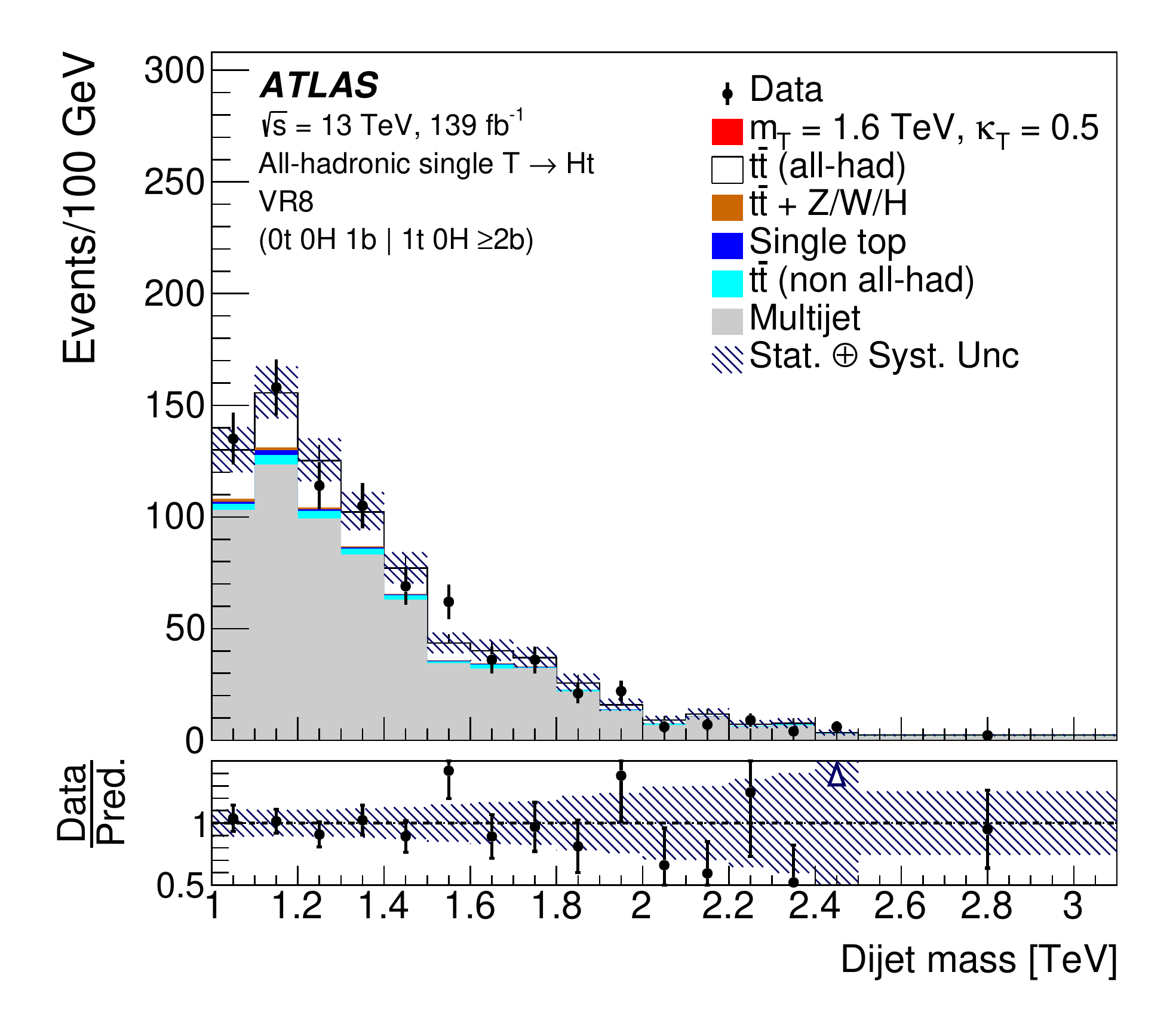}
\label{fig: 49 VLQ_m}
}
 
\caption{Dijet invariant mass distributions for the two \largeR\ jets in
four validation regions:
(a) \VR{5} defined by requiring a leading jet Higgs-boson-tagged with ${\ge}2$~$b$-tags and
second-leading jet top-quark-tagged with no associated $b$-tag,
(b) \VR{6} defined by requiring a leading jet top-quark-tagged with no associated $b$-tag and second-leading jet
Higgs-boson-tagged with ${\ge}2$~$b$-tags,
(c) \VR{7} defined by requiring a leading jet top-quark-tagged with ${\ge}2$~$b$-tags and
second-leading jet neither top-quark-tagged nor Higgs-boson-tagged with 1~$b$-tag, and
(d) \VR{8}\ defined by requiring a leading jet neither top-quark-tagged nor Higgs-boson-tagged with 1~$b$-tag and
second-leading jet top-quark-tagged with ${\ge}2$~$b$-tags.
The predicted distributions include the estimated backgrounds and a hypothetical $T$-quark signal
with $\mTVLQ = 1.6$~\TeV\ and $\kappaT = 0.5$.
The blue hashed lines correspond to the sum in quadrature of the statistical and systematic uncertainties of the prediction in a given bin.
The lower panels show the ratio of the data to the prediction, along with the
uncertainty in the ratio.
A ratio outside the bounds of the axis is represented by a blue arrow.
The last bin includes the event overflows.
Contributions to the predicted yield are stacked in the same order as they appear in the legend.
}
\label{fig: dijet masses VR5 to VR8}
\end{figure}

Further validation of the multijet background estimates is illustrated
in Figure~\ref{fig: dijet masses VR5 to VR8}, where the distributions of the invariant
mass of the two leading jets, or dijet system, are shown for the
four validation regions dominated by multijet backgrounds.
The distributions for events with
a top-quark-tagged jet with no $b$-tags and a Higgs-boson-tagged jet
with ${\ge}2$~$b$-tags (\VR{5}\ and \VR{6}) are shown in Figures~\ref{fig: 83 VLQ_m}\ and~\ref{fig: 38 VLQ_m},
respectively.
Distributions
for events with a jet with 1~$b$-tag, but no top-quark or Higgs-boson tag, and another jet with a
top-quark tag and ${\ge}2$~$b$-tags (\VR{7}\ and \VR{8}) are shown in Figures~\ref{fig: 94 VLQ_m}\ and~\ref{fig: 49 VLQ_m},
respectively.
There is also agreement between the observed and predicted distributions.
 
\FloatBarrier
\section{Systematic uncertainties}
\label{sec:systematics}
 
 
Systematic uncertainties that affect the interpretation of the data are estimated
using data and MC samples.
Variations corresponding
to a $+ 1 \sigma$ and $-1 \sigma$ confidence interval
are derived for each uncertainty.
 
These systematic uncertainties are broken down into
detector-related and modeling uncertainties.
They do not have a significant dependence on the choice
of $T$-quark mass and coupling, so an example of the size of the
systematic uncertainties arising from the likelihood fit described in
Section~\ref{sec:result}\ (the ``post-fit'' results) is provided in
Table~\ref{tab:sys_impact_data} for $\mTVLQ = 1.6$~\TeV\ and $\kappaT = 0.5$.
 
\subsection{Detector-related uncertainties}
 
The most significant detector-related systematic uncertainties arise from
the measurements of jet properties and tagging efficiencies.
 
Uncertainties associated with the \largeR\ jets arise from the jet energy scale (JES), jet mass scale (JMS),
jet mass response (JMR), jet energy resolution (JER), and the JVT requirement.
The uncertainties in the JES, JMS, and JMR are evaluated by using in situ measurements~\cite{JETM-2018-02}.
The JES is measured in events where a \largeR\ jet recoils against well-defined reference objects (photons,
$Z$~bosons, or calibrated \smallR\ jets).
The JMS and JMR uncertainties are measured using both a double-ratio method that compares the
calorimeter-to-tracker response ratio between data and simulation~\cite{JETM-2018-02} and a fit to the
$W$-boson mass peak in high-$\pT$ lepton+jets \ttbar{} events.
The JER uncertainty is measured by studying dijet mass resolution and the effect of
energy flow near the jet radius~\cite{JETM-2018-05}.
The JVT uncertainty arises from the correction factors used to match the efficiencies in the
MC samples to data.
 
The efficiency for tagging $b$-jets is measured in data using dilepton \ttbar{} events \cite{FTAG-2018-01}.
Correction factors are applied to the jets in the MC sample
so that the $b$-jet tagging efficiency as a function of jet $\pT$ in MC events matches that in data events.
Uncertainties arising in the evaluation of the efficiencies are propagated to the correction factors.
The largest source of $b$-jet tagging uncertainty is the extrapolation of tagging efficiencies to
$b$-jets with $\pT>300$~\GeV, as $b$-jet tagging calibrations use data with $\pT < 300$~\GeV.
 
The efficiency and rejection power of the DNN top-quark tagger is measured in data and correction
factors are applied to MC events to match the measured efficiencies~\cite{ATL-PHYS-PUB-2020-017}.
These corrections take into account the correlations between the tagging efficiencies and
other jet observables such as the jet energy and mass.
The uncertainties in these corrections are treated as systematic uncertainties.
 
The efficiency of the $\tau_{21}$\ requirement used for the Higgs-boson tagger
is measured using the calorimeter-to-tracker response double-ratio method~\cite{JETM-2018-02}.
The corresponding uncertainty, which is approximately 2\%,
is included in the uncertainty of the Higgs-boson-tagger efficiency.
 
The relative uncertainty in the integrated luminosity is determined to be 1.7\%~\cite{ATLAS-CONF-2019-021}, obtained using the LUCID-2 detector~\cite{LUCID2}\ for the primary
luminosity measurements.
 
\subsection{Modeling and background uncertainties}
\label{sec:modeling uncertainties}
The most significant modeling uncertainties arise from the MC calculations
of the \ttbar\ production process and decay into the all-hadronic final
states, the modeling of the non-all-hadronic \ttbar\ background,
the cross-sections for processes producing smaller backgrounds involving at least one top quark,
and the multijet background estimates.
 
The $\ttbar$ background estimate has systematic uncertainties from initial/final-state
radiation (ISR/FSR), the
renormalization scale, factorization scale, PDF, parton-shower algorithm,
matrix-element calculation, and \hdamp parameter value.
The effects of ISR/FSR, renormalization scale, and factorization scale uncertainties are evaluated
using the method described in Section~\ref{sec:datamcsamples}.
The PDF uncertainties are evaluated by use of the PDF4LHC15 Hessian uncertainties, where
the 30 variations are combined into one nuisance parameter.
Uncertainties arising from the choice of parton-shower and hadronization algorithms
are evaluated by comparing the nominal \POWPY{8} sample with the
\POWHER sample.
The uncertainty arising from the matrix-element calculation is assessed by
comparing the nominal MC sample with the \MGNLOPY{8} sample.
 
Although the non-all-hadronic \ttbar\ background is relatively small in the \SR\ and \NR,
a 5\%\ excess of predicted events relative to the data is observed in \VR{1}\ defined by the
event-tagging state with the
leading \largeR\ jet top-quark-tagged with 1~$b$-tag and the second-leading jet
Higgs-boson-tagged with 1~$b$-tag.
This validation region is estimated to have a non-all-hadronic \ttbar\ background fraction of
approximately 15\%\ and it is
possible that the observed excess is due to mismodeling of this background.
A conservative uncertainty of
$62\%$, which covers the excess if it is attributed entirely to the
non-all-hadronic \ttbar\ background, is applied to the size of this background in the \SR\ and \NR.
 
The uncertainty in the multijet background estimate is approximately $4\%$,
as described in Section~\ref{subsec: TTbarDiffXs}.
The uncertainty in the predicted single-top-quark background estimate is $75\%$\
while the uncertainty in the predicted
$t\bar{t} + \Wboson/\Zzero/H$ background estimate is $22\%$, as described in
Section~\ref{subsec: MiscBkg}.
 
\begin{table}[t]
\centering
\caption{Size of the post-fit uncertainties in the $T$-quark signal
cross-section for a $T$-quark mass of 1.6~\TeV\ and coupling $\kappaT = 0.5$.
The fitted cross-section is $-10$~fb and is consistent with zero.
The background uncertainty is the sum in quadrature of the systematic uncertainty on the
multijet background and the statistical uncertainties on the MC-derived backgrounds.
The total uncertainty of $25$~fb is the sum in quadrature of the total systematic uncertainty and
statistical uncertainty. The uncertainty arising from simultaneously fitting the \ttbar\ normalization factor is included in the total systematic uncertainty.
The individual uncertainties do not add up in quadrature to the total uncertainty because of their correlations in the fit.
}
\footnotesize
\begin{tabular}{ l  c }
\toprule
Category              & Uncertainty in $\sigma$ ($pp \rightarrow T + X \rightarrow Ht + X$) [fb]   \\
\midrule
\multicolumn{2}{c}{Detector Uncertainties} \\
\midrule
$b$-jet tagging                        & $6.1$          \\
Top-quark jet tagging & $5.9$ \\
Jet mass resolution           & $3.0$         \\
Jet mass scale                & $2.3$        \\
Jet energy scale           & $1.8$          \\
Jet energy resolution           & $1.7$          \\
Higgs-boson tagging & $1.6$ \\
Other detector uncertainties & ~~$0.3$ \\
\midrule
\multicolumn{2}{c}{Modeling Uncertainties} \\
\midrule
Other $t\bar{t}$ modeling uncertainties & $4.9$ \\
$t\bar{t}$ parton shower and hadronization                  & $1.9$     \\
$t\bar{t}$ matrix element                   & $2.4$     \\
\midrule
Background uncertainty & $7.3$     \\
Signal MC statistical uncertainty & $4.9$     \\
\midrule
\ttbar\ normalization ($\alpha^{\mathrm{fit}}$)& $1.5$ \\
Other top-quark-background theory uncertainties & ~~$1.8$ \\
\midrule
\multicolumn{2}{c}{Total Uncertainties} \\
\midrule
Total statistical uncertainty     & $19$~~~~~      \\
Total systematic uncertainty      & $15$~~~~~      \\
\midrule
Total uncertainty                 & $ 25$~~~~~    \\
\bottomrule \\[0.2 cm]
\end{tabular}
\label{tab:sys_impact_data}
\end{table}

\section{Results}
\label{sec:result}
 
\begin{figure}[t]
\centering
\subfloat[ ]{
\includegraphics[width=0.47\linewidth]{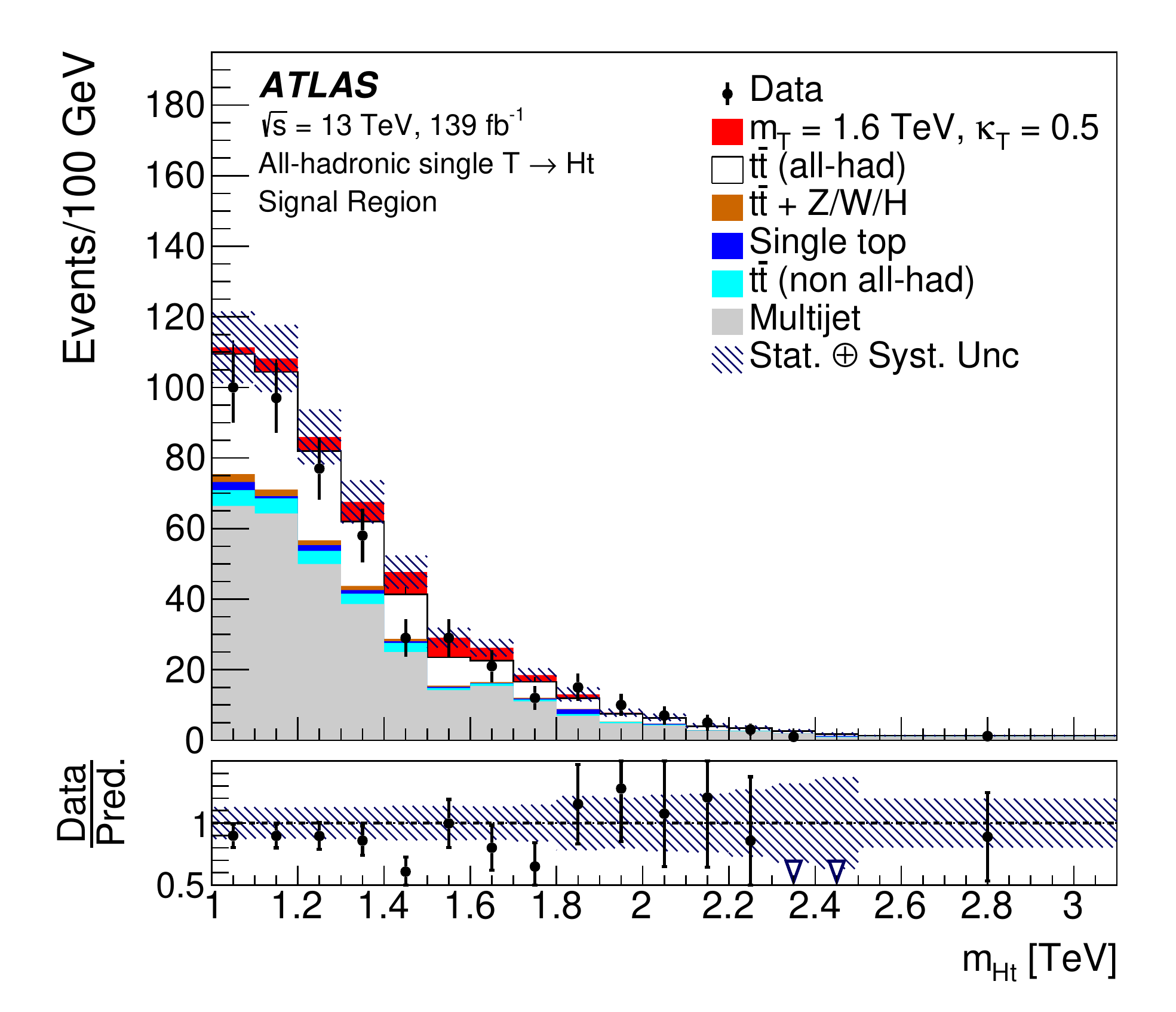}
\label{fig: SR VLQ_m}
}\quad
\subfloat[ ]{
\includegraphics[width=0.47\linewidth]{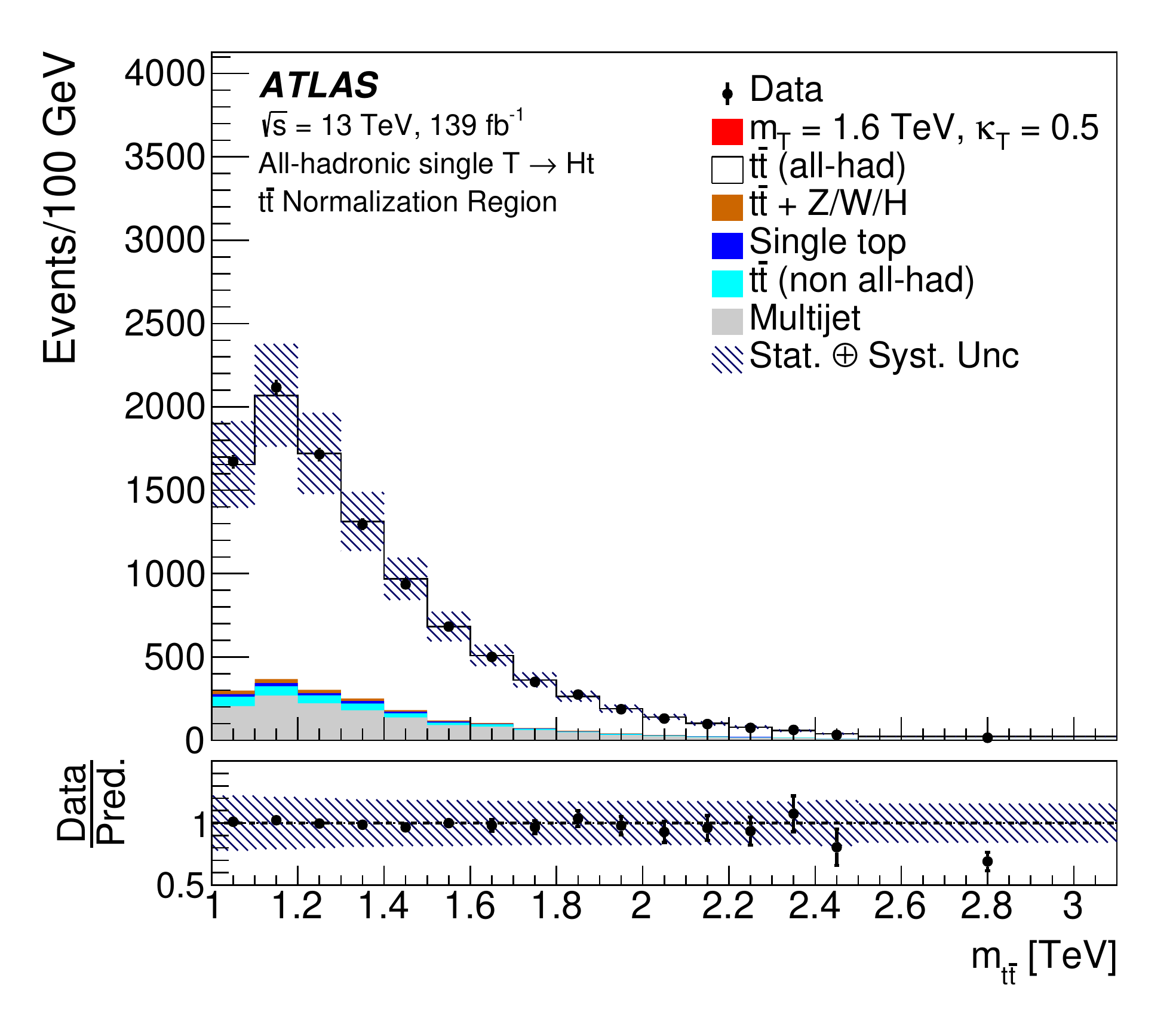}
\label{fig: ttNR VLQ_m}
}
 
\caption{Dijet invariant mass distributions
in (a) the \SR\ and (b) the \NR\
before the fit of the signal and background model to the data.
A $T$-quark hypothesis with $\mTVLQ = 1.6$~\TeV\ and $\kappaT = 0.5$ is used in these plots.
The blue hashed lines correspond to the sum in quadrature of the statistical and systematic uncertainties of the prediction in a given bin.
The lower panels show the ratio of the data to the prediction, along with the
uncertainty in the ratio.
A ratio outside the bounds of the axis is represented by a blue arrow.
The last bin includes the event overflows.
Contributions to the predicted distributions are stacked in the same order as they appear in the legend.
}
\label{fig:Prefit}
\end{figure}

The dijet invariant mass formed from the tagged \largeR\ jets
is interpreted as a combination of the expected SM backgrounds and a $T$-quark signal.
The dijet mass in the \SR\ is the invariant mass of the Higgs-boson and top-quark candidates
while in the \NR\ it is the invariant mass of the two top-quark candidates.
The dijet invariant mass distributions for the \SR\ and \NR\ are shown in
Figure~\ref{fig:Prefit}, assuming a $T$-quark signal contribution with
$m_T=1.6$~\TeV\ and $\kappa_T = 0.5$\ scaled to the theory cross-section of 41~fb.
The overall acceptance times efficiency of $T$-quark detection in the
all-hadronc final state is 1.6\%\ for this choice of mass and
couplings, taking into account the kinematic requirements and
tagging efficiencies.
The predicted background rates and shapes are in good agreement with the observed distributions.
 
\begin{table}[t]
\caption{The event yields are shown for the \ttbar\ normalization region and for the signal region
after the likelihood fit (post-fit).
The requirement of the $\Ht$\ invariant mass being greater than 1 \TeV\ has been imposed on
the signal region.
The post-fit yields incorporate the statistical, detector,
and modeling uncertainties as well as their correlations.
} \label{tab:yields}
\begin{center}
\sisetup{
table-number-alignment=right,
separate-uncertainty=true,
table-figures-integer = 2,
table-figures-decimal = 0
}
\begin{tabular}{
l
S[separate-uncertainty=true,table-figures-uncertainty=1]
c
S[separate-uncertainty=true,table-figures-uncertainty=1]
S[separate-uncertainty=true,table-figures-uncertainty=1]
c
S[separate-uncertainty=true,table-figures-uncertainty=1]
}
\toprule
Region &
\multispan{3}{\hfil \ttbar\ Normalization \hfil}  &
\multispan{3}{\hfil \Ht\ Signal Region \hfil} \\
& \multispan{3}{\hfil  Post-Fit\hfil}  &
\multispan{3}{\hfil  Post-Fit \hfil} \\
 
\hline

\vspace{.10cm}
 
\ttbar{} all-hadronic 		&  8366 & $\pm$ &  216 & 147 & $\pm$ & 17  \\
 
\vspace{.10cm}

\ttbar{} non-all-hadronic 		&  189 &$\pm$&  133 & 14 & $\pm$ & 10 \\
 
\vspace{.10cm}
 
Single top-quark 			&  92 &$\pm$&  49 & 8 & $\pm$ & 6 \\
 
\vspace{.10cm}
 
\ttbar{}$+W/Z/H$			&  117 &$\pm$&  25 & 9 & $\pm$ & 2 \\
 
\vspace{.10cm}
 
Multijet events				&   1452 &$\pm$&  57 & 316 & $\pm$ & 9  \\
 
\hline
 
Signal events (\mTVLQ\ = 1.6 TeV, \kappaT\ = 0.5)  &   & &  & -9 & $\pm$ & 21  \\

\vspace{.10cm}
 
Predicted background ${}^{\phantom{1}^{\phantom{1}}}$	&  10216 &$\pm$ &  150 & 494 & $\pm$ & 22 \\
 
\vspace{.10cm}
 
Data (\lumitot) 	 					& 10231 &  &  & 471 &  &   \\
\bottomrule \\[0.20cm]
\end{tabular}

\end{center}
\end{table}
 
The dijet mass is used as a discriminant in
the \SR\ and \NR\ to test for the presence of a $T$-quark signal.
Two parameters of interest are defined:
$\sigma_{\textrm obs}$, the observed cross-section for single production of a $T$-quark,
and \alphafit, the \SR\ and \NR\ \ttbar\ background normalization.
 
A binned-likelihood fit is performed in which a $T$-quark signal and the background
model is fitted to the \SR\ dijet mass distribution
and simultaneously the \NR\ background model is fitted
to the \NR\ dijet mass distribution.
The fit is performed for events with a dijet mass greater than 1~\TeV.
The fit model in the \SR\ is the sum of the background distributions and a $T$-quark
signal distribution with a
given mass, coupling, and signal cross-section $\sigma_\text{obs}$.
In the \NR\ the very small contribution from the $T$-quark  signal is neglected.
The signal cross-section is allowed to take negative values in the fit
whereas \alphafit\ is constrained to be positive.
The fit of the \ttbar\ background in the \NR\ and \SR\
measures \alphafit\ using both regions and thus provides a scaled \ttbar\ background
contribution in the \SR.
 
The fit incorporates the systematic uncertainties as Gaussian nuisance parameters.
Additional bin-by-bin uncertainties are included to account for the
statistical uncertainties in the predicted multijet and MC backgrounds.
The \ttbar\ contributions to the \NR\ and \SR\ are fully correlated in the fit.
The likelihood is then profiled~\cite{bootstrap_bohm2010}\ as a function of each nuisance
parameter and used as the test statistic to determine the statistical significance of the
fit results.
 
\begin{figure}[t]
\centering
\subfloat[ ]{
\includegraphics[width=0.47\linewidth]{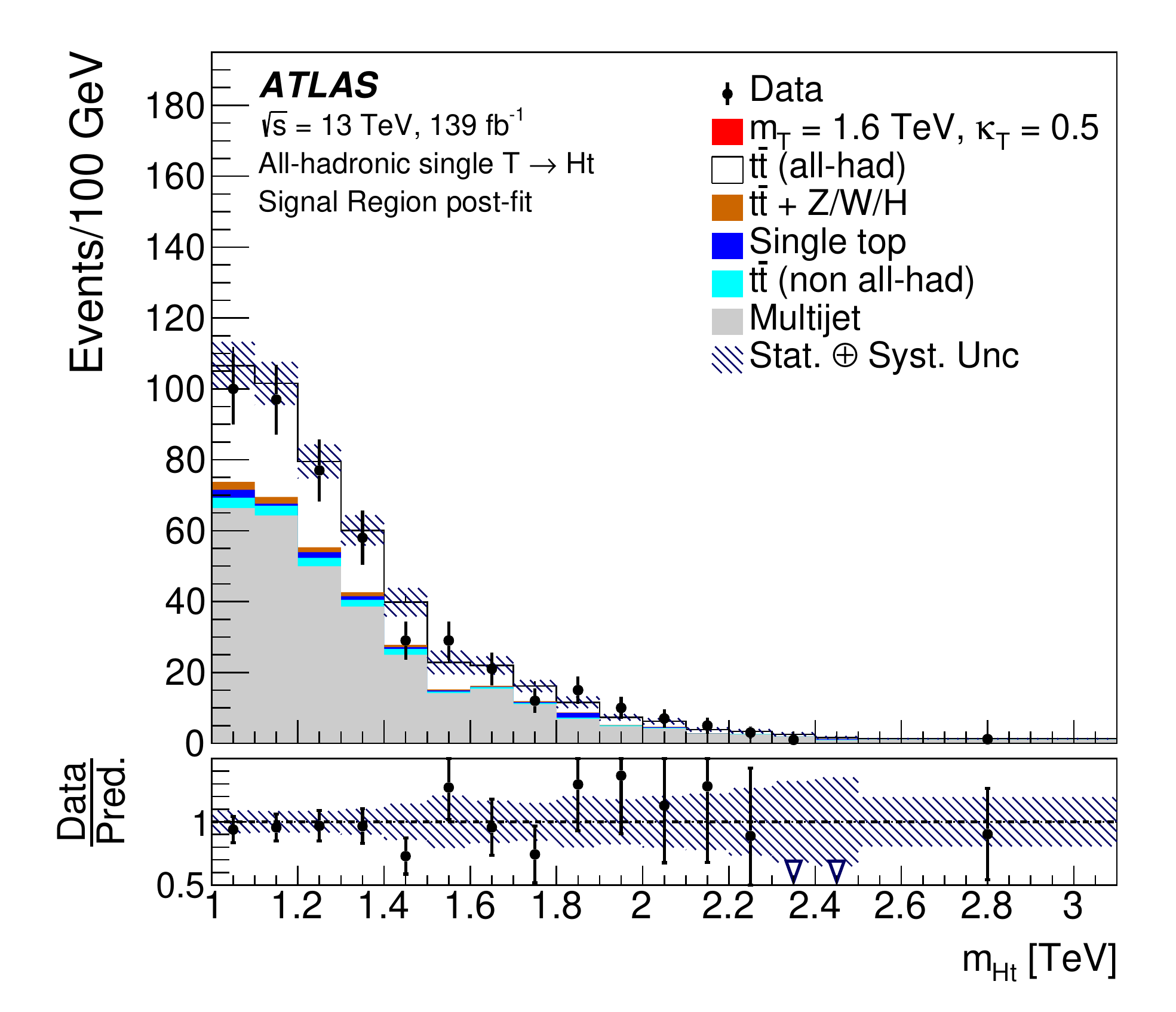}
\label{fig:SRPostFit}}
\quad
\subfloat[ ]{
\includegraphics[width=0.47\linewidth]{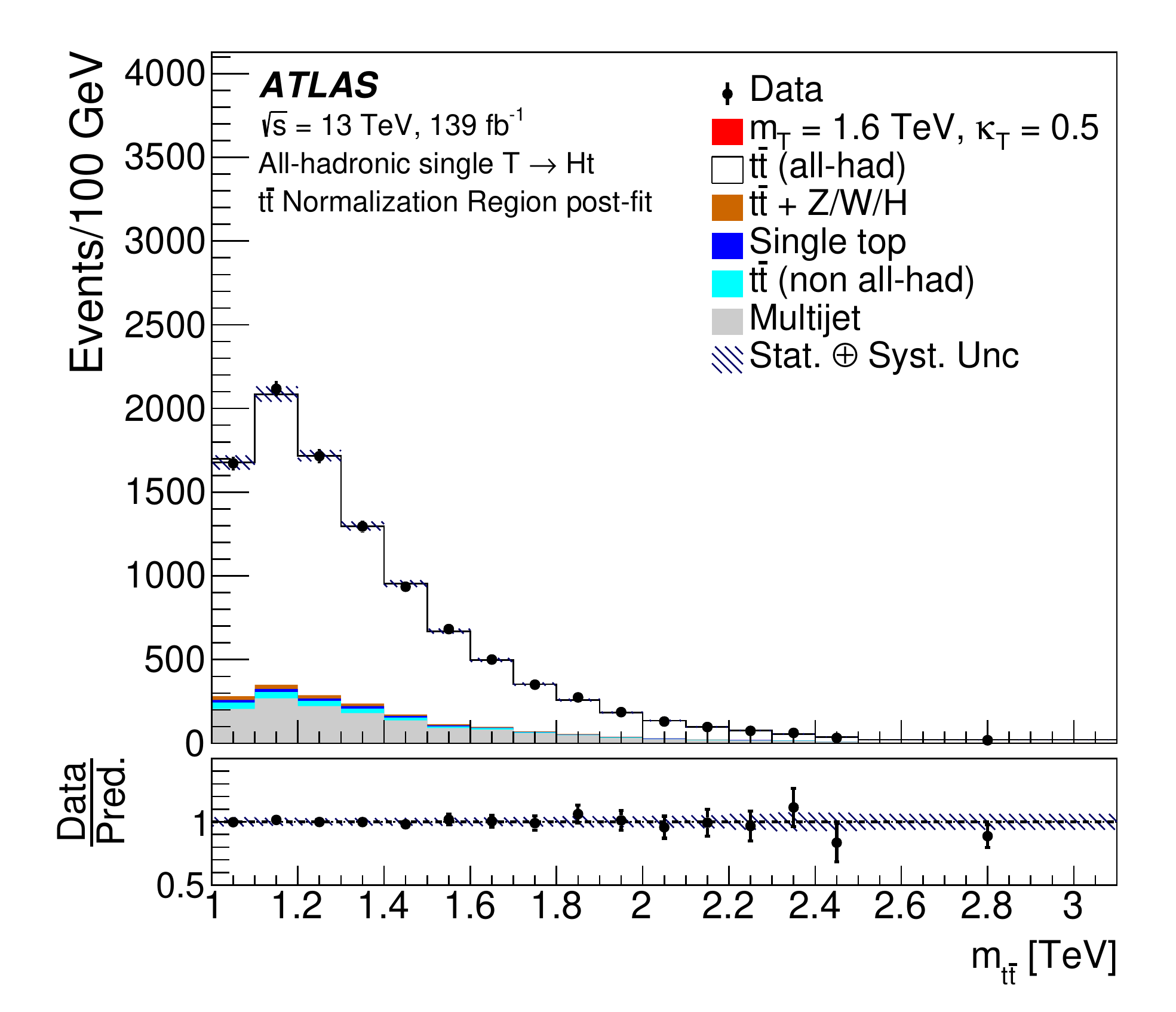}
\label{fig:ttNRPostFit}}
 
\caption{Dijet invariant mass distributions for (a) the \SR\ and (b) the
\NR\ showing the results of the model when
fitted to the data.
A $T$-quark hypothesis with $\mTVLQ = 1.6$~\TeV\ and $\kappaT = 0.5$ is used in the fit.
Since the central value of the fitted $T$-quark cross-section is negative, the predicted
\SR\ mass distribution shows no contribution from the signal.
The blue hashed lines correspond to the sum in quadrature of the statistical and systematic
uncertainties of the prediction.
The lower panels show the ratio of the data to the prediction, along with the
uncertainty in the ratio.
A ratio outside the bounds of the axis is represented by a blue arrow.
The last bin includes the event overflows.
Contributions to the predicted distributions are stacked in the same order as they appear in the legend.
}
\label{fig:PostFit}
\end{figure}

Figure~\ref{fig:PostFit} shows the dijet mass distributions for
the \SR\ and \NR\
after the fit (post-fit) assuming a signal hypothesis with $\mTVLQ = 1.6$~\TeV\ and $\kappaT = 0.5$.
The observed and predicted event yields in the \NR\ and \SR\ are given in Table~\ref{tab:yields}.
The fitted value of $\alphafit = 0.79\pm 0.12$\ is consistent with the \ttbar\ normalization factor
$\alphanorm\ = 0.82 \pm 0.01$\
determined from the background-subtracted event yield in the \NR\
(the uncertainty on \alphanorm\ is statistical only).
There is good agreement between the predicted post-fit signal region background yield of $494 \pm 22$\
events and the observed yield of $471$ events, consistent with no significant excess in data
above SM backgrounds over the entire \Ht\ invariant mass distribution as seen in Figure~\ref{fig:SRPostFit}.
The fit of the $\mTVLQ = 1.6~\TeV$\ and $\kappaT= 0.5$ signal hypothesis
results in $\sigma(pp \rightarrow T+X \rightarrow Ht+X) = -10 \pm 25$~fb,
further confirming no excess of events at $Ht$\ masses around 1.6~\TeV.
 
Similarly, fit results with $T$-quark cross-sections consistent with zero are obtained for
$T$-quark masses between 1.0 and 2.3 \TeV\ and for $\kappaT$\ values from 0.1 to 1.6.
Based on these fit results, for $1.0 < \mTVLQ <  2.3$~\TeV\ there is no significant evidence of a $T$~quark decaying
to the \Ht\ final state.
 
\begin{figure}[t]
\centering
\subfloat[ ]{
\includegraphics[width=0.47\linewidth]{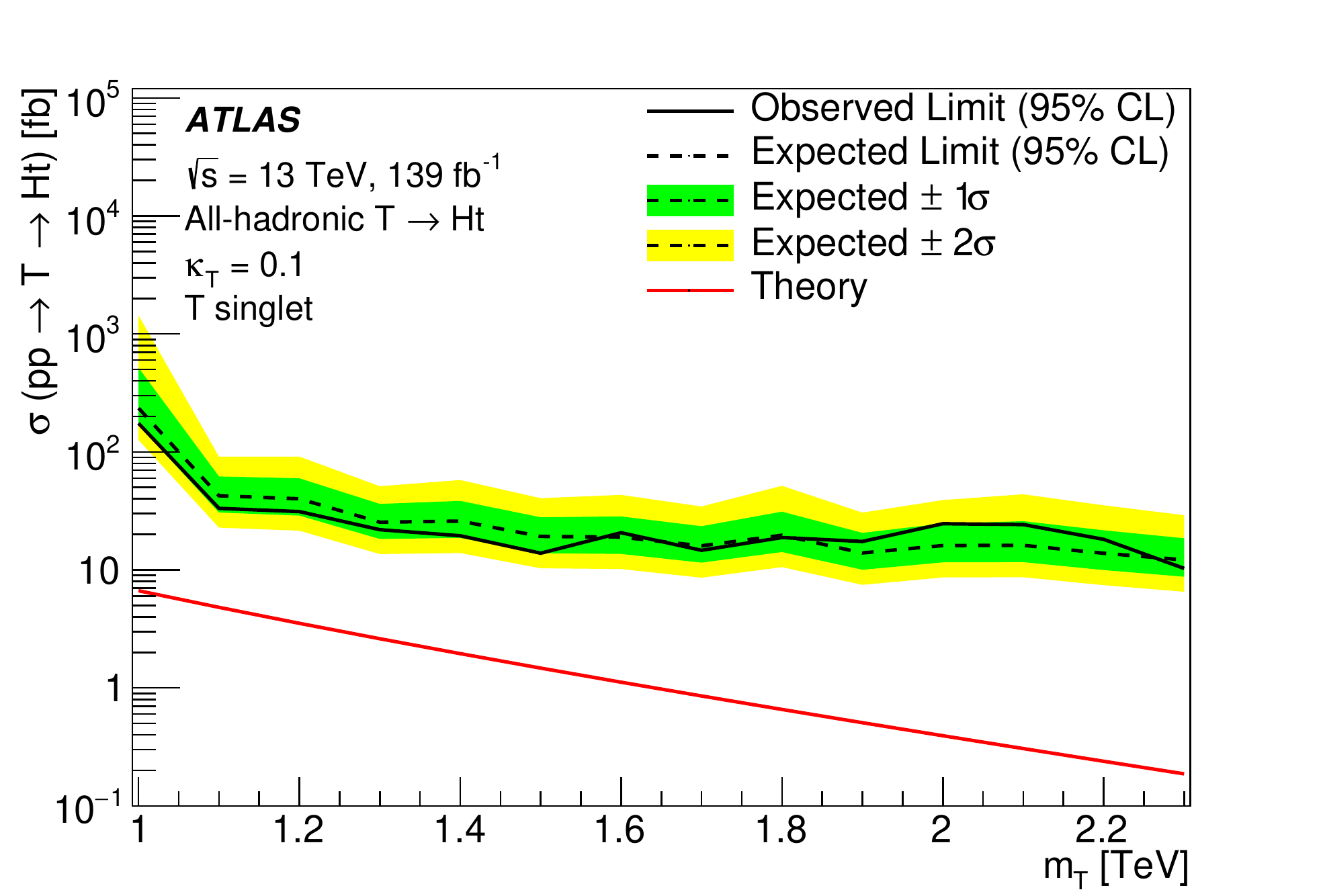}
\label{fig:Kappa01}
}\quad
\subfloat[ ]{
\includegraphics[width=0.47\linewidth]{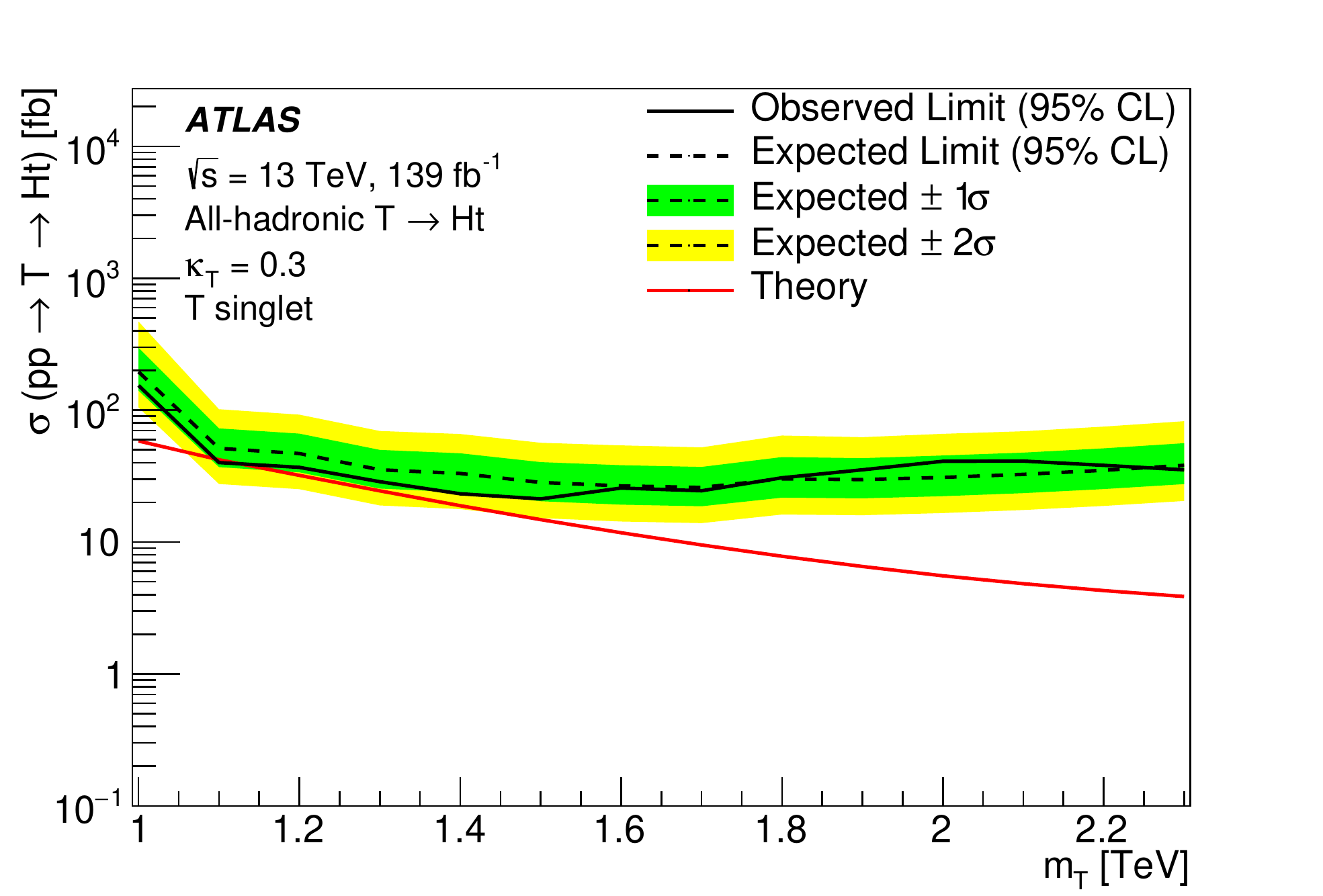}
\label{fig:Kappa03}
}
 
\subfloat[ ]{
\includegraphics[width=0.47\linewidth]{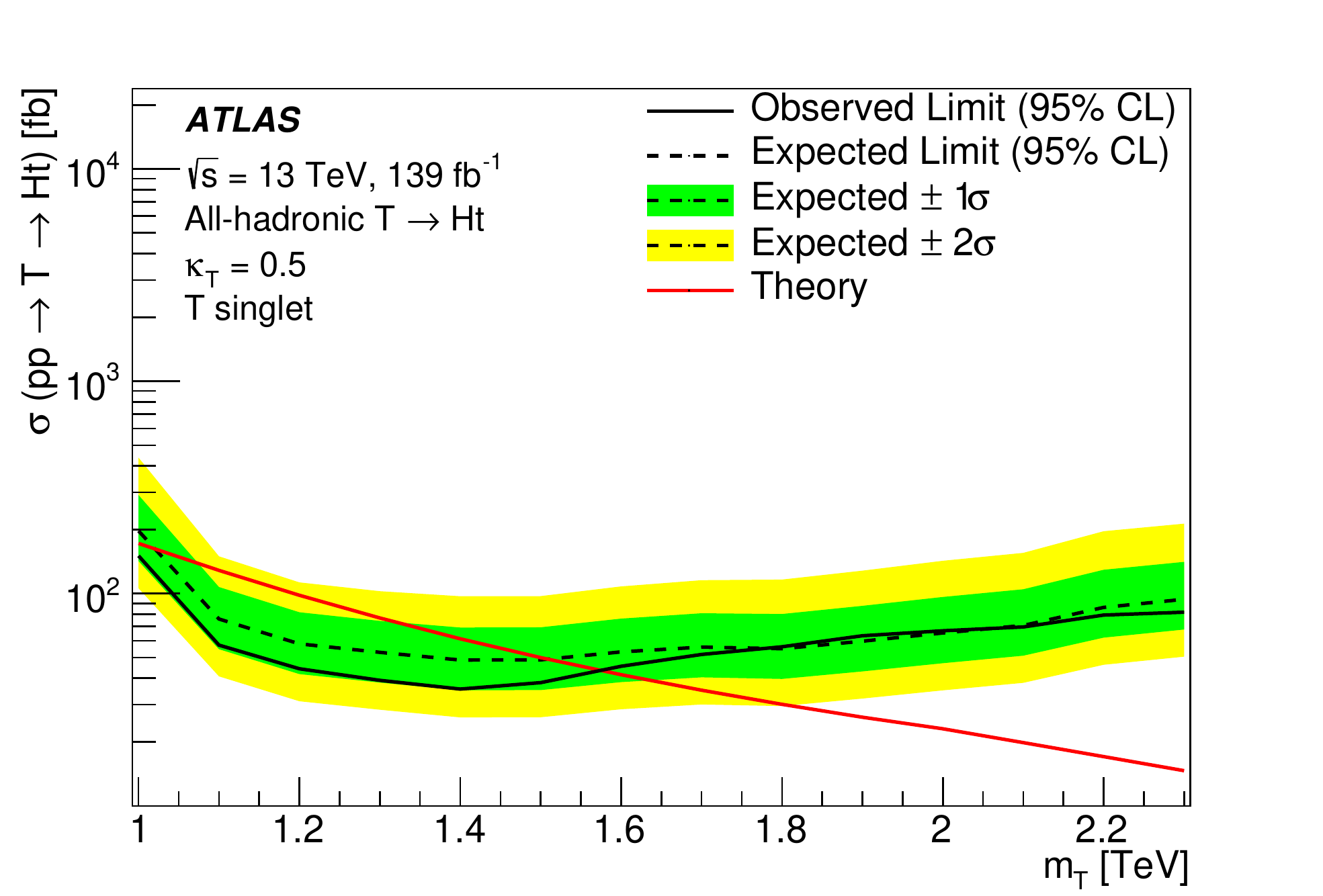}
\label{fig:Kappa05}
}\quad
\subfloat[ ]{
\includegraphics[width=0.47\linewidth]{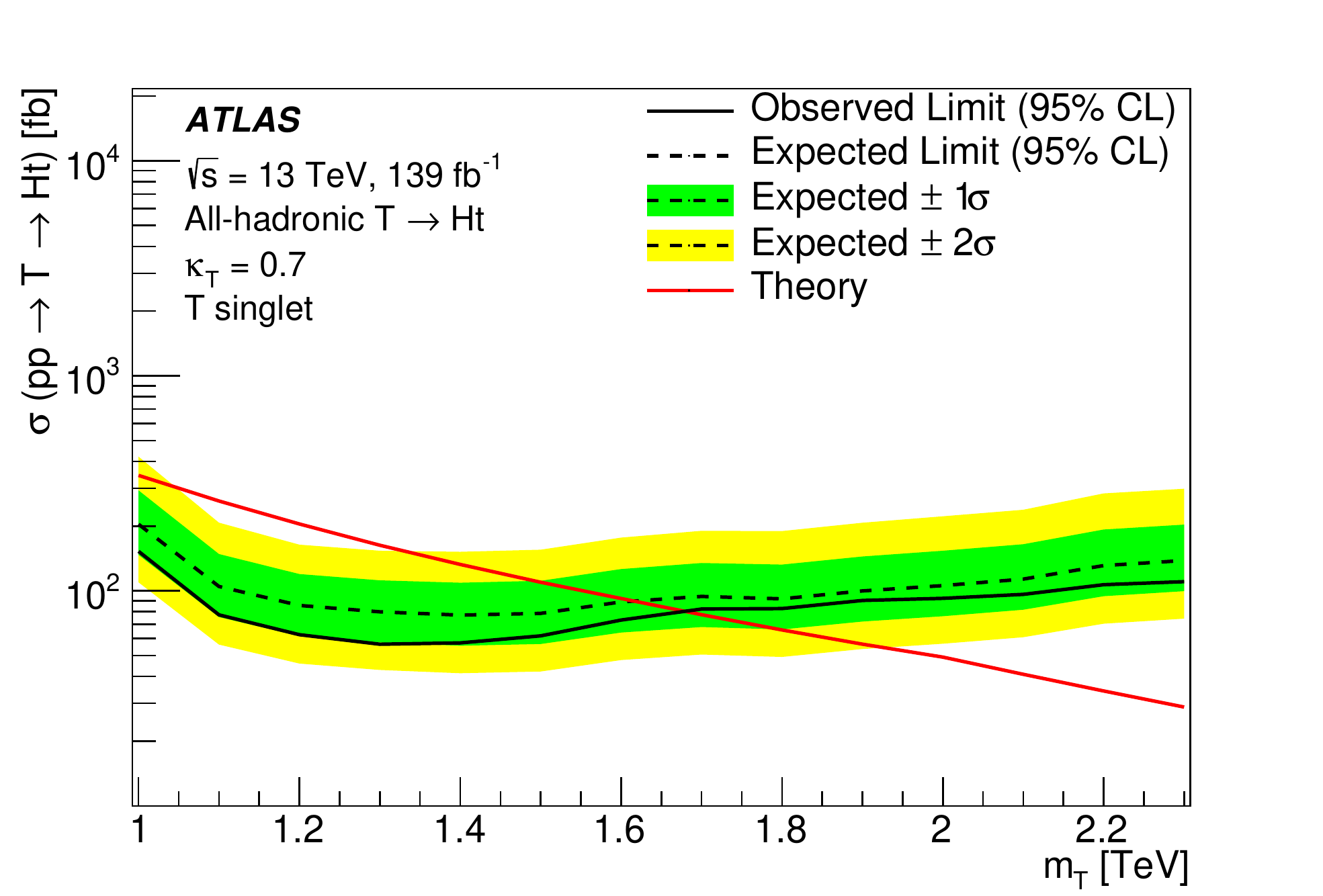}
\label{fig:Kappa07}
}

\subfloat[ ]{
\includegraphics[width=0.47\linewidth]{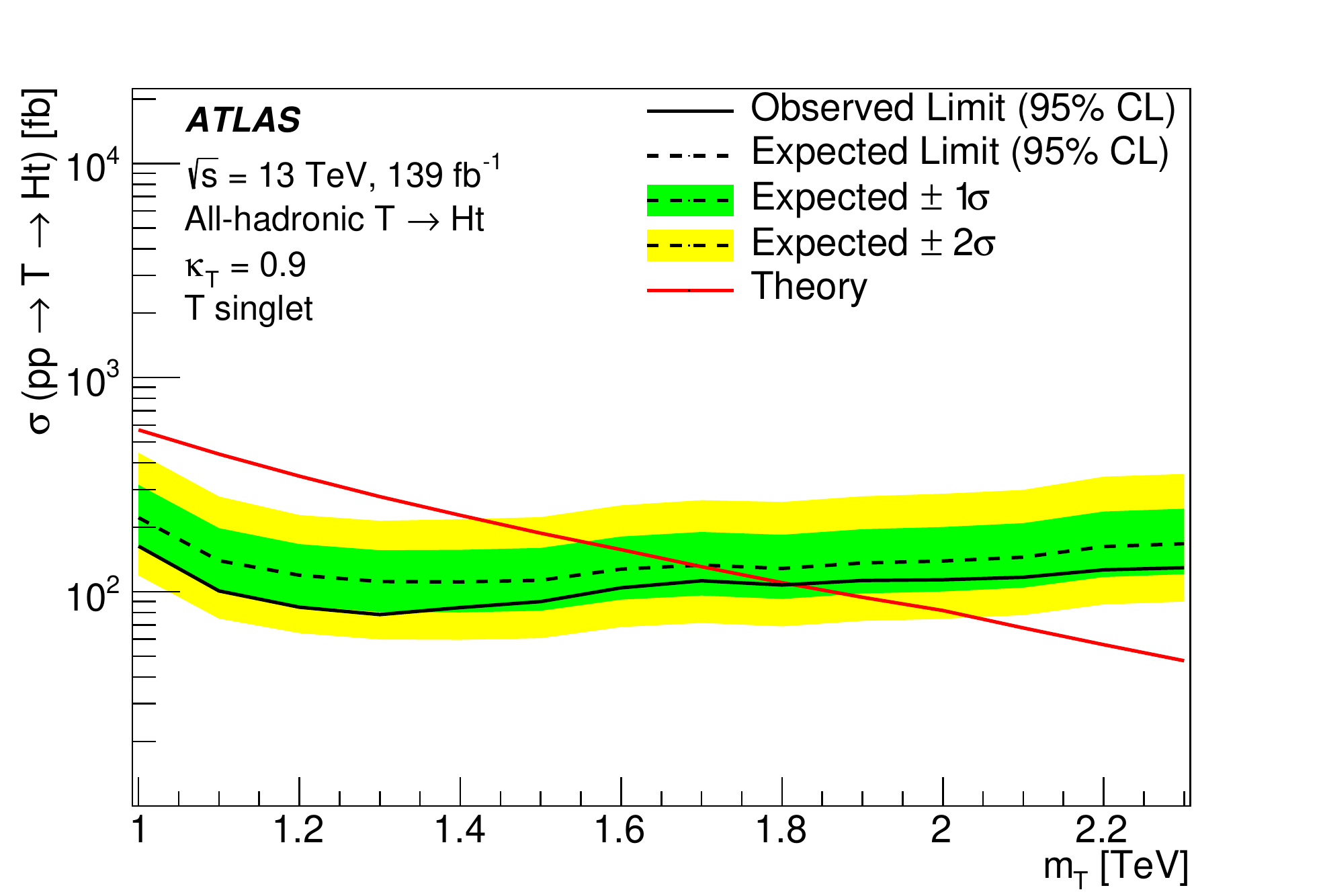}
\label{fig:Kappa09}
}\quad
\subfloat[ ]{
\includegraphics[width=0.47\linewidth]{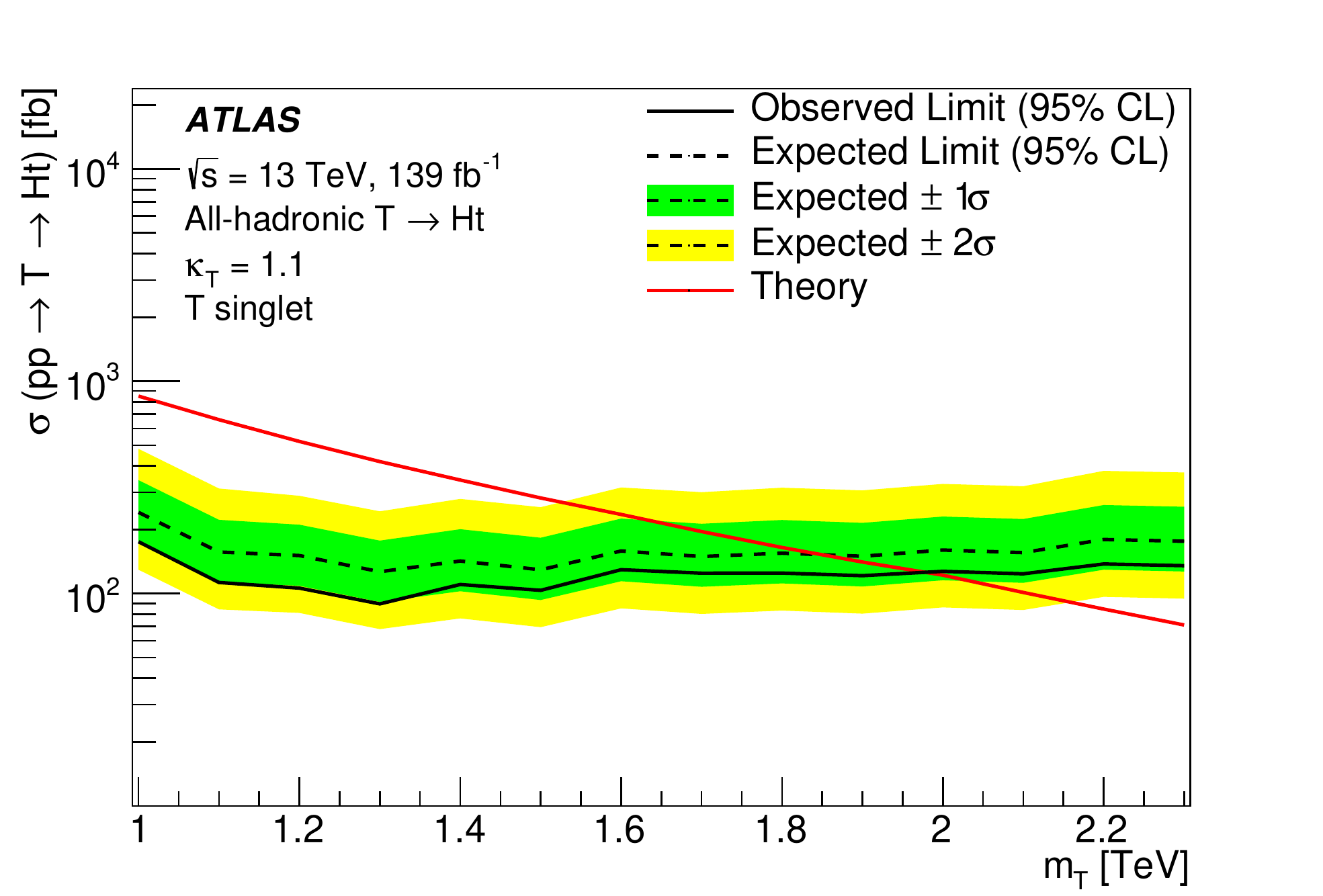}
\label{fig:Kappa11}
}
 
\caption{Observed and expected 95\%\ CL upper limits on the single $T$-quark production cross-section as a
function of $\mTVLQ$\ for values of $\kappa_T$ ranging from 0.1 to 1.1.
The green (yellow) band is the 68\% (95\%) confidence interval
around the median expected limit, as determined using pseudo-experiments.
The predicted cross-sections of single $T$-quark production
are shown in red.
}
\label{fig:MassXVals}
\end{figure}

\begin{figure}[t]
\centering
\subfloat{
\includegraphics[width=\linewidth]{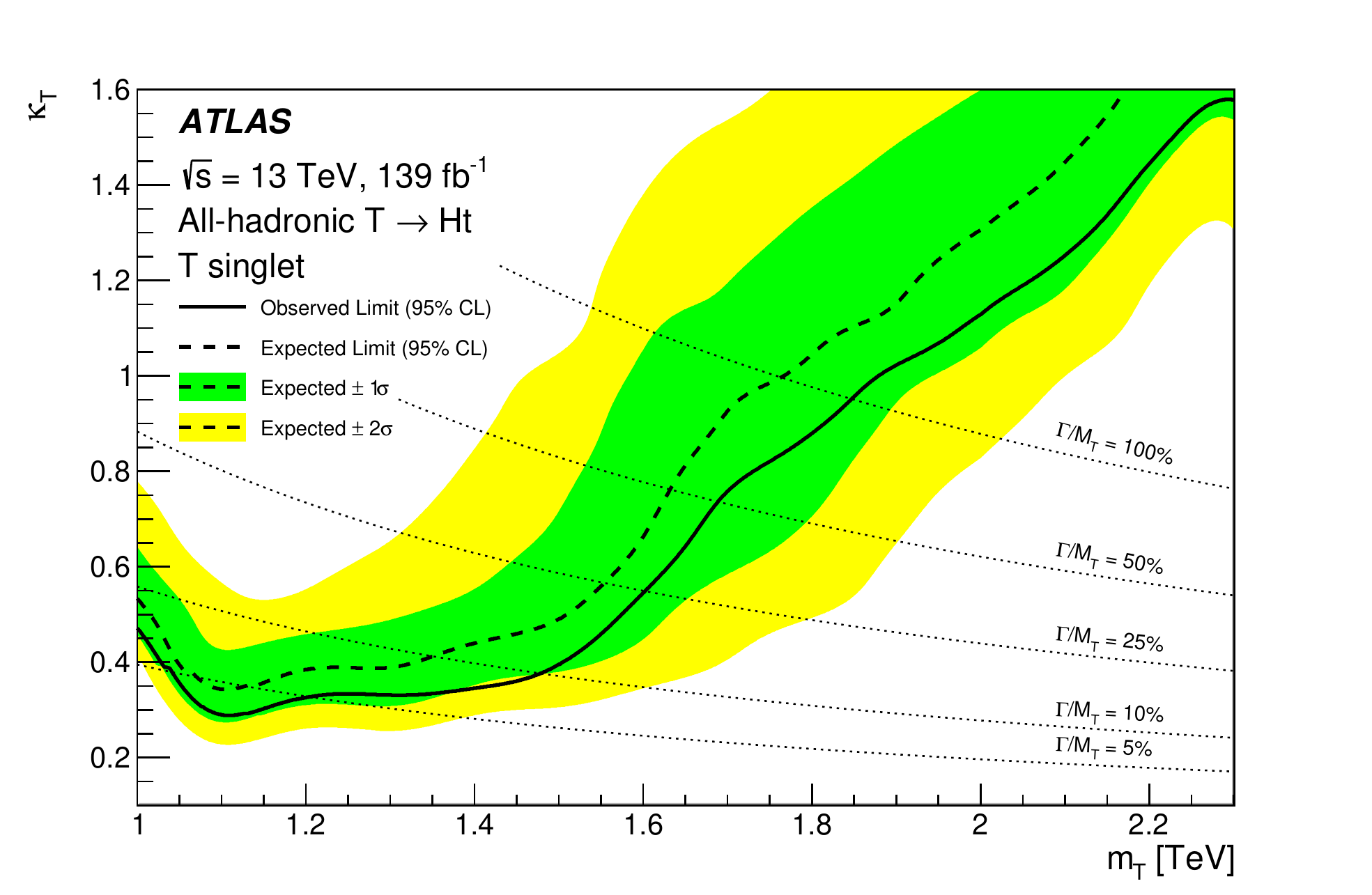}}
\caption{Observed and expected 95\%\ CL upper limits on the single $T$-quark
coupling $\kappa_T$\ as a function of $\mTVLQ$\ are shown as solid and
dashed lines, respectively.
The green (yellow) band is the 68\% (95\%) confidence interval around
the median expected limit, as determined using pseudo-experiments.
All values of $\kappa_T$\ above the solid line are excluded.
The dashed curves represent contours of fixed $\Gamma/\mTVLQ$.
}
\label{fig:MassVsKappaAll}
\end{figure}
 
\begin{figure}[t]
\centering
\subfloat[ ]{
\includegraphics[width=0.49\linewidth]{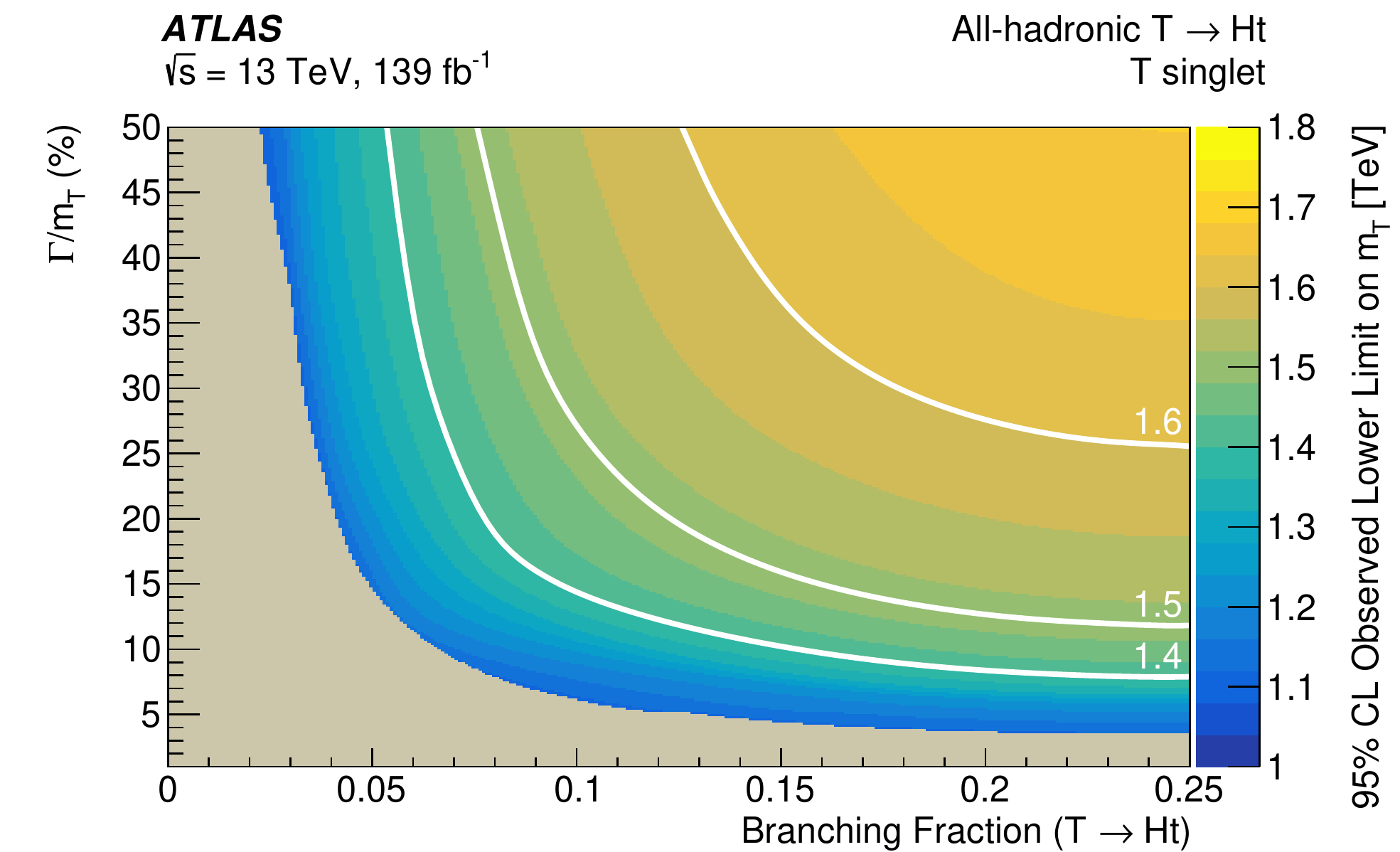}
\label{fig:2DMassLimitHtObs}}
\medskip
\subfloat[ ]{
\includegraphics[width=0.49\linewidth]{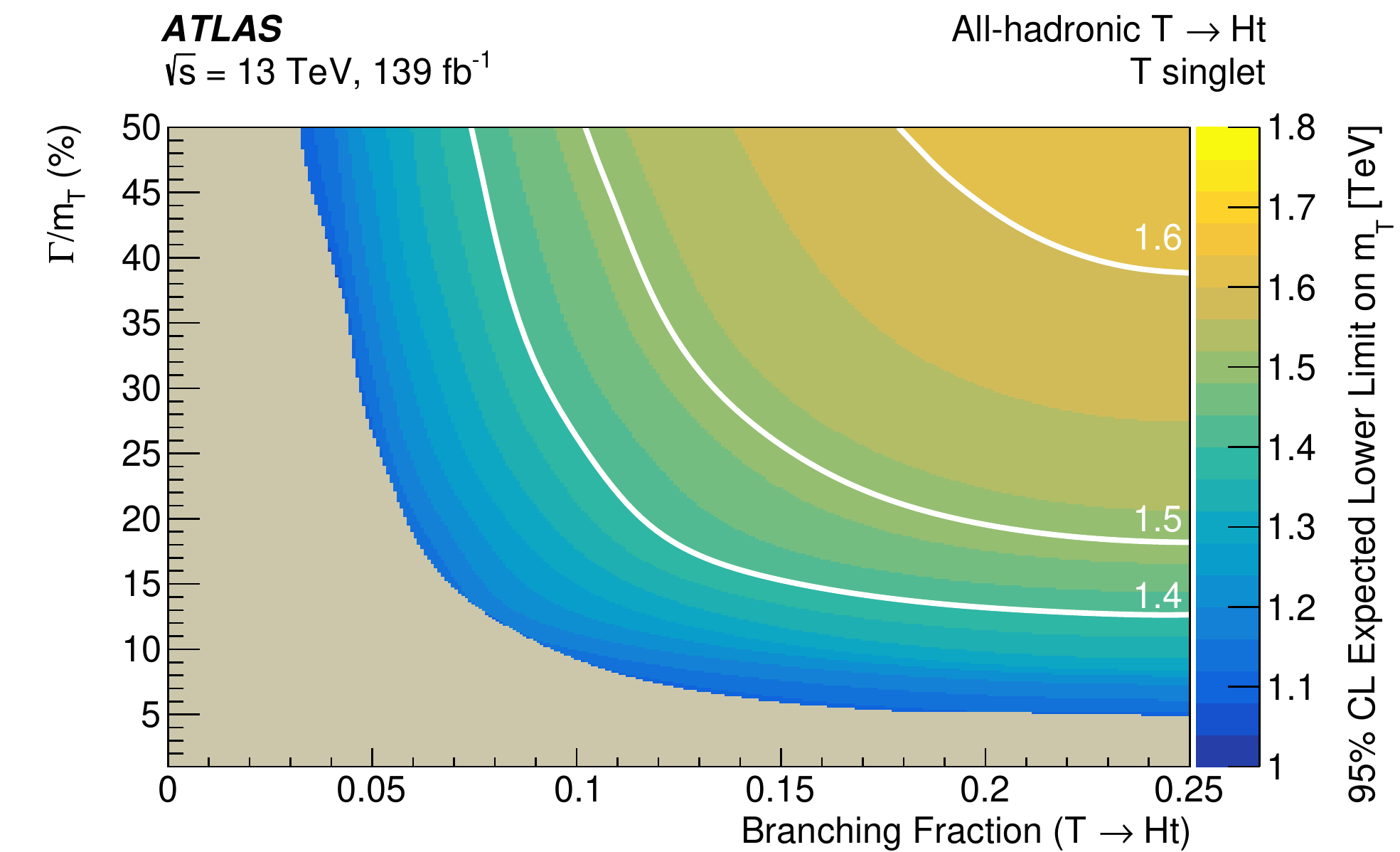}
\label{fig:2DMassLimitHtExp}}
\caption{
Observed (a) and expected (b) 95\%\ CL lower
limits on the $T$-quark mass as a function of the $T$-quark width-to-mass ratio and the branching
fraction of the $T \rightarrow Ht$\ decay ($\Gamma_T$\ is the $T$-quark width) .
The branching fractions (${\cal{B}}$) for $T\rightarrow Ht$\
and $T\rightarrow Zt$\
are kept equal.
The branching fraction for $T\rightarrow Wb$\ is
$1-{\cal{B}}(T\rightarrow Ht)-{\cal{B}}(T\rightarrow Zt)$.
The colour scale on the right side of each plot defines the 95\%\ CL limit on the
$T$-quark mass.
Masses below the observed limit are excluded.
The dashed white contour lines denote isolines of equal exclusion on the mass in units of \TeV.
}
\label{fig:2DMassLimitHt}
\end{figure}

The fit results are used to set 95\% CL upper limits on the single-$T$-quark production cross-section
for $1.0 < \mTVLQ < 2.3$~\TeV\ and $0.1 < \kappaT < 1.6$\
using the CL$_{\mathrm{s}}$ method \cite{Cowan_2011}.
The predicted cross-sections assume a singlet $T$~quark
with a $T\to\Ht$\ branching fraction of $1/4$.
Figure \ref{fig:MassXVals} shows the 95\% CL upper limits as a function of \mTVLQ\ for
different values of \kappaT.
The cross-section limits range from ${\sim}10$~fb to ${\sim}200$~fb, depending
on \kappaT.
The decrease in sensitivity for masses from 1.0 to 1.2~\TeV\ arises from the change
in signal shape due to the \pT\ requirements on the Higgs-boson and top-quark candidate jets.
The \pT\ requirements shape the distribution to peak at roughly 1.2~\TeV,
which can be seen in Figures~\ref{fig:Prefit} and~\ref{fig:PostFit}.
Figure \ref{fig:MassVsKappaAll} shows the exclusion limits as a function of \mTVLQ\ and \kappaT.
Figure~\ref{fig:2DMassLimitHt}\ shows the observed and expected 95\%~CL limits on
the $T$-quark mass as a function of the $T$-quark width-to-mass ratio $\Gamma/\mTVLQ$ and the branching fraction
for $T$-quark decay into a Higgs boson and a top quark.

For the considered mass range of 1.0 to 2.3~\TeV\,
the upper limit on allowed values of \kappaT\ rises from a minimum value of 0.3 starting at $\mTVLQ = 1.1$~\TeV,
up to 1.6 for $\mTVLQ\ = 2.3$~\TeV.
 
At 95\%\ CL, this analysis excludes $T$~quarks with $\Gamma/\mTVLQ \ge 0.05$\ for $1.05 < \mTVLQ < 1.2$~\TeV,
with the mass limits rising with $\Gamma/\mTVLQ$\ to exclude $\mTVLQ < 1.7$~\TeV\ for
$\Gamma/\mTVLQ \ge 0.5$.
 
\FloatBarrier

\section{Conclusion}
\label{sec:conclusion}
 
A search is reported for the single production of a vector-like singlet $T$~quark decaying into a
Higgs boson and a top quark both of which decay hadronically.
The search uses 139~fb$^{-1}$ of 13~\TeV\ proton--proton collision data collected with the ATLAS detector at the LHC.
The final states are fully reconstructed by clustering the decay products into two
\largeR\ jets.
The use of fully hadronic decays allows the
direct reconstruction of the $T$-quark final state, increasing the signal-to-background ratio
for the search.
The results significantly extend the sensitivity for the production of $T$ quarks
decaying fully hadronically.
The search sensitivity is further improved by a larger dataset than used previously,
tagging techniques with
greater background rejection, and a data-driven
multijet background estimate that reduces the uncertainty in the background modeling.
The cross-section upper limits are typically a factor of 2 lower than previous searches.
 
The analysis is performed by searching for an excess above SM backgrounds
in the \Ht\ invariant mass distribution.
This distribution shows no evidence of significant contributions from single $T$-quark
production and is consistent with the expected SM background sources.
Therefore, limits are set at 95\% C.L. on the production cross-section of a $T$~quark decaying to
the \Ht\ final state.
These depend on the $T$-quark mass and coupling to SM particles and range from ${\sim}10$~fb to
${\sim}200$~fb, depending on the assumed $\kappa_T$\ value for the couplings.
In the resonance mass range between 1.0 and 2.3~\TeV, the upper limit on the allowed coupling
values rises with \mTVLQ\ from a minimum value of 0.3 for $\mTVLQ = 1.1$~\TeV\ to
1.6 for $\mTVLQ = 2.3$~\TeV.
This analysis excludes $T$~quarks with $\Gamma/\mTVLQ \ge 0.05$\ for $1.05 < \mTVLQ < 1.2$~\TeV,
with the mass limits rising with $\Gamma/\mTVLQ$\ to exclude $\mTVLQ < 1.7$~\TeV\ for
$\Gamma/\mTVLQ \ge 0.5$.
 
These results provide significantly improved mass and coupling limits on vector-like quark
models involving a $T$~quark decaying into a Higgs boson and a top quark.
The exclusion limits set by this analysis extend the limits set by previous searches.
 
 
\FloatBarrier
 
\section*{Acknowledgments}
 

We thank CERN for the very successful operation of the LHC, as well as the
support staff from our institutions without whom ATLAS could not be
operated efficiently.
 
We acknowledge the support of
ANPCyT, Argentina;
YerPhI, Armenia;
ARC, Australia;
BMWFW and FWF, Austria;
ANAS, Azerbaijan;
SSTC, Belarus;
CNPq and FAPESP, Brazil;
NSERC, NRC and CFI, Canada;
CERN;
ANID, Chile;
CAS, MOST and NSFC, China;
Minciencias, Colombia;
MEYS CR, Czech Republic;
DNRF and DNSRC, Denmark;
IN2P3-CNRS and CEA-DRF/IRFU, France;
SRNSFG, Georgia;
BMBF, HGF and MPG, Germany;
GSRI, Greece;
RGC and Hong Kong SAR, China;
ISF and Benoziyo Center, Israel;
INFN, Italy;
MEXT and JSPS, Japan;
CNRST, Morocco;
NWO, Netherlands;
RCN, Norway;
MEiN, Poland;
FCT, Portugal;
MNE/IFA, Romania;
JINR;
MES of Russia and NRC KI, Russian Federation;
MESTD, Serbia;
MSSR, Slovakia;
ARRS and MIZ\v{S}, Slovenia;
DSI/NRF, South Africa;
MICINN, Spain;
SRC and Wallenberg Foundation, Sweden;
SERI, SNSF and Cantons of Bern and Geneva, Switzerland;
MOST, Taiwan;
TAEK, Turkey;
STFC, United Kingdom;
DOE and NSF, United States of America.
In addition, individual groups and members have received support from
BCKDF, CANARIE, Compute Canada and CRC, Canada;
COST, ERC, ERDF, Horizon 2020 and Marie Sk{\l}odowska-Curie Actions, European Union;
Investissements d'Avenir Labex, Investissements d'Avenir Idex and ANR, France;
DFG and AvH Foundation, Germany;
Herakleitos, Thales and Aristeia programmes co-financed by EU-ESF and the Greek NSRF, Greece;
BSF-NSF and GIF, Israel;
Norwegian Financial Mechanism 2014-2021, Norway;
NCN and NAWA, Poland;
La Caixa Banking Foundation, CERCA Programme Generalitat de Catalunya and PROMETEO and GenT Programmes Generalitat Valenciana, Spain;
G\"{o}ran Gustafssons Stiftelse, Sweden;
The Royal Society and Leverhulme Trust, United Kingdom.
 
The crucial computing support from all WLCG partners is acknowledged gratefully, in particular from CERN, the ATLAS Tier-1 facilities at TRIUMF (Canada), NDGF (Denmark, Norway, Sweden), CC-IN2P3 (France), KIT/GridKA (Germany), INFN-CNAF (Italy), NL-T1 (Netherlands), PIC (Spain), ASGC (Taiwan), RAL (UK) and BNL (USA), the Tier-2 facilities worldwide and large non-WLCG resource providers. Major contributors of computing resources are listed in Ref.~\cite{ATL-SOFT-PUB-2021-003}.
 
 
\clearpage

\printbibliography

@string{jhep = {JHEP}}

@string{epj = {Eur. Phys. J.}}

@string{nim = {Nucl. Instrum. Meth. Phys.}}

@article{Sjostrand:2007gs,
      author         = "{Sj\"ostrand}, Torbjorn and Mrenna, Stephen and Skands, Peter Z.",
      title          = "{A brief introduction to PYTHIA 8.1}",
      journal        = "Comput. Phys. Commun.",
      volume         = "178",
      pages          = "852",
      doi            = "10.1016/j.cpc.2008.01.036",
      year           = "2008",
      eprint         = "0710.3820",
      archivePrefix  = "arXiv",
      primaryClass   = "hep-ph",
      reportNumber   = "CERN-LCGAPP-2007-04, LU-TP-07-28,
                        FERMILAB-PUB-07-512-CD-T",
      SLACcitation   = "%%CITATION = ARXIV:0710.3820;%%"
}

@article{ATL-PHYS-PUB-2014-021,
%    author         = "{ATLAS Collaboration}",
%    title          = "{ATLAS Pythia~8 tunes to $7\;\mbox{TeV}$ data}",
%    journal   = "{ATL-PHYS-PUB-2014-021}",
%    url            = "http://cdsweb.cern.ch/record/1966419",
%    year           = "2014",
%}

@article{Ball:2012cx,
      author         = "Ball, Richard D. and others",
      title          = "{Parton distributions with LHC data}",
      journal        = "Nucl.~Phys.~B",
      volume         = "867",
      year           = "2013",
      pages          = "244-289",
      doi            = "10.1016/j.nuclphysb.2012.10.003",
      eprint         = "1207.1303",
      archivePrefix  = "arXiv",
      primaryClass   = "hep-ph",
      reportNumber   = "EDINBURGH-2012-08, IFUM-FT-997, FR-PHENO-2012-014,
                        RWTH-TTK-12-25, CERN-PH-TH-2012-037, SFB-CPP-12-47\,
                        --CERN-PH-TH-2012-037",
      SLACcitation   = "%%CITATION = ARXIV:1207.1303;%%"
}

@article{akt1,
      author         = "Cacciari, Matteo and Salam, Gavin P. and Soyez, Gregory",
      title          = "{The anti-${k_t}$ jet clustering algorithm}",
      journal        = jhep,
      volume         = "\textmd{04}",
      pages          = "063",
      doi            = "10.1088/1126-6708/2008/04/063",
      year           = "2008",
      eprint         = "0802.1189",
      archivePrefix  = "arXiv",
      primaryClass   = "hep-ph",
}

@article{Cacciari:2011ma, 
  	author     	= "Cacciari, Matteo and Salam, Gavin P. and Soyez, Gregory", 
  	title      	= "{FastJet User Manual}", 
  	journal    	= "Eur. Phys. J. C", 
  	volume     	= "72", 
  	year       	= "2012", 
  	pages      	= "1896", 
  	doi        	= "10.1140/epjc/s10052-012-1896-2", 
  	eprint     	= "1111.6097", 
  	archivePrefix  = "arXiv", 
  	primaryClass   = "hep-ph", 
  	reportNumber   = "CERN-PH-TH-2011-297" 
}

@article{Cacciari:2011hy,
      author         = "Cacciari, Matteo and Czakon, Michal and Mangano,
                        Michelangelo and Mitov, Alexander and Nason, Paolo",
      title          = "{Top-pair production at hadron colliders with
                        next-to-next-to-leading logarithmic soft-gluon
                        resummation}",
      journal        = "Phys. Lett.",
      volume         = "B710",
      pages          = "612-622",
      doi            = "10.1016/j.physletb.2012.03.013",
      year           = "2012",
      eprint         = "1111.5869",
      archivePrefix  = "arXiv",
      primaryClass   = "hep-ph",
      reportNumber   = "CERN-PH-TH-2011-277, TTK-11-54",
      SLACcitation   = "%%CITATION = ARXIV:1111.5869;%%",
}

@article{Beneke:2011mq,
      author         = "Beneke, M. and Falgari, P. and Klein, S. and Schwinn, C.",
      title          = "{Hadronic top-quark pair production with NNLL threshold
                        resummation}",
      journal        = "Nucl.~Phys. B",
      volume         = "855",
      pages          = "695-741",
      doi            = "10.1016/j.nuclphysb.2011.10.021",
      year           = "2012",
      eprint         = "1109.1536",
      archivePrefix  = "arXiv",
      primaryClass   = "hep-ph",
      reportNumber   = "TTK-11-38, ITP-UU-11-26, SPIN-11-19, FR-PHENO-2011-015,
                        SFB-CPP-11-49",
      SLACcitation   = "%%CITATION = ARXIV:1109.1536;%%",
}

@article{Baernreuther:2012ws,
      author         = "Baernreuther, Peter and Czakon, Michal and Mitov,
                        Alexander",
      title          = "{Percent Level Precision Physics at the Tevatron: First
                        Genuine NNLO QCD Corrections to $q \bar{q} \to t \bar{t} +
                        X$}",
      journal        = "Phys. Rev. Lett.",
      volume         = "109",
      pages          = "132001",
      doi            = "10.1103/PhysRevLett.109.132001",
      year           = "2012",
      eprint         = "1204.5201",
      archivePrefix  = "arXiv",
      primaryClass   = "hep-ph",
      SLACcitation   = "%%CITATION = ARXIV:1204.5201;%%",
}

@article{Czakon:2012pz,
      author         = "Czakon, Michal and Mitov, Alexander",
      title          = "{NNLO corrections to top pair production at hadron
                        colliders: the quark-gluon reaction}",
      journal        = jhep,
      volume         = "01",
      pages          = "080",
      doi            = "10.1007/JHEP01(2013)080",
      year           = "2013",
      eprint         = "1210.6832",
      archivePrefix  = "arXiv",
      primaryClass   = "hep-ph",
      SLACcitation   = "%%CITATION = ARXIV:1210.6832;%%",
}

@article{Czakon:2012zr,
      author         = "Czakon, Michal and Mitov, Alexander",
      title          = "{NNLO corrections to top-pair production at hadron
                        colliders: the all-fermionic scattering channels}",
      journal        = jhep,
      volume         = "12",
      pages          = "054",
      doi            = "10.1007/JHEP12(2012)054",
      year           = "2012",
      eprint         = "1207.0236",
      archivePrefix  = "arXiv",
      primaryClass   = "hep-ph",
      SLACcitation   = "%%CITATION = ARXIV:1207.0236;%%",
}

@article{Czakon:2011xx,
      author         = "Czakon, Michal and Mitov, Alexander",
      title          = "{Top++: A program for the calculation of the top-pair
                        cross-section at hadron colliders}",
      journal        = "Comput. Phys. Commun.",
      volume         = "185",
      year           = "2014",
      pages          = "2930",
      doi            = "10.1016/j.cpc.2014.06.021",
      eprint         = "1112.5675",
      archivePrefix  = "arXiv",
      primaryClass   = "hep-ph",
      reportNumber   = "CERN-PH-TH-2011-303, TTK-11-58",
      SLACcitation   = "%%CITATION = ARXIV:1112.5675;%%"
}

@article{Czakon:2013goa,
      author         = "Czakon, Michal and Fiedler, Paul and Mitov, Alexander",
      title          = "{Total Top-Quark Pair-Production Cross Section at
                        Hadron Colliders Through O($\alpha_S^4$)}",
      journal        = "Phys. Rev. Lett.",
      volume         = "110", 
      pages	     = "252004",
      doi            = "10.1103/PhysRevLett.110.252004",
      year           = "2013",
      eprint         = "1303.6254",
      archivePrefix  = "arXiv",
      primaryClass   = "hep-ph",
}

@article{Frixione:2007vw,
      author         = "Frixione, Stefano and Nason, Paolo and Oleari, Carlo",
      title          = "{Matching NLO QCD computations with parton shower
                        simulations: the POWHEG method}",
      journal        = jhep,
      volume         = "11",
      pages          = "070",
      doi            = "10.1088/1126-6708/2007/11/070",
      year           = "2007",
      eprint         = "0709.2092",
      archivePrefix  = "arXiv",
      primaryClass   = "hep-ph",
      reportNumber   = "BICOCCA-FT-07-9, GEF-TH-21-2007",
      SLACcitation   = "%%CITATION = ARXIV:0709.2092;%%",
}

@article{Aliev:2010zk,
      author         = "Aliev, M. and Lacker, H. and Langenfeld, U. and Moch, S.
                        and Uwer, P. and others",
      title          = "{HATHOR: HAdronic Top and Heavy quarks crOss section
                        calculatoR}",
      journal        = "Comput. Phys. Commun.",
      volume         = "182", 
      pages	     = "1034",
      doi            = "10.1016/j.cpc.2010.12.040",
      year           = "2011",
      eprint         = "1007.1327",
      archivePrefix  = "arXiv",
      primaryClass   = "hep-ph",
}

@article{GEANT4,
      collaboration     = "{GEANT4}",
      author         = "Agostinelli, S. and others",
      title          = "{GEANT4: a simulation toolkit}",
      journal        = nim,
      volume         = "\textmd{A} 506", 
      pages	     = "250",
      doi            = "10.1016/S0168-9002(03)01368-8",
      year           = "2003",
}

@article{Kidonakis:2010ux,
      author         = "Kidonakis, Nikolaos",
      title          = "{Two-loop soft anomalous dimensions for single top quark
                        associated production with a $W^{-}$ or $H^{-}$}",
      journal        = "Phys. Rev.",
      volume         = "\textmd{D} 82", 
      pages	     = "054018",
      doi            = "10.1103/PhysRevD.82.054018",
      year           = "2010",
      eprint         = "1005.4451",
      archivePrefix  = "arXiv",
      primaryClass   = "hep-ph",
}

@Book{bootstrap_bohm2010,
 author = {Bohm, Gerhard and Zech, G\"unther},
 title = {Introduction to statistics and data analysis for physicists},
 publisher = {Verl. Dt. Elektronen-Synchrotron},
 pages ="336",
 year = {2010},
 address = {Hamburg},
 isbn = "\href{http://www-library.desy.de/preparch/books/vstatmp_engl.pdf}{978-3-935702-41-6}",
 doi  = "10.3204/DESY-BOOK/statisics (e-book)",
 note = {\href{http://www-library.desy.de/preparch/books/vstatmp_engl.pdf}{http://www-library.desy.de/preparch/books/vstatmp\_engl.pdf}}
}

@article{Martin:2009bu,
      author         = "Martin, A. D. and Stirling, W. J. and Thorne, R. S. and
                        Watt, G.",
      title          = "{Uncertainties on alpha(S) in global PDF analyses and
                        implications for predicted hadronic cross sections}",
      journal        = "Eur. Phys. J. C",
      volume         = "64",
      year           = "2009",
      pages          = "653-680",
      doi            = "10.1140/epjc/s10052-009-1164-2",
      eprint         = "0905.3531",
      archivePrefix  = "arXiv",
      primaryClass   = "hep-ph",
      reportNumber   = "IPPP-09-33, DCPT-09-66, CAVENDISH-HEP-09-06",
      SLACcitation   = "%%CITATION = ARXIV:0905.3531;%%"
}

@article{Gao:2013xoa,
      author         = "Gao, Jun and Guzzi, Marco and Huston, Joey and Lai,
                        Hung-Liang and Li, Zhao and others",
      title          = "{CT10 next-to-next-to-leading order global analysis of
                        QCD}",
      journal        = "Phys.~Rev. D",
      number         = "3",
      volume         = "89",
      pages          = "033009",
      doi            = "10.1103/PhysRevD.89.033009",
      year           = "2014",
      eprint         = "1302.6246",
      archivePrefix  = "arXiv",
      primaryClass   = "hep-ph",
      reportNumber   = "SMU-HEP-12-23",
      SLACcitation   = "%%CITATION = ARXIV:1302.6246;%%",
}

@article{Alwall:2011uj,
      author         = "Alwall, Johan and Herquet, Michel and Maltoni, Fabio and
                        Mattelaer, Olivier and Stelzer, Tim",
      title          = "{MadGraph 5 : going beyond}",
      journal        = jhep,
      volume         = "06",
      year           = "2011",
      pages          = "128",
      doi            = "10.1007/JHEP06(2011)128",
      eprint         = "1106.0522",
      archivePrefix  = "arXiv",
      primaryClass   = "hep-ph",
      reportNumber   = "FERMILAB-PUB-11-448-T",
      SLACcitation   = "%%CITATION = ARXIV:1106.0522;%%"
}

@Article{ATLAS-TDR-19,
     author       = "{ATLAS Collaboration}",
     title        = "ATLAS Insertable B-Layer Technical Design Report",
     journal = "ATLAS-TDR-19",
     year         = "2010",
     url          = "https://cds.cern.ch/record/1291633",
     addendum     = "{ATLAS Insertable B-Layer Technical Design Report Addendum}, ATLAS-TDR-19-ADD-1 (2012), {\scriptsize{URL}}: \url{http://cds.cern.ch/record/1451888}",
}

@article{thaler_2011,
      author         = "Thaler, J. and Tilburg, K. Van",
      title          = "{Identifying boosted objects with N-subjettiness}",
      year           = "2011",
      journal        = "JHEP",
      volume         = "03",
      pages          = "015",
      eprint         = "1011.2268",
      archivePrefix  = "arXiv",
      primaryClass   = "hep-ph",
      doi = "10.1007/JHEP03(2011)015",
}

@article{thaler_2012,
      author         = "Thaler, J. and Tilburg, K. Van",
      title          = "{Maximizing boosted top identification by minimizing N-subjettiness}",
      year           = "2012",
      journal        = "JHEP",
      volume         = "02",
      pages          = "093",
      eprint         = "1108.2701",
      archivePrefix  = "arXiv",
      primaryClass   = "hep-ph",
      doi            = "10.1007/JHEP02(2012)093",
}

@article{Susskind:1978ms,
      author         = "Susskind, Leonard",
      title          = "{Dynamics of spontaneous symmetry breaking in the
                        Weinberg-Salam Theory}",
      journal        = "Phys. Rev. D",
      volume         = "20",
      year           = "1979",
      pages          = "2619-2625",
      doi            = "10.1103/PhysRevD.20.2619",
      reportNumber   = "SLAC-PUB-2142",
      SLACcitation   = "%%CITATION = PHRVA,D20,2619;%%"
}

@article{ArkaniHamed:2002qy,
      author         = "Arkani-Hamed, N. and Cohen, A. G. and Katz, E. and
                        Nelson, A. E.",
      title          = "{The Littlest Higgs}",
      journal        = "JHEP",
      volume         = "07",
      year           = "2002",
      pages          = "034",
      doi            = "10.1088/1126-6708/2002/07/034",
      eprint         = "hep-ph/0206021",
      archivePrefix  = "arXiv",
      reportNumber   = "UW-PT-01-07, HUTP-02-A017, BUHEP-02-23",
      SLACcitation   = "%%CITATION = HEP-PH/0206021;%%"
}

@article{Schmaltz:2005ky,
      author         = "Schmaltz, Martin and Tucker-Smith, David",
      title          = "{Little Higgs Theories}",
      journal        = "Ann. Rev. Nucl. Part. Sci.",
      volume         = "55",
      year           = "2005",
      pages          = "229-270",
      doi            = "10.1146/annurev.nucl.55.090704.151502",
      eprint         = "hep-ph/0502182",
      archivePrefix  = "arXiv",
      SLACcitation   = "%%CITATION = HEP-PH/0502182;%%"
}

@article{Kaplan:1983sm,
      author         = "Kaplan, David B. and Georgi, Howard and Dimopoulos,
                        Savas",
      title          = "{Composite Higgs scalars}",
      journal        = "Phys. Lett. B",
      volume         = "136",
      year           = "1984",
      pages          = "187-190",
      doi            = "10.1016/0370-2693(84)91178-X",
      reportNumber   = "HUTP-83/A079",
      SLACcitation   = "%%CITATION = PHLTA,B136,187;%%"
}

@article{Agashe:2004rs,
      author         = "Agashe, Kaustubh and Contino, Roberto and Pomarol, Alex",
      title          = "{The minimal composite Higgs model}",
      journal        = "Nucl. Phys. B",
      volume         = "719",
      year           = "2005",
      pages          = "165-187",
      doi            = "10.1016/j.nuclphysb.2005.04.035",
      eprint         = "hep-ph/0412089",
      archivePrefix  = "arXiv",
      reportNumber   = "UAB-FT-567",
      SLACcitation   = "%%CITATION = HEP-PH/0412089;%%"
}

@article{Hill:2002ap,
      author         = "Hill, Christopher T. and Simmons, Elizabeth H.",
      title          = "{Strong dynamics and electroweak symmetry breaking}",
      journal        = "Phys. Rept.",
      volume         = "381",
      year           = "2003",
      pages          = "235-402",
      doi            = "10.1016/S0370-1573(03)00140-6",
      note           = "[Erratum: \href{https://doi.org/10.1016/j.physrep.2003.10.002}{Phys. Rept. \textbf{390} (2004) 553]}",
      eprint         = "hep-ph/0203079",
      archivePrefix  = "arXiv",
      reportNumber   = "FERMILAB-PUB-02-045-T, BUHEP-01-09, FERMI-PUB-02-045-T",
      SLACcitation   = "%%CITATION = HEP-PH/0203079;%%"
}

@article{delAguila:1982fs,
      author         = "del Aguila, F. and Bowick, Mark J.",
      title          = "{The possibility of new fermions with $\Delta$ I = 0
                        Mass}",
      journal        = "Nucl. Phys. B",
      volume         = "224",
      year           = "1983",
      pages          = "107",
      doi            = "10.1016/0550-3213(83)90316-4",
      reportNumber   = "UFTP-82-19",
      SLACcitation   = "%%CITATION = NUPHA,B224,107;%%"
}

@article{AguilarSaavedra:2009es,
      author         = "Aguilar-Saavedra, J. A.",
      title          = "{Identifying top partners at LHC}",
      journal        = "JHEP",
      volume         = "11",
      year           = "2009",
      pages          = "030",
      doi            = "10.1088/1126-6708/2009/11/030",
      eprint         = "0907.3155",
      archivePrefix  = "arXiv",
      primaryClass   = "hep-ph",
      SLACcitation   = "%%CITATION = ARXIV:0907.3155;%%"
}

@article{Aguilar-Saavedra:2013wba,
      author         = "Aguilar-Saavedra, J. A.",
      title          = "{Mixing with vector-like quarks: constraints and
                        expectations}",
      journal        = "EPJ Web of Conferences",
      year           = "2013",
      volume         = "60",
      pages          = "16012",
      doi            = "10.1051/epjconf/20136016012",
      eprint         = "1306.4432",
      archivePrefix  = "arXiv",
      primaryClass   = "hep-ph",
      SLACcitation   = "%%CITATION = ARXIV:1306.4432;%%"
}

@article{Aguilar-Saavedra:2013qpa,
      author         = "Aguilar-Saavedra, J. A. and Benbrik, R. and Heinemeyer,
                        S. and P\'erez-Victoria, M.",
      title          = "{Handbook of vectorlike quarks: Mixing and single
                        production}",
      journal        = "Phys. Rev. D",
      volume         = "88",
      year           = "2013",
      number         = "9",
      pages          = "094010",
      doi            = "10.1103/PhysRevD.88.094010",
      eprint         = "1306.0572",
      archivePrefix  = "arXiv",
      primaryClass   = "hep-ph",
      SLACcitation   = "%%CITATION = ARXIV:1306.0572;%%"
}

@article{Atre:2008iu,
      author         = "Atre, Anupama and Carena, Marcela and Han, Tao and
                        Santiago, Jose",
      title          = "{Heavy quarks above the top at the Tevatron}",
      journal        = "Phys. Rev. D",
      volume         = "79",
      year           = "2009",
      pages          = "054018",
      doi            = "10.1103/PhysRevD.79.054018",
      eprint         = "0806.3966",
      archivePrefix  = "arXiv",
      primaryClass   = "hep-ph",
      reportNumber   = "FERMILAB-PUB-08-098-T, NSF-KITP-08-64, MADPH-08-1511",
      SLACcitation   = "%%CITATION = ARXIV:0806.3966;%%"
}

@article{Atre:2011ae,
      author         = "Atre, Anupama and Azuelos, Georges and Carena, Marcela
                        and Han, Tao and Ozcan, Erkcan and Santiago, Jose and
                        Unel, Gokhan",
      title          = "{Model-independent searches for new quarks at the LHC}",
      journal        = "JHEP",
      volume         = "08",
      year           = "2011",
      pages          = "080",
      doi            = "10.1007/JHEP08(2011)080",
      eprint         = "1102.1987",
      archivePrefix  = "arXiv",
      primaryClass   = "hep-ph",
      reportNumber   = "FERMILAB-PUB-11-910-T",
      SLACcitation   = "%%CITATION = ARXIV:1102.1987;%%"
}

@Article{CMS-HIG-19-001,
      author        = "{CMS Collaboration}",
      title         = "{Measurements of properties of the Higgs boson in the
                       four-lepton final state in proton-proton collisions at
                       $\sqrt{s}=13~\mathrm{TeV}$}",
      institution   = "CERN",
      collaboration = "CMS Collaboration",
      address       = "Geneva",
      number        = "CMS-PAS-HIG-19-001",
      year          = "2019",
      reportNumber  = "CMS-PAS-HIG-19-001",
      url           = "https://cds.cern.ch/record/2668684"
}

@article{Roy_2020,
   title={Novel interpretation strategy for searches of singly produced vectorlike quarks at the LHC},
   volume={101},
   eprint = "2003.00640",
   archivePrefix = "arXiv",
   number={11},
   journal={Physical Review D},
   publisher={American Physical Society (APS)},
   author={Roy, Avik and Nikiforou, Nikiforos and Castro, Nuno and Andeen, Timothy},
   year={2020},
   month={Jun},
   doi="10.1103/PhysRevD.101.115027"
}

@article{deFlorian:2016spz,
      author         = "de Florian, D. and others",
      title          = "{Handbook of LHC Higgs Cross Sections: 4. Deciphering the
                        Nature of the Higgs Sector}",
      collaboration  = "LHC Higgs Cross Section Working Group",
      doi            = "10.23731/CYRM-2017-002",
      year           = "2016",
      eprint         = "1610.07922",
      archivePrefix  = "arXiv",
      primaryClass   = "hep-ph",
      reportNumber   = "FERMILAB-FN-1025-T, CERN-2017-002-M",
      SLACcitation   = "%%CITATION = ARXIV:1610.07922;%%"
}

@article{Buchkremer:2013bha,
    author = "Buchkremer, Mathieu and Cacciapaglia, Giacomo and Deandrea, Aldo and Panizzi, Luca",
    title = "{Model-independent framework for searches of top partners}",
    eprint = "1305.4172",
    archivePrefix = "arXiv",
    primaryClass = "hep-ph",
    reportNumber = "LYCEN-2013-03, SHEP-13-10, CP3-13-22",
    doi = "10.1016/j.nuclphysb.2013.08.010",
    journal = "Nucl. Phys. B",
    volume = "876",
    pages = "376--417",
    year = "2013"
}

@article{ATLAS_IBL,
      author         = "Abbott, B. and others",
      title          = "{Production and integration of the ATLAS Insertable
                        B-Layer}",
      collaboration  = "ATLAS IBL",
      journal        = "JINST",
      volume         = "13",
      year           = "2018",
      number         = "05",
      pages          = "T05008",
      doi            = "10.1088/1748-0221/13/05/T05008",
      eprint         = "1803.00844",
      archivePrefix  = "arXiv",
      primaryClass   = "physics.ins-det",
      SLACcitation   = "%%CITATION = ARXIV:1803.00844;%%"
}

@article{Cowan_2011,
   title={Asymptotic formulae for likelihood-based tests of new physics},
   volume={71},
   ISSN={1434-6052},
   url={http://dx.doi.org/10.1140/epjc/s10052-011-1554-0},
   DOI={10.1140/epjc/s10052-011-1554-0},
   number={2},
   journal={Eur. Phys. J. C},
   publisher={Springer Science and Business Media LLC},
   author={Cowan, Glen and Cranmer, Kyle and Gross, Eilam and Vitells, Ofer},
   year={2011},
   month={Feb}
}

@Article{JETM-2018-06,
    author         = "{ATLAS Collaboration}",
    title          = "{Optimisation of large-radius jet reconstruction for the ATLAS detector in \(13\,\text{TeV}\) proton--proton collisions}",
    year           = "2020",
    reportNumber   = "CERN-EP-2020-134",
    eprint         = "2009.04986",
    archivePrefix  = "arXiv",
    primaryClass   = "hep-ex"
}

@Article{JETM-2018-05,
    author         = "{ATLAS Collaboration}",
    title          = "{Jet energy scale and resolution measured in proton--proton collisions at \(\sqrt{s} = 13\,\text{TeV}\) with the ATLAS detector}",
    year           = "2020",
    reportNumber   = "CERN-EP-2020-083",
    eprint         = "2007.02645",
    archivePrefix  = "arXiv",
    primaryClass   = "hep-ex"
}

@article{VRjets,
    author = "Krohn, David and Thaler, Jesse and Wang, Lian-Tao",
    title = "{Jets with variable R}",
    eprint = "0903.0392",
    archivePrefix = "arXiv",
    primaryClass = "hep-ph",
    doi = "10.1088/1126-6708/2009/06/059",
    journal = "JHEP",
    volume = "06",
    pages = "059",
    year = "2009"
}

@article{LUCID2,
  author={G. Avoni and others},
  title={The new LUCID-2 detector for luminosity measurement and monitoring in ATLAS},
  journal={JINST},
  volume={13},
  number={07},
  pages={P07017},
  doi="10.1088/1748-0221/13/07/P07017",
  year={2018}
}

@article{VLQatNLO,
   title={Next-to-leading-order predictions for single vector-like quark production at the LHC},
   volume={793},
   eprint = "1811.05055",
   archivePrefix = "arXiv",
   journal={Physics Letters B},
   publisher={Elsevier BV},
   author={Cacciapaglia, Giacomo and Carvalho, Alexandra and Deandrea, Aldo and Flacke, Thomas and Fuks, Benjamin and Majumder, Devdatta and Panizzi, Luca and Shao, Hua-Sheng},
   year={2019},
   month={Jun},
   pages={206–211},
   doi="10.1016/j.physletb.2019.04.056",
}

@Article{CMS-B2G-18-005,
    author         = "{CMS Collaboration}",
    title          = "{Search for pair production of vectorlike quarks in the fully hadronic final state}",
    journal        = "Phys. Rev. D",
    volume         = "100",
    year           = "2019",
    pages          = "072001",
    doi            = "10.1103/PhysRevD.100.072001",
    reportNumber   = "CERN-EP-2019-129",
    eprint         = "1906.11903",
    archivePrefix  = "arXiv",
    primaryClass   = "hep-ex",
}

@Article{Butterworth:2015oua,
      author         = "Butterworth, Jon and others",
      title          = "{PDF4LHC recommendations for LHC Run II}",
      journal        = "J. Phys. G",
      volume         = "43",
      year           = "2016",
      pages          = "023001",
      doi            = "10.1088/0954-3899/43/2/023001",
      eprint         = "1510.03865",
      archivePrefix  = "arXiv",
      primaryClass   = "hep-ph",
      reportNumber   = "OUTP-15-17P, SMU-HEP-15-12, TIF-UNIMI-2015-14,
                        LCTS-2015-27, CERN-PH-TH-2015-249",
      SLACcitation   = "%%CITATION = ARXIV:1510.03865;%%"
}

@article{Lai:2010vv,
      author         = "Lai, H.-L. and others",
      title          = "{New parton distributions for collider physics}",
      journal        = "Phys. Rev. D",
      volume         = "82",
      pages          = "074024",
      doi            = "10.1103/PhysRevD.82.074024",
      year           = "2010",
      eprint         = "1007.2241",
      archivePrefix  = "arXiv",
      primaryClass   = "hep-ph",
      reportNumber   = "MSUHEP-100707, SMU-HEP-10-10",
      SLACcitation   = "%%CITATION = ARXIV:1007.2241;%%",
}

@article{Martin:2009iq,
      author         = "Martin, A. D. and Stirling, W. J. and Thorne, R. S. and Watt, G.",
      title          = "{Parton distributions for the LHC}",
      journal        = "Eur. Phys. J. C",
      volume         = "63",
      year           = "2009",
      pages          = "189",
      doi            = "10.1140/epjc/s10052-009-1072-5",
      eprint         = "0901.0002",
      archivePrefix  = "arXiv",
      primaryClass   = "hep-ph",
      reportNumber   = "IPPP-08-95, DCPT-08-190, CAVENDISH-HEP-08-16",
      SLACcitation   = "%%CITATION = ARXIV:0901.0002;%%"
}

@article{Harland-Lang:2014zoa,
      author         = "Harland-Lang, L.A. and Martin, A.D. and Motylinski, P.
                        and Thorne, R.S.",
      title          = "{Parton distributions in the LHC era: MMHT 2014 PDFs}",
      journal        = "Eur. Phys. J. C",
      number         = "5",
      volume         = "75",
      pages          = "204",
      doi            = "10.1140/epjc/s10052-015-3397-6",
      year           = "2015",
      eprint         = "1412.3989",
      archivePrefix  = "arXiv",
      primaryClass   = "hep-ph",
      reportNumber   = "LCTS-2014-47, IPPP-14-97, DCPT-14-194",
      SLACcitation   = "%%CITATION = ARXIV:1412.3989;%%",
}

@article{Ball:2014uwa,
      author         = "Ball, Richard D. and others",
      title          = "{Parton distributions for the LHC run II}",
      collaboration  = "NNPDF",
      journal        = "JHEP",
      volume         = "04",
      year           = "2015",
      pages          = "040",
      doi            = "10.1007/JHEP04(2015)040",
      eprint         = "1410.8849",
      archivePrefix  = "arXiv",
      primaryClass   = "hep-ph",
      reportNumber   = "EDINBURGH-2014-15, IFUM-1034-FT, CERN-PH-TH-2013-253,
                        OUTP-14-11P, CAVENDISH-HEP-14-11",
      SLACcitation   = "%%CITATION = ARXIV:1410.8849;%%"
}

@article{Sjostrand:2014zea,
      author         = "Sj{\"o}strand, Torbj{\"o}rn and Ask, Stefan and Christiansen,
                        Jesper R. and Corke, Richard and Desai, Nishita and Ilten,
                        Philip and Mrenna, Stephen and Prestel, Stefan and
                        Rasmussen, Christine O. and Skands, Peter Z.",
      title          = "{An introduction to PYTHIA 8.2}",
      journal        = "Comput. Phys. Commun.",
      volume         = "191",
      year           = "2015",
      pages          = "159",
      doi            = "10.1016/j.cpc.2015.01.024",
      eprint         = "1410.3012",
      archivePrefix  = "arXiv",
      primaryClass   = "hep-ph",
      reportNumber   = "LU-TP-14-36, MCNET-14-22, CERN-PH-TH-2014-190,
                        FERMILAB-PUB-14-316-CD, DESY-14-178, SLAC-PUB-16122,
                        --FERMILAB-PUB-14-316-CD",
      SLACcitation   = "%%CITATION = ARXIV:1410.3012;%%"
}

@Article{Lange:2001uf,
      author         = "Lange, D. J.",
      title          = "{The EvtGen particle decay simulation package}",
      booktitle      = "{Proceedings, 7th International Conference on B physics
                        at hadron machines (BEAUTY 2000)}",
      journal        = "Nucl. Instrum. Meth. A",
      volume         = "462",
      year           = "2001",
      pages          = "152",
      doi            = "10.1016/S0168-9002(01)00089-4",
      SLACcitation   = "%%CITATION = NUIMA,A462,152;%%"
}

@Article{Frixione:2008yi,
     author    = "Frixione, Stefano and Laenen, Eric and Motylinski, Patrick and White, Chris and Webber, Bryan R.",
     title     = "{Single-top hadroproduction in association with a \(W\) boson}",
     journal   = "JHEP",
     volume    = "07",
     year      = "2008",
     pages     = "029",
     eprint    = "0805.3067",
     archivePrefix = "arXiv",
     primaryClass  =  "hep-ph",
     doi       = "10.1088/1126-6708/2008/07/029"
}

@Article{Bahr:2008pv,
      author         = "B{\"a}hr, M. and others",
      title          = "{Herwig++ physics and manual}",
      journal        = "Eur. Phys. J. C",
      volume         = "58",
      year           = "2008",
      pages          = "639",
      doi            = "10.1140/epjc/s10052-008-0798-9",
      eprint         = "0803.0883",
      archivePrefix  = "arXiv",
      primaryClass   = "hep-ph",
      reportNumber   = "CERN-PH-TH-2008-038, CAVENDISH-HEP-08-03, KA-TP-05-2008,
                        DCPT-08-22, IPPP-08-11, CP3-08-05",
      SLACcitation   = "%%CITATION = ARXIV:0803.0883;%%"
}

@Article{Bellm:2015jjp,
      author         = "Bellm, Johannes and others",
      title          = "{Herwig 7.0/Herwig++ 3.0 release note}",
      journal        = "Eur. Phys. J. C",
      volume         = "76",
      year           = "2016",
      number         = "4",
      pages          = "196",
      doi            = "10.1140/epjc/s10052-016-4018-8",
      eprint         = "1512.01178",
      archivePrefix  = "arXiv",
      primaryClass   = "hep-ph",
      reportNumber   = "CERN-PH-TH-2015-289, MAN-HEP-2015-15, IFJPAN-IV-2015-13,
                        HERWIG-2015-01, KA-TP-18-2015, DCPT-15-142, MCNET-15-28,
                        IPPP-15-71, --HERWIG-2015-01",
      SLACcitation   = "%%CITATION = ARXIV:1512.01178;%%"
}

@article{Bellm:2017jjp,
      author         = "Bellm, Johannes and others",
      title          = "{Herwig 7.1 Release Note}",
      eprint         = "1705.06919",
      year           = "2017",
      archivePrefix  = "arXiv",
      primaryClass   = "hep-ph",
      reportNumber   = "CERN-PH-TH-2017-109, MAN/HEP/2017/08, UWTHPH-2017-10, 
                        IFJPAN-IV-2017-7, NIKHEF 2017-026, HERWIG-2017-02, 
                        KA-TP-19-2017, MCnet-17-08, IPPP/17/40",
      SLACcitation   = "%%CITATION = ARXIV:1705.06919;%%"
}

@Article{Nason:2004rx,
      author         = "Nason, Paolo",
      title          = "{A new method for combining NLO QCD with shower Monte Carlo algorithms}",
      journal        = "JHEP",
      volume         = "11",
      pages          = "040",
      doi            = "10.1088/1126-6708/2004/11/040",
      year           = "2004",
      eprint         = "hep-ph/0409146",
      archivePrefix  = "arXiv",
}

@Article{Alioli:2010xd,
      author         = "Alioli, Simone and Nason, Paolo and Oleari, Carlo and Re,
                        Emanuele",
      title          = "{A general framework for implementing NLO calculations in
                        shower Monte Carlo programs: the POWHEG BOX}",
      journal        = "JHEP",
      volume         = "06",
      pages          = "043",
      doi            = "10.1007/JHEP06(2010)043",
      year           = "2010",
      eprint         = "1002.2581",
      archivePrefix  = "arXiv",
      primaryClass   = "hep-ph",
}

@article{Hartanto:2015uka,
      author         = "Hartanto, Heribertus B. and J{\"a}ger, Barbara and Reina,
                        Laura and Wackeroth, Doreen",
      title          = "{Higgs boson production in association with top quarks in
                        the POWHEG BOX}",
      journal        = "Phys. Rev. D",
      volume         = "91",
      year           = "2015",
      number         = "9",
      pages          = "094003",
      doi            = "10.1103/PhysRevD.91.094003",
      eprint         = "1501.04498",
      archivePrefix  = "arXiv",
      primaryClass   = "hep-ph",
      SLACcitation   = "%%CITATION = ARXIV:1501.04498;%%"
}

@article{Frederix:2012dh,
      author         = "Frederix, Rikkert and Re, Emanuele and Torrielli, Paolo",
      title          = "{Single-top \(t\)-channel hadroproduction in the four-flavour
                        scheme with POWHEG and aMC@NLO}",
      journal        = "JHEP",
      volume         = "09",
      year           = "2012",
      pages          = "130",
      doi            = "10.1007/JHEP09(2012)130",
      eprint         = "1207.5391",
      archivePrefix  = "arXiv",
      primaryClass   = "hep-ph",
      reportNumber   = "ZU-TH-14-12, IPPP-12-54, DCPT-12-108,
                        CERN-PH-TH-2012-206, MCNET-12-10, LPN12-084",
      SLACcitation   = "%%CITATION = ARXIV:1207.5391;%%"
}

@article{Re:2010bp,
      author         = "Re, Emanuele",
      title          = "{Single-top \(Wt\)-channel production matched with parton
                        showers using the POWHEG method}",
      journal        = "Eur. Phys. J. C",
      volume         = "71",
      year           = "2011",
      pages          = "1547",
      doi            = "10.1140/epjc/s10052-011-1547-z",
      eprint         = "1009.2450",
      archivePrefix  = "arXiv",
      primaryClass   = "hep-ph",
      reportNumber   = "IPPP-10-74, DCPT-10-148",
      SLACcitation   = "%%CITATION = ARXIV:1009.2450;%%"
}

@article{Kant:2014oha,
      author         = "Kant, P. and Kind, O. M. and Kintscher, T. and Lohse, T.
                        and Martini, T. and Mölbitz, S. and Rieck, P. and Uwer, P.",
      title          = "{HatHor for single top-quark production: Updated
                        predictions and uncertainty estimates for single top-quark
                        production in hadronic collisions}",
      journal        = "Comput. Phys. Commun.",
      volume         = "191",
      year           = "2015",
      pages          = "74-89",
      doi            = "10.1016/j.cpc.2015.02.001",
      eprint         = "1406.4403",
      archivePrefix  = "arXiv",
      primaryClass   = "hep-ph",
      reportNumber   = "HU-EP-14-22",
      SLACcitation   = "%%CITATION = ARXIV:1406.4403;%%"
}

@inproceedings{Kidonakis:2013zqa,
      author         = "Kidonakis, Nikolaos",
      title          = "{Top Quark Production}",
      booktitle      = "{Proceedings, Helmholtz International Summer School on
                        Physics of Heavy Quarks and Hadrons (HQ 2013)}",
      eventdate      = "2013-07-15/2013-07-28",
      venue          = "JINR, Dubna, Russia",
      pages          = "139-168",
      doi            = "10.3204/DESY-PROC-2013-03/Kidonakis",
      eprint         = "1311.0283",
      archivePrefix  = "arXiv",
      primaryClass   = "hep-ph",
      reportNumber   = "KSU-HEP-110113",
      SLACcitation   = "%%CITATION = ARXIV:1311.0283;%%"
}

@article{Frixione:2007nw,
      author         = "Frixione, Stefano and Nason, Paolo and Ridolfi, Giovanni",
      title          = "{A positive-weight next-to-leading-order Monte Carlo for
                        heavy flavour hadroproduction}",
      journal        = "JHEP",
      volume         = "09",
      pages          = "126",
      doi            = "10.1088/1126-6708/2007/09/126",
      year           = "2007",
      eprint         = "0707.3088",
      archivePrefix  = "arXiv",
      primaryClass   = "hep-ph",
      SLACcitation   = "%%CITATION = ARXIV:0707.3088;%%",
}

@Article{PERF-2007-01,
    author         = "{ATLAS Collaboration}",
    title          = "{The ATLAS Experiment at the CERN Large Hadron Collider}",
    journal        = "JINST",
    volume         = "3",
    year           = "2008",
    pages          = "S08003",
    doi            = "10.1088/1748-0221/3/08/S08003",
    primaryClass   = "hep-ex",
}

@Article{SOFT-2010-01,
    author         = "{ATLAS Collaboration}",
    title          = "{The ATLAS Simulation Infrastructure}",
    journal        = "Eur. Phys. J. C",
    volume         = "70",
    year           = "2010",
    pages          = "823",
    doi            = "10.1140/epjc/s10052-010-1429-9",
    eprint         = "1005.4568",
    archivePrefix  = "arXiv",
    primaryClass   = "physics.ins-det",
}

@Article{EXOT-2011-22,
    author         = "{ATLAS Collaboration}",
    title          = "{Search for Pair Production of a New \(b'\) Quark that Decays into a \(Z\) Boson and a Bottom Quark with the ATLAS Detector}",
    journal        = "Phys. Rev. Lett.",
    volume         = "109",
    year           = "2012",
    pages          = "071801",
    doi            = "10.1103/PhysRevLett.109.071801",
    reportNumber   = "CERN-PH-EP-2012-073",
    eprint         = "1204.1265",
    archivePrefix  = "arXiv",
    primaryClass   = "hep-ex",
}

@Article{TOPQ-2011-07,
    author         = "{ATLAS Collaboration}",
    title          = "{Measurements of top quark pair relative differential cross-sections with ATLAS in \(pp\) collisions at \(\sqrt{s} = 7\,\text{TeV}\)}",
    journal        = "Eur. Phys. J. C",
    volume         = "73",
    year           = "2013",
    pages          = "2261",
    doi            = "10.1140/epjc/s10052-012-2261-1",
    reportNumber   = "CERN-PH-EP-2012-165",
    eprint         = "1207.5644",
    archivePrefix  = "arXiv",
    primaryClass   = "hep-ex",
}

@Article{HIGG-2012-27,
    author         = "{ATLAS Collaboration}",
    title          = "{Observation of a new particle in the search for the Standard Model Higgs boson with the ATLAS detector at the LHC}",
    journal        = "Phys. Lett. B",
    volume         = "716",
    year           = "2012",
    pages          = "1",
    doi            = "10.1016/j.physletb.2012.08.020",
    reportNumber   = "CERN-PH-EP-2012-218",
    eprint         = "1207.7214",
    archivePrefix  = "arXiv",
    primaryClass   = "hep-ex",
}

@Article{PERF-2012-02,
    author         = "{ATLAS Collaboration}",
    title          = "{Performance of jet substructure techniques for large-\(R\) jets in proton--proton collisions at \(\sqrt{s} = 7\,\text{TeV}\) using the ATLAS detector}",
    journal        = "JHEP",
    volume         = "09",
    year           = "2013",
    pages          = "076",
    doi            = "10.1007/JHEP09(2013)076",
    reportNumber   = "CERN-PH-EP-2013-069",
    eprint         = "1306.4945",
    archivePrefix  = "arXiv",
    primaryClass   = "hep-ex",
}

@Article{TOPQ-2012-08,
    author         = "{ATLAS Collaboration}",
    title          = "{Measurements of normalized differential cross-sections for  \(t\bar{t}\)  production in \(pp\) collisions at \(\sqrt{s} = 7\,\text{TeV}\) using the ATLAS detector}",
    journal        = "Phys. Rev. D",
    volume         = "90",
    year           = "2014",
    pages          = "072004",
    doi            = "10.1103/PhysRevD.90.072004",
    reportNumber   = "CERN-PH-EP-2014-099",
    eprint         = "1407.0371",
    archivePrefix  = "arXiv",
    primaryClass   = "hep-ex",
}

@Article{TOPQ-2012-15,
    author         = "{ATLAS Collaboration}",
    title          = "{Search for resonances decaying into top-quark pairs using fully hadronic decays in \(pp\) collisions with ATLAS at \(\sqrt{s} = 7\,\text{TeV}\)}",
    journal        = "JHEP",
    volume         = "01",
    year           = "2013",
    pages          = "116",
    doi            = "10.1007/JHEP01(2013)116",
    reportNumber   = "CERN-PH-EP-2012-291",
    eprint         = "1211.2202",
    archivePrefix  = "arXiv",
    primaryClass   = "hep-ex",
}

@Article{EXOT-2013-18,
    author         = "{ATLAS Collaboration}",
    title          = "{Search for production of vector-like quark pairs and of four top quarks in the lepton-plus-jets final state in \(pp\) collisions at \(\sqrt{s} = 8\,\text{TeV}\) with the ATLAS detector}",
    journal        = "JHEP",
    volume         = "08",
    year           = "2015",
    pages          = "105",
    doi            = "10.1007/JHEP08(2015)105",
    reportNumber   = "CERN-PH-EP-2015-095",
    eprint         = "1505.04306",
    archivePrefix  = "arXiv",
    primaryClass   = "hep-ex",
}

@Article{TOPQ-2013-07,
    author         = "{ATLAS Collaboration}",
    title          = "{Differential top-antitop cross-section measurements as a function of observables constructed from final-state particles using \(pp\) collisions at \(\sqrt{s} = 7\,\text{TeV}\) in the ATLAS detector}",
    journal        = "JHEP",
    volume         = "06",
    year           = "2015",
    pages          = "100",
    doi            = "10.1007/JHEP06(2015)100",
    reportNumber   = "CERN-PH-EP-2014-295",
    eprint         = "1502.05923",
    archivePrefix  = "arXiv",
    primaryClass   = "hep-ex",
}

@Article{EXOT-2014-10,
    author         = "{ATLAS Collaboration}",
    title          = "{Search for pair production of a new heavy quark that decays into a \(W\) boson and a light quark in \(pp\) collisions at \(\sqrt{s} = 8\,\text{TeV}\) with the ATLAS detector}",
    journal        = "Phys. Rev. D",
    volume         = "92",
    year           = "2015",
    pages          = "112007",
    doi            = "10.1103/PhysRevD.92.112007",
    reportNumber   = "CERN-PH-EP-2015-212",
    eprint         = "1509.04261",
    archivePrefix  = "arXiv",
    primaryClass   = "hep-ex",
}

@Article{EXOT-2014-12,
    author         = "{ATLAS Collaboration}",
    title          = "{Search for single production of vector-like quarks decaying into \(Wb\) in \(pp\) collisions at \(\sqrt{s} = 8\,\text{TeV}\) with the ATLAS detector}",
    journal        = "Eur. Phys. J. C",
    volume         = "76",
    year           = "2016",
    pages          = "442",
    doi            = "10.1140/epjc/s10052-016-4281-8",
    reportNumber   = "CERN-PH-EP-2015-319",
    eprint         = "1602.05606",
    archivePrefix  = "arXiv",
    primaryClass   = "hep-ex",
}

@Article{EXOT-2014-13,
    author         = "{ATLAS Collaboration}",
    title          = "{Search for the production of single vector-like and excited quarks in the \(Wt\) final state in \(pp\) collisions at \(\sqrt{s} = 8\,\text{TeV}\) with the ATLAS detector}",
    journal        = "JHEP",
    volume         = "02",
    year           = "2016",
    pages          = "110",
    doi            = "10.1007/JHEP02(2016)110",
    reportNumber   = "CERN-PH-EP-2015-235",
    eprint         = "1510.02664",
    archivePrefix  = "arXiv",
    primaryClass   = "hep-ex",
}

@Article{PERF-2014-03,
    author         = "{ATLAS Collaboration}",
    title          = "{Performance of pile-up mitigation techniques for jets in \(pp\) collisions at \(\sqrt{s} = 8\,\text{TeV}\) using the ATLAS detector}",
    journal        = "Eur. Phys. J. C",
    volume         = "76",
    year           = "2016",
    pages          = "581",
    doi            = "10.1140/epjc/s10052-016-4395-z",
    reportNumber   = "CERN-PH-EP-2015-206",
    eprint         = "1510.03823",
    archivePrefix  = "arXiv",
    primaryClass   = "hep-ex",
}

@Article{TOPQ-2014-15,
    author         = "{ATLAS Collaboration}",
    title          = "{Measurement of the differential cross-section of highly boosted top quarks as a function of their transverse momentum in \(\sqrt{s} = 8\,\text{TeV}\) proton--proton collisions using the ATLAS detector}",
    journal        = "Phys. Rev. D",
    volume         = "93",
    year           = "2016",
    pages          = "032009",
    doi            = "10.1103/PhysRevD.93.032009",
    reportNumber   = "CERN-PH-EP-2015-237",
    eprint         = "1510.03818",
    archivePrefix  = "arXiv",
    primaryClass   = "hep-ex",
}

@Article{PERF-2015-09,
    author         = "{ATLAS Collaboration}",
    title          = "{Jet reconstruction and performance using particle flow with the ATLAS Detector}",
    journal        = "Eur. Phys. J. C",
    volume         = "77",
    year           = "2017",
    pages          = "466",
    doi            = "10.1140/epjc/s10052-017-5031-2",
    reportNumber   = "CERN-EP-2017-024",
    eprint         = "1703.10485",
    archivePrefix  = "arXiv",
    primaryClass   = "hep-ex",
}

@Article{PERF-2015-10,
    author         = "{ATLAS Collaboration}",
    title          = "{Muon reconstruction performance of the ATLAS detector in proton--proton collision data at \(\sqrt{s} = 13\,\text{TeV}\)}",
    journal        = "Eur. Phys. J. C",
    volume         = "76",
    year           = "2016",
    pages          = "292",
    doi            = "10.1140/epjc/s10052-016-4120-y",
    reportNumber   = "CERN-EP-2016-033",
    eprint         = "1603.05598",
    archivePrefix  = "arXiv",
    primaryClass   = "hep-ex",
}

@Article{TOPQ-2015-06,
    author         = "{ATLAS Collaboration}",
    title          = "{Measurements of top-quark pair differential cross-sections in the lepton+jets channel in \(pp\) collisions at \(\sqrt{s} = 8\,\text{TeV}\) using the ATLAS detector}",
    journal        = "Eur. Phys. J. C",
    volume         = "76",
    year           = "2016",
    pages          = "538",
    doi            = "10.1140/epjc/s10052-016-4366-4",
    reportNumber   = "CERN-PH-EP-2015-239",
    eprint         = "1511.04716",
    archivePrefix  = "arXiv",
    primaryClass   = "hep-ex",
}

@Article{TOPQ-2015-07,
    author         = "{ATLAS Collaboration}",
    title          = "{Measurement of top quark pair differential cross sections in the dilepton channel in \(pp\) collisions at \(\sqrt{s} = 7\) and \(8\,\text{TeV}\) with ATLAS}",
    journal        = "Phys. Rev. D",
    volume         = "94",
    year           = "2016",
    pages          = "092003",
    doi            = "10.1103/PhysRevD.94.092003",
    reportNumber   = "CERN-EP-2016-144",
    eprint         = "1607.07281",
    archivePrefix  = "arXiv",
    primaryClass   = "hep-ex",
}

@Article{EXOT-2016-13,
    author         = "{ATLAS Collaboration}",
    title          = "{Search for pair production of up-type vector-like quarks and for four-top-quark events in final states with multiple \(b\)-jets with the ATLAS detector}",
    journal        = "JHEP",
    volume         = "07",
    year           = "2018",
    pages          = "089",
    doi            = "10.1007/JHEP07(2018)089",
    reportNumber   = "CERN-EP-2018-031",
    eprint         = "1803.09678",
    archivePrefix  = "arXiv",
    primaryClass   = "hep-ex",
}

@Article{EXOT-2016-14,
    author         = "{ATLAS Collaboration}",
    title          = "{Search for pair production of heavy vector-like quarks decaying to high-\(p_{\text{T}}\) \(W\) bosons and \(b\) quarks in the lepton-plus-jets final state in \(pp\) collisions at \(\sqrt{s} = 13\,\text{TeV}\) with the ATLAS detector}",
    journal        = "JHEP",
    volume         = "10",
    year           = "2017",
    pages          = "141",
    doi            = "10.1007/JHEP10(2017)141",
    reportNumber   = "CERN-EP-2017-094",
    eprint         = "1707.03347",
    archivePrefix  = "arXiv",
    primaryClass   = "hep-ex",
}

@Article{EXOT-2016-15,
    author         = "{ATLAS Collaboration}",
    title          = "{Search for pair production of vector-like top quarks in events with one lepton, jets, and missing transverse momentum in \(\sqrt{s} = 13\,\text{TeV}\) \(pp\) collisions with the ATLAS detector}",
    journal        = "JHEP",
    volume         = "08",
    year           = "2017",
    pages          = "052",
    doi            = "10.1007/JHEP08(2017)052",
    reportNumber   = "CERN-EP-2017-075",
    eprint         = "1705.10751",
    archivePrefix  = "arXiv",
    primaryClass   = "hep-ex",
}

@Article{TOPQ-2016-01,
    author         = "{ATLAS Collaboration}",
    title          = "{Measurements of top-quark pair differential cross-sections in the lepton+jets channel in \(pp\) collisions at \(\sqrt{s} = 13\,\text{TeV}\) using the ATLAS detector}",
    journal        = "JHEP",
    volume         = "11",
    year           = "2017",
    pages          = "191",
    doi            = "10.1007/JHEP11(2017)191",
    reportNumber   = "CERN-EP-2017-058",
    eprint         = "1708.00727",
    archivePrefix  = "arXiv",
    primaryClass   = "hep-ex",
}

@Article{TOPQ-2016-04,
    author         = "{ATLAS Collaboration}",
    title          = "{Measurements of top-quark pair differential cross-sections in the \(e\mu\) channel in \(pp\) collisions at \(\sqrt{s} = 13\,\text{TeV}\) using the ATLAS detector}",
    journal        = "Eur. Phys. J. C",
    volume         = "77",
    year           = "2017",
    pages          = "292",
    doi            = "10.1140/epjc/s10052-017-4821-x",
    eprint         = "1612.05220",
    archivePrefix  = "arXiv",
    primaryClass   = "hep-ex",
}

@Article{TOPQ-2016-09,
    author         = "{ATLAS Collaboration}",
    title          = "{Measurements of \(t\bar{t}\) differential cross-sections of highly boosted top quarks decaying to all-hadronic final states in \(pp\) collisions at \(\sqrt{s} = 13\,\text{TeV}\) using the ATLAS detector}",
    journal        = "Phys. Rev. D",
    volume         = "98",
    year           = "2018",
    pages          = "012003",
    doi            = "10.1103/PhysRevD.98.012003",
    reportNumber   = "CERN-EP-2017-226",
    eprint         = "1801.02052",
    archivePrefix  = "arXiv",
    primaryClass   = "hep-ex",
}

@Article{TRIG-2016-01,
    author         = "{ATLAS Collaboration}",
    title          = "{Performance of the ATLAS trigger system in 2015}",
    journal        = "Eur. Phys. J. C",
    volume         = "77",
    year           = "2017",
    pages          = "317",
    doi            = "10.1140/epjc/s10052-017-4852-3",
    reportNumber   = "CERN-EP-2016-241",
    eprint         = "1611.09661",
    archivePrefix  = "arXiv",
    primaryClass   = "hep-ex",
}

@Article{EXOT-2017-17,
    author         = "{ATLAS Collaboration}",
    title          = "{Combination of the searches for pair-produced vector-like partners of the third-generation quarks at \(\sqrt{s} = 13\,\text{TeV}\) with the ATLAS detector}",
    journal        = "Phys. Rev. Lett.",
    volume         = "121",
    year           = "2018",
    pages          = "211801",
    doi            = "10.1103/PhysRevLett.121.211801",
    reportNumber   = "CERN-EP-2018-205",
    eprint         = "1808.02343",
    archivePrefix  = "arXiv",
    primaryClass   = "hep-ex",
}

@Article{EXOT-2017-34,
    author         = "{ATLAS Collaboration}",
    title          = "{Search for pair production of heavy vector-like quarks decaying into high-\(p_{\text{T}}\) \(W\) bosons and top quarks in the lepton-plus-jets final state in \(pp\) collisions at \(\sqrt{s} = 13\,\text{TeV}\) with the ATLAS detector}",
    journal        = "JHEP",
    volume         = "08",
    year           = "2018",
    pages          = "048",
    doi            = "10.1007/JHEP08(2018)048",
    reportNumber   = "CERN-EP-2018-088",
    eprint         = "1806.01762",
    archivePrefix  = "arXiv",
    primaryClass   = "hep-ex",
}

@Article{EGAM-2018-01,
    author         = "{ATLAS Collaboration}",
    title          = "{Electron and photon performance measurements with the ATLAS detector using the 2015--2017 LHC proton--proton collision data}",
    journal        = "JINST",
    volume         = "14",
    year           = "2019",
    pages          = "P12006",
    doi            = "10.1088/1748-0221/14/12/P12006",
    reportNumber   = "CERN-EP-2019-145",
    eprint         = "1908.00005",
    archivePrefix  = "arXiv",
    primaryClass   = "hep-ex",
}

@Article{FTAG-2018-01,
    author         = "{ATLAS Collaboration}",
    title          = "{ATLAS \(b\)-jet identification performance and efficiency measurement with \(t\bar{t}\) events in \(pp\) collisions at \(\sqrt{s} = 13\,\text{TeV}\)}",
    journal        = "Eur. Phys. J. C",
    volume         = "79",
    year           = "2019",
    pages          = "970",
    doi            = "10.1140/epjc/s10052-019-7450-8",
    reportNumber   = "CERN-EP-2019-132",
    eprint         = "1907.05120",
    archivePrefix  = "arXiv",
    primaryClass   = "hep-ex",
}

@Article{HIGG-2018-57,
    author         = "{ATLAS Collaboration}",
    title          = "{Combined measurements of Higgs boson production and decay using up to \(80\,\text{fb}^{-1}\) of proton--proton collision data at \(\sqrt{s} = 13\,\text{TeV}\) collected with the ATLAS experiment}",
    journal        = "Phys. Rev. D",
    volume         = "101",
    year           = "2020",
    pages          = "012002",
    doi            = "10.1103/PhysRevD.101.012002",
    reportNumber   = "CERN-EP-2019-097",
    eprint         = "1909.02845",
    archivePrefix  = "arXiv",
    primaryClass   = "hep-ex",
}

@Article{JETM-2018-02,
    author         = "{ATLAS Collaboration}",
    title          = "{In situ calibration of large-radius jet energy and mass in \(13\,\text{TeV}\) proton--proton collisions with the ATLAS detector}",
    journal        = "Eur. Phys. J. C",
    volume         = "79",
    year           = "2019",
    pages          = "135",
    doi            = "10.1140/epjc/s10052-019-6632-8",
    reportNumber   = "CERN-EP-2018-191",
    eprint         = "1807.09477",
    archivePrefix  = "arXiv",
    primaryClass   = "hep-ex",
}

@Article{JETM-2018-03,
    author         = "{ATLAS Collaboration}",
    title          = "{Performance of top-quark and \(W\)-boson tagging with ATLAS in Run~2 of the LHC}",
    journal        = "Eur. Phys. J. C",
    volume         = "79",
    year           = "2019",
    pages          = "375",
    doi            = "10.1140/epjc/s10052-019-6847-8",
    reportNumber   = "CERN-EP-2018-192",
    eprint         = "1808.07858",
    archivePrefix  = "arXiv",
    primaryClass   = "hep-ex",
}

@Article{MUON-2018-03,
    author         = "{ATLAS Collaboration}",
    title          = "{Muon reconstruction and identification efficiency in ATLAS using the full Run~2 \(pp\) collision data set at \(\sqrt{s} = 13\,\text{TeV}\)}",
    journal        = "Eur. Phys. J. C",
    volume         = "81",
    year           = "2020",
    pages          = "578",
    doi            = "10.1140/epjc/s10052-021-09233-2",
    reportNumber   = "CERN Preprint ID: CERN-EP-2020-199",
    eprint         = "2012.00578",
    archivePrefix  = "arXiv",
    primaryClass   = "hep-ex",
}

@Article{STDM-2018-34,
    author         = "{ATLAS Collaboration}",
    title          = "{Measurements of \(W^{+}W^{-} + \ge 1\) jet production cross-sections in \(pp\) collisions at \(\sqrt{s} = 13\,\text{TeV}\) with the ATLAS detector}",
    journal        = "JHEP",
    volume         = "06",
    year           = "2021",
    pages          = "003",
    doi            = "10.1007/JHEP06(2021)003",
    reportNumber   = "CERN-EP-2021-030",
    eprint         = "2103.10319",
    archivePrefix  = "arXiv",
    primaryClass   = "hep-ex",
}

@Booklet{ATL-SOFT-PUB-2021-001,
    author         = "{ATLAS Collaboration}",
    title          = "{The ATLAS Collaboration Software and Firmware}",
    howpublished   = "{ATL-SOFT-PUB-2021-001}",
    url            = "https://cds.cern.ch/record/2767187",
    year           = "2021",
}

@Booklet{ATL-SOFT-PUB-2021-003,
    author         = "{ATLAS Collaboration}",
    title          = "{ATLAS Computing Acknowledgements}",
    howpublished   = "{ATL-SOFT-PUB-2021-003}",
    url            = "https://cds.cern.ch/record/2776662",
}

@Article{CMS-EXO-11-005,
    author         = "{CMS Collaboration}",
    title          = "{Search for a Vectorlike Quark with Charge 2/3 in \(t\) + \(Z\) Events from \(pp\) Collisions at \(\sqrt{s} = 7\,\text{TeV}\)}",
    journal        = "Phys. Rev. Lett.",
    volume         = "107",
    year           = "2011",
    pages          = "271802",
    doi            = "10.1103/PhysRevLett.107.271802",
    reportNumber   = "CERN-PH-EP-2011-139",
    eprint         = "1109.4985",
    archivePrefix  = "arXiv",
    primaryClass   = "hep-ex",
}

@Article{CMS-EXO-11-099,
    author         = "{CMS Collaboration}",
    title          = "{Search for pair produced fourth-generation up-type quarks in \(pp\) collisions at \(\sqrt{s} = 7\,\text{TeV}\) with a lepton in the final state}",
    journal        = "Phys. Lett. B",
    volume         = "718",
    year           = "2012",
    pages          = "307",
    doi            = "10.1016/j.physletb.2012.10.038",
    reportNumber   = "CERN-PH-EP-2012-244",
    eprint         = "1209.0471",
    archivePrefix  = "arXiv",
    primaryClass   = "hep-ex",
}

@Article{CMS-TOP-11-013,
    author         = "{CMS Collaboration}",
    title          = "{Measurement of differential top-quark-pair production cross sections in \(pp\) colisions at \(\sqrt{s} = 7\,\text{TeV}\)}",
    journal        = "Eur. Phys. J. C",
    volume         = "73",
    year           = "2013",
    pages          = "2339",
    doi            = "10.1140/epjc/s10052-013-2339-4",
    reportNumber   = "CERN-PH-EP-2012-322",
    eprint         = "1211.2220",
    archivePrefix  = "arXiv",
    primaryClass   = "hep-ex",
}

@Article{CMS-B2G-12-015,
    author         = "{CMS Collaboration}",
    title          = "{Inclusive search for a vector-like \(T\) quark with charge \(\frac{2}{3}\) in \(pp\) collisions at \(\sqrt{s} = 8\,\text{TeV}\)}",
    journal        = "Phys. Lett. B",
    volume         = "729",
    year           = "2014",
    pages          = "149",
    doi            = "10.1016/j.physletb.2014.01.006",
    reportNumber   = "CERN-PH-EP-2013-215",
    eprint         = "1311.7667",
    archivePrefix  = "arXiv",
    primaryClass   = "hep-ex",
}

@Article{CMS-HIG-12-028,
    author         = "{CMS Collaboration}",
    title          = "{Observation of a new boson at a mass of 125 GeV with the CMS experiment at the LHC}",
    journal        = "Phys. Lett. B",
    volume         = "716",
    year           = "2012",
    pages          = "30",
    doi            = "10.1016/j.physletb.2012.08.021",
    reportNumber   = "CERN-PH-EP-2012-220",
    eprint         = "1207.7235",
    archivePrefix  = "arXiv",
    primaryClass   = "hep-ex",
}

@Article{CMS-TOP-12-028,
    author         = "{CMS Collaboration}",
    title          = "{Measurement of the differential cross section for top quark pair production in \(pp\) collisions at \(\sqrt{s} = 8\,\text{TeV}\)}",
    journal        = "Eur. Phys. J. C",
    volume         = "75",
    year           = "2015",
    pages          = "542",
    doi            = "10.1140/epjc/s10052-015-3709-x",
    reportNumber   = "CERN-PH-EP-2015-117",
    eprint         = "1505.04480",
    archivePrefix  = "arXiv",
    primaryClass   = "hep-ex",
}

@Article{CMS-B2G-13-005,
    author         = "{CMS Collaboration}",
    title          = "{Search for vectorlike charge 2/3 \(T\) quarks in proton--proton collisions at \(\sqrt{(s)} = 8\,\text{TeV}\)}",
    journal        = "Phys. Rev. D",
    volume         = "93",
    year           = "2016",
    pages          = "012003",
    doi            = "10.1103/PhysRevD.93.012003",
    reportNumber   = "CERN-PH-EP-2015-228",
    eprint         = "1509.04177",
    archivePrefix  = "arXiv",
    primaryClass   = "hep-ex",
}

@Article{CMS-B2G-13-006,
    author         = "{CMS Collaboration}",
    title          = "{Search for pair-produced vectorlike \(B\) quarks in proton--proton collisions at \(\sqrt{s} = 8\,\text{TeV}\)}",
    journal        = "Phys. Rev. D",
    volume         = "93",
    year           = "2016",
    pages          = "112009",
    doi            = "10.1103/PhysRevD.93.112009",
    reportNumber   = "CERN-PH-EP-2015-170",
    eprint         = "1507.07129",
    archivePrefix  = "arXiv",
    primaryClass   = "hep-ex",
}

@Article{CMS-B2G-14-002,
    author         = "{CMS Collaboration}",
    title          = "{Search for vector-like \(T\) quarks decaying to top quarks and Higgs bosons in the all-hadronic channel using jet substructure}",
    journal        = "JHEP",
    volume         = "06",
    year           = "2015",
    pages          = "080",
    doi            = "10.1007/JHEP06(2015)080",
    reportNumber   = "CERN-PH-EP-2015-034",
    eprint         = "1503.01952",
    archivePrefix  = "arXiv",
    primaryClass   = "hep-ex",
}

@Article{CMS-TOP-14-008,
    author         = "{CMS Collaboration}",
    title          = "{Measurement of the semileptonic \(t\bar{t} + \gamma\) production cross section in \(pp\) collisions at \(\sqrt{s} = 8\,\text{TeV}\)}",
    journal        = "JHEP",
    volume         = "10",
    year           = "2017",
    pages          = "006",
    doi            = "10.1007/JHEP10(2017)006",
    reportNumber   = "CERN-EP-2017-112",
    eprint         = "1706.08128",
    archivePrefix  = "arXiv",
    primaryClass   = "hep-ex",
}

@Article{CMS-TOP-14-012,
    author         = "{CMS Collaboration}",
    title          = "{Measurement of the integrated and differential \(t\bar{t}\) production cross sections for high-\(p_{\text{T}}\) top quarks in \(pp\) collisions at \(\sqrt{s} = 8\,\text{TeV}\)}",
    journal        = "Phys. Rev. D",
    volume         = "94",
    year           = "2016",
    pages          = "072002",
    doi            = "10.1103/PhysRevD.94.072002",
    reportNumber   = "CERN-EP-2016-078",
    eprint         = "1605.00116",
    archivePrefix  = "arXiv",
    primaryClass   = "hep-ex",
}

@Article{CMS-TOP-14-013,
    author         = "{CMS Collaboration}",
    title          = "{Measurement of double-differential cross sections for top quark pair production in \(pp\) collisions at \(\sqrt{s} = 8\,\text{TeV}\) and impact on parton distribution functions}",
    journal        = "Eur. Phys. J. C",
    volume         = "77",
    year           = "2017",
    pages          = "459",
    doi            = "10.1140/epjc/s10052-017-4984-5",
    reportNumber   = "CERN-EP-2017-030",
    eprint         = "1703.01630",
    archivePrefix  = "arXiv",
    primaryClass   = "hep-ex",
}

@Article{CMS-B2G-15-008,
    author         = "{CMS Collaboration}",
    title          = "{Search for single production of a heavy vector-like \(T\) quark decaying to a Higgs boson and a top quark with a lepton and jets in the final state}",
    journal        = "Phys. Lett. B",
    volume         = "771",
    year           = "2017",
    pages          = "80",
    doi            = "10.1016/j.physletb.2017.05.019",
    reportNumber   = "CERN-EP-2016-279",
    eprint         = "1612.00999",
    archivePrefix  = "arXiv",
    primaryClass   = "hep-ex",
}

@Article{CMS-B2G-16-001,
    author         = "{CMS Collaboration}",
    title          = "{Search for single production of vector-like quarks decaying to a \(Z\) boson and a top or a bottom quark in proton--proton collisions at \(\sqrt{s} = 13\,\text{TeV}\)}",
    journal        = "JHEP",
    volume         = "05",
    year           = "2017",
    pages          = "029",
    doi            = "10.1007/JHEP05(2017)029",
    reportNumber   = "CERN-EP-2016-326",
    eprint         = "1701.07409",
    archivePrefix  = "arXiv",
    primaryClass   = "hep-ex",
}

@Article{CMS-B2G-16-005,
    author         = "{CMS Collaboration}",
    title          = "{Search for electroweak production of a vector-like quark decaying to a top quark and a Higgs boson using boosted topologies in fully hadronic final states}",
    journal        = "JHEP",
    volume         = "04",
    year           = "2017",
    pages          = "136",
    doi            = "10.1007/JHEP04(2017)136",
    reportNumber   = "CERN-EP-2016-290",
    eprint         = "1612.05336",
    archivePrefix  = "arXiv",
    primaryClass   = "hep-ex",
}

@Article{CMS-B2G-16-013,
    author         = "{CMS Collaboration}",
    title          = "{Search for a heavy resonance decaying to a top quark and a vector-like top quark at \(\sqrt{s} = 13\,\text{TeV}\)}",
    journal        = "JHEP",
    volume         = "09",
    year           = "2017",
    pages          = "053",
    doi            = "10.1007/JHEP09(2017)053",
    reportNumber   = "CERN-EP-2017-035",
    eprint         = "1703.06352",
    archivePrefix  = "arXiv",
    primaryClass   = "hep-ex",
}

@Article{CMS-B2G-16-024,
    author         = "{CMS Collaboration}",
    title          = "{Search for pair production of vector-like \(T\) and \(B\) quarks in single-lepton final states using boosted jet substructure techniques at \(\sqrt{s} = 13\,\text{TeV}\)}",
    journal        = "JHEP",
    volume         = "11",
    year           = "2017",
    pages          = "085",
    doi            = "10.1007/JHEP11(2017)085",
    reportNumber   = "CERN-EP-2017-107",
    eprint         = "1706.03408",
    archivePrefix  = "arXiv",
    primaryClass   = "hep-ex",
}

@Article{CMS-TOP-16-007,
    author         = "{CMS Collaboration}",
    title          = "{Measurement of normalized differential \(t\bar{t}\) cross sections in the dilepton channel from \(pp\) collisions at \(\sqrt{s} = 13\,\text{TeV}\)}",
    journal        = "JHEP",
    volume         = "04",
    year           = "2018",
    pages          = "060",
    doi            = "10.1007/JHEP04(2018)060",
    reportNumber   = "CERN-EP-2017-120",
    eprint         = "1708.07638",
    archivePrefix  = "arXiv",
    primaryClass   = "hep-ex",
}

@Article{CMS-TOP-16-008,
    author         = "{CMS Collaboration}",
    title          = "{Measurement of differential cross sections for top quark pair production using the lepton+jets final state in proton--proton collisions at \(13\,\text{TeV}\)}",
    journal        = "Phys. Rev. D",
    volume         = "95",
    year           = "2017",
    pages          = "092001",
    doi            = "10.1103/PhysRevD.95.092001",
    reportNumber   = "CERN-EP-2016-227",
    eprint         = "1610.04191",
    archivePrefix  = "arXiv",
    primaryClass   = "hep-ex",
}

@Article{CMS-TOP-16-014,
    author         = "{CMS Collaboration}",
    title          = "{Measurements of differential cross sections of top quark pair production as a function of kinematic event variables in proton--proton collisions at \(\sqrt{s} = 13\,\text{TeV}\)}",
    journal        = "JHEP",
    volume         = "06",
    year           = "2018",
    pages          = "002",
    doi            = "10.1007/JHEP06(2018)002",
    reportNumber   = "CERN-EP-2018-013",
    eprint         = "1803.03991",
    archivePrefix  = "arXiv",
    primaryClass   = "hep-ex",
}

@Article{CMS-B2G-17-003,
    author         = "{CMS Collaboration}",
    title          = "{Search for pair production of vector-like quarks in the \(bW\bar{b}W\) channel from proton--proton collisions at \(\sqrt{s} = 13\,\text{TeV}\)}",
    journal        = "Phys. Lett. B",
    volume         = "779",
    year           = "2018",
    pages          = "82",
    doi            = "10.1016/j.physletb.2018.01.077",
    reportNumber   = "CERN-EP-2017-224",
    eprint         = "1710.01539",
    archivePrefix  = "arXiv",
    primaryClass   = "hep-ex",
}

@Article{CMS-B2G-17-007,
    author         = "{CMS Collaboration}",
    title          = "{Search for single production of a vector-like \(T\) quark decaying to a \(Z\) boson and a top quark in proton--proton collisions at \(\sqrt{s} = 13\,\text{TeV}\)}",
    journal        = "Phys. Lett. B",
    volume         = "781",
    year           = "2018",
    pages          = "574",
    doi            = "10.1016/j.physletb.2018.04.036",
    reportNumber   = "CERN-EP-2017-155",
    eprint         = "1708.01062",
    archivePrefix  = "arXiv",
    primaryClass   = "hep-ex",
}

@Article{CMS-B2G-17-011,
    author         = "{CMS Collaboration}",
    title          = "{Search for vector-like \(T\) and \(B\) quark pairs in final states with leptons at \(\sqrt{s} = 13\,\text{TeV}\)}",
    journal        = "JHEP",
    volume         = "08",
    year           = "2018",
    pages          = "177",
    doi            = "10.1007/JHEP08(2018)177",
    reportNumber   = "CERN-EP-2018-069",
    eprint         = "1805.04758",
    archivePrefix  = "arXiv",
    primaryClass   = "hep-ex",
}

@Article{CMS-B2G-17-012,
    author         = "{CMS Collaboration}",
    title          = "{Search for vector-like quarks in events with two oppositely charged leptons and jets in proton--proton collisions at \(\sqrt{s} = 13\,\text{TeV}\)}",
    journal        = "Eur. Phys. J. C",
    volume         = "79",
    year           = "2019",
    pages          = "364",
    doi            = "10.1140/epjc/s10052-019-6855-8",
    reportNumber   = "CERN-EP-2018-290",
    eprint         = "1812.09768",
    archivePrefix  = "arXiv",
    primaryClass   = "hep-ex",
}

@Article{CMS-B2G-17-018,
    author         = "{CMS Collaboration}",
    title          = "{Search for single production of vector-like quarks decaying to a top quark and a \(W\) boson in proton--proton collisions at \(\sqrt{s} = 13\,\text{TeV}\)}",
    journal        = "Eur. Phys. J. C",
    volume         = "79",
    year           = "2019",
    pages          = "90",
    doi            = "10.1140/epjc/s10052-019-6556-3",
    reportNumber   = "CERN-EP-2018-230",
    eprint         = "1809.08597",
    archivePrefix  = "arXiv",
    primaryClass   = "hep-ex",
}

@Article{CMS-HIG-17-031,
    author         = "{CMS Collaboration}",
    title          = "{Combined measurements of Higgs boson couplings in proton--proton collisions at \(\sqrt{s} = 13\,\text{TeV}\)}",
    journal        = "Eur. Phys. J. C",
    volume         = "79",
    year           = "2019",
    pages          = "421",
    doi            = "10.1140/epjc/s10052-019-6909-y",
    reportNumber   = "CERN-EP-2018-263",
    eprint         = "1809.10733",
    archivePrefix  = "arXiv",
    primaryClass   = "hep-ex",
}

@Article{CMS-TOP-17-002,
    author         = "{CMS Collaboration}",
    title          = "{Measurement of differential cross sections for the production of top quark pairs and of additional jets in lepton+jets events from \(pp\) collisions at \(\sqrt{s} = 13\,\text{TeV}\)}",
    journal        = "Phys. Rev. D",
    volume         = "97",
    year           = "2018",
    pages          = "112003",
    doi            = "10.1103/PhysRevD.97.112003",
    reportNumber   = "CERN-EP-2018-039",
    eprint         = "1803.08856",
    archivePrefix  = "arXiv",
    primaryClass   = "hep-ex",
}

@Article{CMS-B2G-18-003,
    author         = "{CMS Collaboration}",
    title          = "{Search for electroweak production of a vector-like \(T\) quark using fully hadronic final states}",
    journal        = "JHEP",
    volume         = "01",
    year           = "2020",
    pages          = "036",
    doi            = "10.1007/JHEP01(2020)036",
    reportNumber   = "CERN-EP-2019-174",
    eprint         = "1909.04721",
    archivePrefix  = "arXiv",
    primaryClass   = "hep-ex",
}

@Article{CMS-TOP-18-004,
    author         = "{CMS Collaboration}",
    title          = "{Measurement of \(t\bar{t}\) normalised multi-differential cross sections in \(pp\) collisions at \(\sqrt{s} = 13\,\text{TeV}\), and simultaneous determination of the strong coupling strength, top quark pole mass, and parton distribution functions}",
    journal        = "Eur. Phys. J. C",
    volume         = "80",
    year           = "2020",
    pages          = "658",
    doi            = "10.1140/epjc/s10052-020-7917-7",
    reportNumber   = "CERN-EP-2019-028",
    eprint         = "1904.05237",
    archivePrefix  = "arXiv",
    primaryClass   = "hep-ex",
}

@Article{CMS-TOP-18-013,
    author         = "{CMS Collaboration}",
    title          = "{Measurement of differential \(t\bar{t}\) production cross sections using top quarks at large transverse momenta in \(pp\) collisions at \(\sqrt{s} = 13\,\text{TeV}\)}",
    journal        = "Phys. Rev. D",
    volume         = "103",
    year           = "2021",
    pages          = "052008",
    doi            = "10.1103/PhysRevD.103.052008",
    reportNumber   = "CERN-EP-2020-121",
    eprint         = "2008.07860",
    archivePrefix  = "arXiv",
    primaryClass   = "hep-ex",
}

@Article{CMS-HIG-19-002,
    author         = "{CMS Collaboration}",
    title          = "{Measurement of the inclusive and differential Higgs boson production cross sections in the leptonic \(WW\) decay mode at \(\sqrt{s} = 13\,\text{TeV}\)}",
    journal        = "JHEP",
    volume         = "03",
    year           = "2021",
    pages          = "003",
    doi            = "10.1007/JHEP03(2021)003",
    reportNumber   = "CERN-EP-2020-106",
    eprint         = "2007.01984",
    archivePrefix  = "arXiv",
    primaryClass   = "hep-ex",
}

@Booklet{ATLAS-CONF-2019-021,
    author         = "{ATLAS Collaboration}",
    title          = "{Luminosity determination in \(pp\) collisions at \(\sqrt{s} = 13\,\text{TeV}\) using the ATLAS detector at the LHC}",
    howpublished   = "{ATLAS-CONF-2019-021}",
    url            = "https://cds.cern.ch/record/2677054",
    year           = "2019",
}

@Booklet{ATLAS-CONF-2020-022,
    author         = "{ATLAS Collaboration}",
    title          = "{Measurement of the ATLAS Detector Jet Mass Response using Forward Folding with \(80\,\text{fb}^{-1}\) of \(\sqrt{s} = 13\,\text{TeV}\) \(pp\) data}",
    howpublished   = "{ATLAS-CONF-2020-022}",
    url            = "https://cds.cern.ch/record/2724442",
    year           = "2020",
}

@Booklet{ATL-PHYS-PUB-2015-026,
    author         = "{ATLAS Collaboration}",
    title          = "{Vertex Reconstruction Performance of the ATLAS Detector at \(\sqrt{s} = 13~\text{TeV}\)}",
    howpublished   = "{ATL-PHYS-PUB-2015-026}",
    url            = "https://cds.cern.ch/record/2037717",
    year           = "2015",
}

@Booklet{ATL-PHYS-PUB-2016-017,
    author         = "{ATLAS Collaboration}",
    title          = "{The Pythia~8 A3 tune description of ATLAS minimum bias and inelastic measurements incorporating the Donnachie--Landshoff diffractive model}",
    howpublished   = "{ATL-PHYS-PUB-2016-017}",
    url            = "https://cds.cern.ch/record/2206965",
    year           = "2016",
}

@Booklet{ATL-PHYS-PUB-2016-020,
    author         = "{ATLAS Collaboration}",
    title          = "{Studies on top-quark Monte Carlo modelling for Top2016}",
    howpublished   = "{ATL-PHYS-PUB-2016-020}",
    url            = "https://cds.cern.ch/record/2216168",
    year           = "2016",
}

@Booklet{ATL-PHYS-PUB-2020-017,
    author         = "{ATLAS Collaboration}",
    title          = "{Boosted hadronic vector boson and top quark tagging with ATLAS using Run 2 data}",
    howpublished   = "{ATL-PHYS-PUB-2020-017}",
    url            = "https://cds.cern.ch/record/2724149",
    year           = "2020",
}
 
\clearpage
 
\begin{flushleft}
\hypersetup{urlcolor=black}
{\Large The ATLAS Collaboration}

\bigskip

\AtlasOrcid[0000-0002-6665-4934]{G.~Aad}$^\textrm{\scriptsize 100}$,    
\AtlasOrcid[0000-0002-5888-2734]{B.~Abbott}$^\textrm{\scriptsize 126}$,    
\AtlasOrcid[0000-0002-7248-3203]{D.C.~Abbott}$^\textrm{\scriptsize 101}$,    
\AtlasOrcid[0000-0002-2788-3822]{A.~Abed~Abud}$^\textrm{\scriptsize 36}$,    
\AtlasOrcid[0000-0002-1002-1652]{K.~Abeling}$^\textrm{\scriptsize 53}$,    
\AtlasOrcid[0000-0002-2987-4006]{D.K.~Abhayasinghe}$^\textrm{\scriptsize 93}$,    
\AtlasOrcid[0000-0002-8496-9294]{S.H.~Abidi}$^\textrm{\scriptsize 29}$,    
\AtlasOrcid[0000-0002-9987-2292]{A.~Aboulhorma}$^\textrm{\scriptsize 35e}$,    
\AtlasOrcid[0000-0001-5329-6640]{H.~Abramowicz}$^\textrm{\scriptsize 158}$,    
\AtlasOrcid[0000-0002-1599-2896]{H.~Abreu}$^\textrm{\scriptsize 157}$
% ...
\bigskip
\\

% ...
$^{29}$Physics Department, Brookhaven National Laboratory, Upton NY; United States of America.\\
% ...
$^{35}$$^{(a)}$Facult\'e des Sciences Ain Chock, R\'eseau Universitaire de Physique des Hautes Energies - Universit\'e Hassan II, Casablanca;$^{(b)}$Facult\'{e} des Sciences, Universit\'{e} Ibn-Tofail, K\'{e}nitra;$^{(c)}$Facult\'e des Sciences Semlalia, Universit\'e Cadi Ayyad, LPHEA-Marrakech;$^{(d)}$LPMR, Facult\'e des Sciences, Universit\'e Mohamed Premier, Oujda;$^{(e)}$Facult\'e des sciences, Universit\'e Mohammed V, Rabat;$^{(f)}$Mohammed VI Polytechnic University, Ben Guerir; Morocco.\\
$^{36}$CERN, Geneva; Switzerland.\\
% ...
$^{53}$II. Physikalisches Institut, Georg-August-Universit\"{a}t G\"ottingen, G\"ottingen; Germany.\\
% ...
$^{93}$Department of Physics, Royal Holloway University of London, Egham; United Kingdom.\\
% ...
$^{100}$CPPM, Aix-Marseille Universit\'e, CNRS/IN2P3, Marseille; France.\\
$^{101}$Department of Physics, University of Massachusetts, Amherst MA; United States of America.\\
% ...
$^{126}$Homer L. Dodge Department of Physics and Astronomy, University of Oklahoma, Norman OK; United States of America.\\
% ...
$^{157}$Department of Physics, Technion, Israel Institute of Technology, Haifa; Israel.\\
$^{158}$Raymond and Beverly Sackler School of Physics and Astronomy, Tel Aviv University, Tel Aviv; Israel.\\
% ...
$^{a}$ Also at Borough of Manhattan Community College, City University of New York, New York NY; United States of America.\\
$^{b}$ Also at Bruno Kessler Foundation, Trento; Italy.\\
$^{c}$ Also at Center for High Energy Physics, Peking University; China.\\
$^{d}$ Also at Centro Studi e Ricerche Enrico Fermi; Italy.\\
$^{e}$ Also at CERN, Geneva; Switzerland.\\
$^{f}$ Also at D\'epartement de Physique Nucl\'eaire et Corpusculaire, Universit\'e de Gen\`eve, Gen\`eve; Switzerland.\\
% ...

\end{flushleft}


\end{document}